\documentclass{article}

\usepackage{dilab_arxiv}

\usepackage{amsfonts}
\usepackage{mathtools}
\usepackage{multirow}

\usepackage[capitalise,nameinlink]{cleveref}

\graphicspath{{figures/}}

\theoremstyle{plain}
\newtheorem{theorem}{Theorem}
\newtheorem{lemma}{Lemma}
\newtheorem{proposition}{Proposition}
\newtheorem{corollary}{Corollary}
\theoremstyle{definition}
\newtheorem{definition}{Definition}
\newtheorem{example}{Example}
\newtheorem{assumption}{Assumption}
\theoremstyle{remark}
\newtheorem{remark}{Remark}
\crefname{assumption}{Assumption}{Assumptions}

\newcommand{\Vul}{\mathrm{Gap}}          

\newcommand{\accTwoKuhnStd}{41\%}

\newcommand{\accTwoLeducStd}{86\%}

\newcommand{\accZeroKuhnAll}{0\%}

\newcommand{\accZeroLeducAll}{50\%}

\newcommand{\costCachedMax}{0.65}
\newcommand{\costCachedMin}{0.13}
\newcommand{\costUncachedMax}{1.52}
\newcommand{\costUncachedMin}{0.76}
\newcommand{\eFourCells}{40}

\newcommand{\eOneNashSafeVulHarm}{66}

\newcommand{\eOneVulSafeNashHarm}{0}
\newcommand{\fnTotal}{0}

\newcommand{\gatedMaxKuhnStd}{+0.0182}

\newcommand{\gatedMaxLeducStd}{+0.0135}

\newcommand{\harmKuhnBelief}{20\%}
\newcommand{\harmKuhnStd}{29\%}
\newcommand{\harmKuhnUnsafe}{80\%}
\newcommand{\harmLeducBelief}{2\%}
\newcommand{\harmLeducStd}{2\%}
\newcommand{\harmLeducUnsafe}{12\%}
\newcommand{\leducDecisionNodes}{38{,}641}
\newcommand{\leducInfosets}{1{,}134}
\newcommand{\nSixRecords}{96}
\newcommand{\nUpdKuhn}{168}
\newcommand{\nUpdLeduc}{1620}
\newcommand{\rakeThetaKuhnMax}{8.4\%}
\newcommand{\rakeThetaLeducMax}{0.7\%}

\newcommand{\shrinkHarmKuhnQuarter}{2\%}
\newcommand{\sixCertSec}{27.2}

\newcommand{\sixHarmStd}{36\%}
\newcommand{\sixHarmUnsafe}{100\%}
\newcommand{\sixResolveMin}{170}

\newcommand{\violSix}{0}
\newcommand{\violTotal}{0}

\newcommand{\dinf}{\delta_{\infty}}      
\newcommand{\mhat}{\hat{m}}              
\newcommand{\Mdef}[1]{M_{#1}}            
\newcommand{\Act}{A_i^{S}}               
\newcommand{\tax}{\Theta}                
\newcommand{\zprof}{\zeta_{\mathrm{prof}}} 
\newcommand{\zsub}{\zeta_{S}}            
\newcommand{\epsi}{\varepsilon_i}        
\newcommand{\bp}{\sigma^{0}}             
\newcommand{\rs}{\sigma^{1}}             

\title{Safe Subgame Resolving in Multiplayer Games}
\runningtitle{Safe Subgame Resolving in Multiplayer Games}
\date{\today}

\paperlogo{\includegraphics[height=1.5cm]{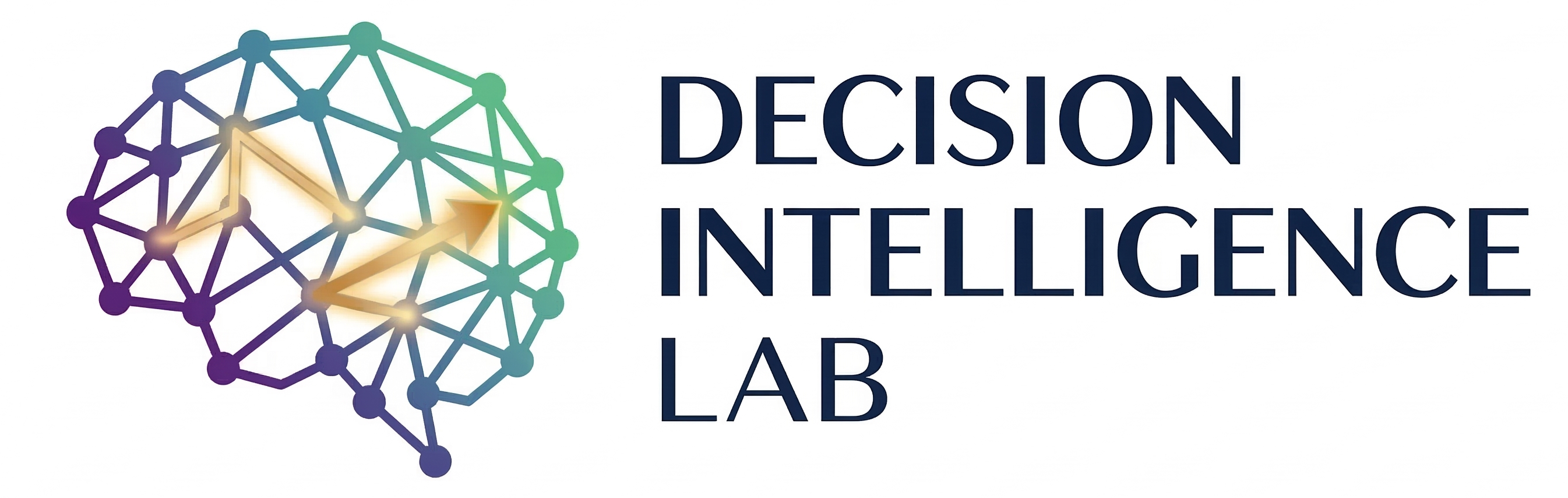}}

\author{
  Boning Li$^{1}$ and Longbo Huang$^{1\,\text{\faEnvelope}}$
  \\[0.3em]\normalfont
  $^1$Institute for Interdisciplinary Information Sciences, Tsinghua University
  \\
  \text{\faEnvelope}\ Correspondence: longbohuang@tsinghua.edu.cn
}

\hypersetup{
  pdftitle={Safe Subgame Resolving in Multiplayer Games},
  pdfauthor={Boning Li, Longbo Huang},
}

\begin{document}
\maketitle
\thispagestyle{fancy}

\begin{abstract}
Poker and other games with hidden information are too large to solve in one pass. Strong systems compute a complete \emph{blueprint} strategy offline and refine it in real time inside the reached \emph{subgame}. In two-player zero-sum games, a safety theorem guarantees that subgame resolving never lets an adversary gain more by exploiting the resolved strategy than by exploiting the blueprint. With three or more players acting independently (uncorrelated profiles), no such guarantee exists yet. Heuristic resolving can increase the profile's \emph{best-response gap}, the largest gain any single player obtains by deviating. We prove the first conditional safety theorem for uncorrelated multiplayer subgame resolving, giving a post-hoc certificate for any delivered update. The resolved profile's best-response gap exceeds the blueprint's by at most four terms. Two concern the resolver, namely the subgame value the blueprint promised without realizing ($\Act$) and the resolver's subgame solve error ($\epsi$). The other two are the opponents' best-response gains inside the subgame over their promised values ($\sum_q \Mdef{q}$) and a structural tax ($\tax$) measuring how much total player utility can change. The tax is zero in constant-sum games, including every rake-free poker variant. Constructions attain each term with equality. On three-player Leduc poker and Kuhn poker with up to six players, resolving without a certificate raises the gap beyond a small tolerance in \harmLeducUnsafe{} to \sixHarmUnsafe{} of the updates. The certificate flags every such update and accepts \accTwoKuhnStd{} of standard resolving updates on Kuhn and \accTwoLeducStd{} on Leduc. Keeping the blueprint whenever the certificate exceeds the tolerance delivers no harmful update. The certificate needs only one best-response pass per opponent inside the subgame, so real-time multiplayer solvers that refine a blueprint can attach it as a safety check. Code is available at \url{https://github.com/lbn187/MultiPlayerSolver}.

\end{abstract}

\section{Introduction}\label{sec:intro}

\begin{figure*}[t]
\centering
\includegraphics[width=\textwidth]{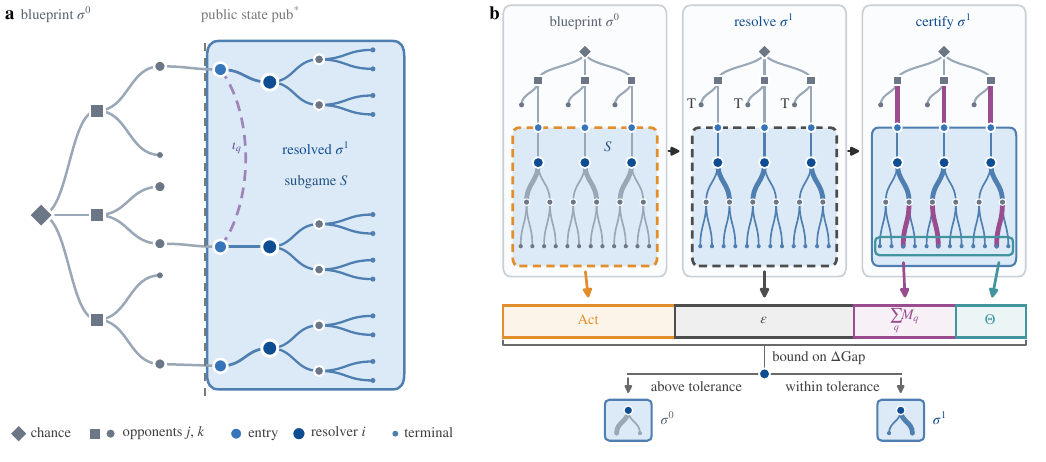}
\caption{Subgame resolving and its certified guarantee.
\textbf{(a)} Three-player game tree. The diamond is the chance node, squares
and small circles are decision nodes of the opponents $j$ and $k$, and dots
are terminals. The blueprint $\bp$ (CFR) is kept
outside the public state $\mathrm{pub}^{*}$. The resolver $i$ updates it to
$\rs$ inside the subgame $S$, where the entries $E$ are grouped into entry
information sets $\iota_q$. With two players, $j$'s gain inside $S$ is $i$'s
loss. With three, value can move between $j$ and $k$ (\cref{ex:crossopp}).
\textbf{(b)} The certificate on the same subgame. Thick edges are the actions
the resolver plays, grey under $\bp$ and blue under $\rs$. The blueprint
leaves the value $\Act(\bp)$ unrealized in $S$ (orange). The gadget solve,
where each opponent may terminate (T) for its blueprint value, has error
$\epsi$ (dark). The opponents' best responses to $\rs$ (purple) give
$\sum_q\Mdef{q}$, and the terminals' total utility (teal) gives $\tax$. These
four terms bound the rise of the gap, $\Delta\Vul$. The resolver plays $\rs$
when the bound is within the tolerance and keeps $\bp$ otherwise.}
\label{fig:method}
\end{figure*}

Imperfect-information extensive-form games (IIEFGs) model sequential
decision making with hidden information, from poker
\citep{bowling2015heads,brown2018superhuman} to Mahjong \citep{li2020suphx}
and Stratego \citep{perolat2022mastering}. Counterfactual regret
minimization (CFR) \citep{zinkevich2007regret} and its variants
\citep{lanctot2009monte,tammelin2014solving,brown2019solving,brown2019deep}
are the standard route to a two-player zero-sum equilibrium, but their cost
grows with the game tree. Large games are therefore solved once offline under
abstraction, a coarsening that groups similar states
\citep{brown2015hierarchical}. The resulting full-game strategy, the
\emph{blueprint}, is then refined in real time inside the subgame actually
reached \citep{burch2014solving,brown2018depth}, as in \cref{fig:method}(a).
In two-player zero-sum games, subgame resolving is safe
\citep{burch2014solving,moravcik2016refining,brown2017safe}. Refinement never
increases exploitability, the amount an adversary gains by best-responding.
This pipeline underpins DeepStack \citep{moravvcik2017deepstack}, Libratus
\citep{brown2018superhuman}, and Student of Games \citep{schmid2023student}.

Beyond two players, no such guarantee is known. Pluribus plays six-player poker using heuristic resolving without a safety guarantee \citep{brown2019superhuman}. Two-player safety relies on an opponent's gain equaling the resolver's loss \citep{von2007theory}. When the number of players $n$ exceeds $2$, this identity disappears and its per-opponent weakening also fails. Value can transfer between opponents without registering in the harmed opponent's margin (\cref{ex:crossopp}). Moreover, the standard resolving gadget \citep{burch2014solving,moravvcik2017deepstack} comes directly from the two-player pipeline. It reports negative margins even for a vacuous update, because its subgame entry weights omit the other opponents' reach (\cref{prop:mismatch}).

We measure a profile by its \emph{best-response gap} $\Vul_G$ in game
$G$, the largest gain any single player obtains by deviating to a best
response \citep{johanson2011accelerating}. A profile whose
gap is at most $\varepsilon$ is an $\varepsilon$-Nash equilibrium
\citep{shoham2008multiagent}, and safety
means that resolving does not raise this gap. We condition safety on two computable
quantities. The structural tax $\tax$ is an oscillation of the total utility,
computable before resolving, which vanishes on every constant-sum game,
including rake-free poker. The per-opponent best-response margin deficit
$\Mdef{q}$ for each opponent $q$ is obtained one subgame best-response pass later. Margins and
solve accuracy alone cannot certify safety, because resolving can raise
$\Vul_G$ with nonnegative margins and zero suboptimality already in
two-player zero-sum games. \Cref{thm:main} states the post-hoc bound. For any
resolving update $\rs$ of a blueprint $\bp$ by resolver $i$ in subgame $S$,
\[
\Vul_G(\rs) \le \Vul_G(\bp) + \Act(\bp) + \tax + \textstyle\sum_q \Mdef{q} + \epsi,
\]
where $\Act(\bp)$ is the activation term, the subgame value the blueprint
promised but did not realize, and $\epsi$ is the resolver's suboptimality
against the blueprint opponents inside $S$. Summing the per-player gaps
gives the NashConv form of the same bound,
without the activation and suboptimality terms.
\Cref{fig:method} shows where each term arises.

The certificate also serves as a decision rule in multiplayer subgame resolving. After each resolve, the resolver runs one best-response pass per opponent inside the subgame to compute the certificate. If the bound exceeds the tolerance, the resolver keeps the blueprint. On three-player Kuhn and Leduc poker, resolving without a certificate raises $\Vul_G$ beyond a small tolerance in \harmKuhnUnsafe{} and \harmLeducUnsafe{} of the updates. The standard gadget does so in \harmKuhnStd{} and \harmLeducStd{} (\cref{sec:experiments}). This rule rejects every harmful update while accepting \accTwoLeducStd{} of the Leduc gadget updates. At six players, certification costs \sixCertSec{} seconds against \sixResolveMin{} minutes for the resolve itself.

Our contributions:

\begin{itemize}
\item We prove the first conditional safety theorem for
  uncorrelated multiplayer subgame resolving. When resolver $i$ updates
  a blueprint $\bp$ to $\rs$ in a subgame $S$, the best-response gap
  rises by at most $\Act(\bp) + \tax + \sum_q \Mdef{q} + \epsi$
  (\cref{thm:main}). On pointwise constant-sum games, the tax vanishes.
  An exact form charges the realized total-utility leak, while the
  NashConv form drops two of the four terms. The bound thus separates
  harm into activation, structural tax, and per-opponent margin deficit.
\item We construct instances that pin down the bound's form. In two-player
  CFR with decomposition (CFR-D, \cref{ex:activation}), the activation term is already exact. A
  three-player transfer forces the per-opponent margin term, and realized
  negative margins carry harm. A certificate too quiet to explain a realized
  increase defeats margin-only bounds. For $n>2$, the standard gadget's root
  measure misweights opponents' beliefs.
\item We show that the certificate turns multiplayer subgame resolving
  into a safe procedure. Gating updates with the certificate delivers no harmful update
  on three-player Leduc poker and on Kuhn poker with up to six players,
  with zero theorem violations. Because the check applies to any
  delivered update using only best responses inside the subgame, it attaches
  directly to multiplayer real-time search solvers such as the superhuman
  multi-player poker system Pluribus \citep{brown2019superhuman}.
\end{itemize}

\Cref{sec:prelim} defines the best-response gap we measure and the opponents'
margins the resolver computes inside the subgame. \Cref{sec:main} uses three
small games to show how a resolve raises the gap. Either the resolve collects
value the blueprint promised without delivering, value moves between
opponents, or the total utility changes or the solve is inexact. In
\cref{thm:main}, we show these are the only causes and bound the rise by one
term for each. Its corollaries collapse the bound into a single number from
one best-response pass per opponent. We prove the theorem in \cref{sec:proof}
for constant total utility and then the general case, and
\cref{sec:experiments} tests this number as a gate in poker. The resolver
plays the update when the bound falls within a tolerance and keeps the
blueprint otherwise.

\section{Related Work}\label{sec:related}

\textbf{Two-player safe resolving.}
CFR-D introduced game decomposition, the resolving gadget, and the first
safety guarantee for two-player zero-sum subgame refinement
\citep{burch2014solving}. Maxmargin resolving made the margin the
optimization objective \citep{moravcik2016refining}. From there,
safe-and-nested solving \citep{brown2017safe} scaled these guarantees
through depth-limited solving \citep{brown2018depth}. Extensions cover
endgame solving \citep{ganzfried2015endgame}, limited lookahead
\citep{kroer2020limited}, subgame solving without common knowledge
\citep{zhang2021subgame}, continual resolving \citep{vsustr2019monte}, sound
online solving \citep{vsustr2021sound}, and efficient refinement
\citep{ge2023efficient}. Every such guarantee relies on two-player zero-sum
structure, where the opponent's gain equals the resolver's loss. Carrying
this margin-certificate language to the original multiplayer game bounds
$\Vul_G$ directly.

\textbf{Multiplayer resolving in practice and refinement theory.}
Pluribus achieved strong six-player play through heuristic real-time
resolving without a safety guarantee \citep{brown2019superhuman}, and no
multiplayer analogue of solved heads-up limit hold'em
\citep{bowling2015heads} exists. Beyond two-player zero-sum play, provable
guarantees rely on restored two-sided or correlated structure. Zero-sum team
games provide two-sided structure
\citep{zhang2020computing,zhang2021computing}, while resolving a correlation
plan is safe in two-player general-sum games \citep{ling2022safe}. Even the
equilibrium concept that subgame solving should target remains open
\citep{kubicek2026equilibrium}. For general $n$-player play without teams or
correlation devices, we give a guarantee conditioned on quantities computed
for the delivered update, requiring no two-sided reduction.

Safe exploitation, adversarial team games, blueprint computation, and
polymatrix games are discussed in
\cref{app:related}.

\section{Preliminaries}\label{sec:prelim}

This section specifies the game model and defines the two quantities in the bound, namely the opponents' margins at the subgame entry and the best-response gap of a profile. We consider a finite $n$-player extensive-form game (EFG) $G$, a game tree with chance moves, hidden information, and perfect recall, meaning no player forgets what it knew or did. The player set is $N=\{1,\dots,n\}$, $Z$ is the set of terminal histories, and $u_p(z)$ is player $p$'s payoff at $z$. A behavioral strategy $\sigma_p$ assigns a distribution over actions to each of $p$'s information sets, the sets of histories $p$ cannot tell apart. A profile $\sigma$ induces expected payoff $u_p(\sigma)$. For a history $h$, $\pi_p^{\sigma}(h)$ is the product of $p$'s action probabilities on the path to $h$. Keeping chance separate, we define the others' joint reach as $\pi_{-p}^{\sigma}(h) := \prod_{q \neq p} \pi_q^{\sigma}(h)$. Every non-terminal history carries a public label $\mathrm{pub}(h)$ representing the information visible to all players, and we write $(x)_+ := \max(x,0)$.

\textbf{Resolving and gadget games.} Subgame resolving starts from a \emph{blueprint} $\bp$ computed offline and refines the strategy of one player, the \emph{resolver} $i$, inside a subgame reached during play. The remaining players are the \emph{opponents} $O := N \setminus \{i\}$. A \emph{public subgame} is specified by a public state $\mathrm{pub}^*$. Its \emph{entry set} $E$ collects the earliest histories $e$ with $\mathrm{pub}(e)=\mathrm{pub}^*$, each with chance probability $c_e$. The subgame $S$ consists of the descendants of $E$, and the rest of the game is the \emph{trunk}. Entries are grouped into \emph{entry information sets} $\iota_q(e) := (\mathrm{priv}_q(e), \mathrm{pub}^*)$, collected in $I_q$, where $\mathrm{priv}_q(e)$ is $q$'s private information at $e$, and $Z_S\subseteq Z$ denotes the terminals inside $S$. Each set carries counterfactual entry mass $\bar{W}_q(\iota) := \sum_{e \in \iota} c_e\, \pi_{-q}^{\bp}(e)$ and blueprint counterfactual value $\mathrm{CFV}_q(\iota) := \bar{W}_q(\iota)^{-1} \sum_{e \in \iota} c_e\, \pi_{-q}^{\bp}(e)\, \mathrm{EV}_q^{\bp}(e)$, where $\mathrm{EV}_q^{\bp}(e)$ is $q$'s expected payoff from entry $e$ under $\bp$ (counterfactual: weighted by the others' reach, never by $q$'s own). Resolving solves a CFR-D/maxmargin-style \emph{gadget game} \citep{burch2014solving,moravcik2016refining,brown2017safe}. A root chance node enters $e\in E$ with probability proportional to $c_e \pi_i^{\bp}(e)$, the gadget's \emph{root measure}. Each opponent $j$ then chooses Terminate or Follow at an information set keyed by $\iota_j$. Terminating pays $\mathrm{CFV}_j(\iota)$, while following lets $j$ play freely. The output is an update $\rs$ \emph{confined to the resolver's subgame}. Here $\rs_p = \bp_p$ for $p \neq i$, and $\rs_i$ differs from $\bp_i$ only on $i$'s information sets inside $S$.

\textbf{Best-response (BR) certified margins.}
The Terminate option pays opponent $j$ what the blueprint promised at $\iota$. The Follow option lets $j$ exploit the update. After the resolve, the certificate asks how much $j$ could gain over that promised value by following and best-responding inside $S$. We evaluate this margin against an exact subgame best response, the payoff-maximizing reply to the others' fixed strategies. For $j \in O$, $\iota \in I_j$, and a subgame strategy $x_i$ of the resolver, $\mathrm{SBV}_j(\iota;\, x_i) := \max_{\tau_j^S} \sum_{e \in \iota} c_e\, \pi_{-j}^{\bp}(e)\; \mathbb{E}[u_j \mid e,\ (\tau_j^S, x_i, \bp_{-ij}) \text{ inside } S]$ is $j$'s best-response value when the resolver plays $x_i$ and all others follow the blueprint $\bp_{-ij}$.

\begin{definition}[BR-certified margin]\label{def:margin}
For an opponent $j \in O$ and entry information set $\iota \in I_j$, the
\emph{BR-certified margin} of the update $\rs$ is
$\mhat_j(\iota) := \mathrm{CFV}_j(\iota) - \mathrm{SBV}_j(\iota;\, \rs_i)/\bar{W}_j(\iota)$.
\end{definition}

\begin{definition}[Margin deficit]\label{def:deficit}
The \emph{margin deficit} of opponent $q \in O$ under $\rs$ is the
mass-weighted sum $\Mdef{q} := \sum_{\iota \in I_q} \bar{W}_q(\iota)\,
\bigl(-\mhat_q(\iota)\bigr)_+$.
\end{definition}

A nonnegative margin at $\iota$ means $q$ cannot gain at $\iota$ by
exploiting the update, whereas a negative margin measures the gain per unit of
counterfactual mass. The deficit sums these violations weighted by that mass.
Rarely reached entry sets therefore contribute little, and only entry groups
with $\bar{W}_q(\iota) > 0$ enter the sum. When every opponent's own action
sequence coincides across $E$, as in poker-style games where all actions
before $\mathrm{pub}^*$ are public, $\sum_\iota \bar{W}_q(\iota) \le 1$ and
$\Mdef{q} \le (-\min_\iota \mhat_q(\iota))_+$. Gadget implementations report
a margin that values the Follow branch under the opponent's gadget average
strategy, an optimistic surrogate for the best response. That reported margin
exceeds $\mhat$ on 33 of 40 entry information sets in our benchmark games
(\cref{app:calibers} lists the conventions). All margins in this paper are
BR-certified.

\begin{definition}[Best-response gap]\label{def:vul}
For a profile $\sigma$ in $G$, player $p$'s best-response gain is
$A_p(\sigma) := \max_{\tau_p} u_p(\tau_p, \sigma_{-p}) - u_p(\sigma)$, and
the \emph{best-response gap} of $\sigma$ is
$\Vul_G(\sigma) := \max_{p \in N} A_p(\sigma)$.
\end{definition}

Beyond two players, two aggregates of the per-player gains are in use. Their sum is $\mathrm{NashConv}(\sigma) := \sum_p A_p(\sigma)$ \citep{lanctot2019openspiel}, and their maximum $\Vul_G$ is the smallest $\varepsilon$ for which $\sigma$ is an $\varepsilon$-Nash equilibrium. To obtain a guarantee for every player, we bound the maximum. A small $\Vul_G$ leaves no single player exploitable, whereas $\mathrm{NashConv}$ can stay flat while one player's gain doubles (\cref{ex:activation}). \Cref{thm:main} yields the $\mathrm{NashConv}$ bound as a corollary. At $n=2$, $\Vul_G$ is the larger of the two gains, distinct from the exploitability convention of \citet{lanctot2019openspiel}, $\tfrac12(A_1+A_2)$. It also differs from the vulnerability of \citet{macqueen2023guarantees}, the payoff a strategy loses when the other players deviate.

\begin{assumption}[Public-state subgame regularity]\label{asm:F1}
The subgame $S$ satisfies: (P1) every information set of every player is
either entirely contained in $S$ or disjoint from it; (P2) the gadget's
entry grouping $\iota_q = (\mathrm{priv}_q, \mathrm{pub}^*)$ coincides with
$q$'s actual information at entry, and $q$'s own reach $\pi_q^{\tau}(e)$ is
constant across $e \in \iota$ for every strategy $\tau$ of $q$.
\end{assumption}

\Cref{asm:F1} restricts the subgame's geometry. (P1) holds automatically for
subgames cut at public states, and (P2) holds under perfect recall whenever
entries follow a public signal (\cref{app:nested}).

\section{Main Theorem and How to Read It}\label{sec:main}

This section defines the three remaining terms of the bound and shows each
cause of harm on its own small game. We then state the theorem and the forms
we compute.

\textbf{Structural tax.} The tax $\tax$ bounds how much an update can change
the players' total utility. When every terminal has the same total, the tax
is zero. With $U(s)=\sum_p u_p(s)$ for pure-strategy profiles $s$ (one action chosen at every information set) and
$U(z):=\sum_p u_p(z)$ at terminals $z$, let
\begin{equation*}
\zprof := \min_c \max_s |U(s)-c| , \qquad
\zsub := \min_c \max_{z\in Z_S} |U(z)-c| , \qquad
\tax := 2\,\min(\zprof,\ \rho_S\,\zsub) ,
\end{equation*}
where $\rho_S := \sum_{e\in E} c_e \prod_p \pi_p^{\bp}(e)$ is the
blueprint's probability of reaching the subgame. Here, $\zprof$ and $\zsub$
are half the range of the total utility over all profiles and over the
terminals of $S$.

\textbf{Activation term.} Define
$\Act(\bp) := \max_{\tau_i:\,\tau_i=\bp_i \text{ outside } S} u_i(\tau_i,\bp_{-i})
- u_i(\bp)$ as the subgame payoff the blueprint promises the resolver but
leaves unrealized. Against the blueprint opponents, this term equals the
resolver's maximum gain from changing its strategy inside $S$. We can compute
it from $\bp$ alone, and $\Act(\bp)\le A_i(\bp)\le \Vul_G(\bp)$.

\textbf{Solve error.} The solve error
$\epsi := \Act(\bp) - [u_i(\rs)-u_i(\bp)] \ge 0$ is the part of that payoff
the delivered update misses. It measures the suboptimality of $\rs_i$ against
the blueprint opponents inside $S$. We
call $\rs$ \emph{margin-certified} when $\mhat_q(\iota)\ge 0$ for all $q$ and all
$\iota$ with $\bar W_q(\iota)>0$. The theorem charges violations through $\Mdef{q}$.
Throughout, $\Delta\Vul := \Vul_G(\rs)-\Vul_G(\bp)$.

Three small games show the channels through which a resolve can raise the
gap, one channel each. First, the activation term survives a certificate
with every margin nonnegative, even in two-player zero-sum games.

\begin{example}[Activation: margin-certified, $\epsi=0$, yet $\Delta\Vul=\Act$]\label{ex:activation}
Two players $\{i,j\}$, zero-sum. Player $j$ moves first:
action $d$ ends the game with $(-1,+1)$; \emph{enter} reaches the subgame
$S$, where $i$ alone acts: $a'$ gives $(0,0)$, $a$ gives $(+1,-1)$. The
blueprint is $\bp=(a',\textit{enter})$: $A_i(\bp)=\Act(\bp)=1$, so
$\Vul_G(\bp)=1$. Resolving delivers $\rs_i=a$:
$\mathrm{CFV}_j=0$, $\mathrm{SBV}_j(\rs_i)=-1$, so $\mhat_j=+1>0$ and
$\epsi=0$; yet $A_j(\rs)=2$, so $\Vul_G(\rs)=2=\Vul_G(\bp)+\Act(\bp)$.
\end{example}

Resolving to $a$ realizes the unit of subgame value the blueprint left
unrealized. No margin flags the move ($\mhat_j=+1>0$), yet $\Vul_G$ rises
$1\to2$. The half-sum convention $\tfrac12(A_1+A_2)$ stays at $1$, as does
$\mathrm{NashConv}$, because $A_i$ falls from $1$ to $0$ while $A_j$ rises
from $1$ to $2$. CFR-D and maxmargin resolving
\citep{burch2014solving,moravcik2016refining} bound only the resolver's own
gain.

Second, a per-pair guarantee keyed to the victim's own margin is false.

\begin{example}[Cross-opponent transfer: the victim's margin is silent]\label{ex:crossopp}
Three players $\{i,j,k\}$, one simultaneous move (the subgame is the whole
game): $i$ chooses in $\{a',a\}$, $j$ in $\{\textit{stay},d\}$, $k$ has no
action. Payoffs are sums of pairwise terms: $u_{ij}(a,\textit{stay})=+1$,
$u_{ij}(a,d)=0$, $u_{ij}(a',\cdot)=0$; $u_{ik}(a,\cdot)=-1$,
$u_{ik}(a',\cdot)=0$; $u_{jk}\equiv 0$, each pair constant-sum. The blueprint is
$\bp=(a',\textit{stay})$: $\Vul_G(\bp)=\Act(\bp)=0$. Resolving delivers
$\rs_i=a$: $\mhat_j=0$, $\mhat_k=-1$, $\epsi=0$. Then
$\Delta\mathrm{PairVul}_{ji}=1>0=(-\mhat_j)_+$, where $\mathrm{PairVul}_{ji}$
is $j$'s best-response gain on the pair utility $u_{ji}$ (\cref{app:perpair}),
refuting the per-pair form,
and $A_j(\rs)=1$, so $\Delta\Vul=1=\sum_q \Mdef{q}$.
\end{example}

The victim's margin stays silent, and the alarm fires at the
\emph{beneficiary} $k$ (\cref{app:perpair} gives the per-pair statement).
Third, a change in the total utility and the solve error each raise the gap on their own.

\begin{example}[Pure leak and self-harm]\label{ex:leak}
\emph{Leak:} two players; $j$ moves first, $d$ ends the game with $(0,0)$,
\emph{enter} reaches $S$, where $i$ alone acts: $a'$ gives $(0,0)$, $a$
gives $(0,-1)$. From $\bp=(a',\textit{enter})$ the update $\rs_i=a$ has
$\Act=\epsi=0$ and $\mhat_j=+1$, yet $A_j(\rs)=1$: $\Delta\Vul=1=\tax$
($\zprof=\zsub=\tfrac12$, $\rho_S=1$). \emph{Self-harm:} two players,
zero-sum, $i$ alone acts: $a$ gives $(1,-1)$, $a'$ gives $(0,0)$. From
$\bp_i=a$ the update $\rs_i=a'$ has $\Act=0$, $\epsi=1$, $\Mdef{j}=1$, and
$\Delta\Vul=A_i(\rs)=1=\epsi$ (\cref{app:L3:ex34}).
\end{example}

In the leak the resolver destroys one unit of $j$'s utility at no cost to
itself, so no margin sees it and only the structural tax $\tax$ prices it.

\begin{theorem}[Conditional safety of multiplayer resolving]\label{thm:main}
Let $G$ be an $n$-player perfect-recall EFG satisfying \cref{asm:F1}, let
$\bp$ be an arbitrary blueprint, and let $\rs$ be any update confined to
resolver $i$'s subgame strategy ($\rs_p = \bp_p$ for all $p \neq i$, and
$\rs_i$ differs from $\bp_i$ only on $i$'s information sets inside $S$).
Let $\mhat$, $\Mdef{q}$, $\Act(\bp)$, and $\epsi$ be defined with respect
to the delivered $\rs$ (\cref{def:margin,def:deficit} and the displays
above), and let $\tax = 2 \min(\zprof, \rho_S \zsub)$. Then
\begin{itemize}
\item[(i)] $A_i(\rs) - A_i(\bp) \;=\; \epsi - \Act(\bp) \;\le\; \epsi$,
\item[(ii)] for every opponent $j \neq i$, $A_j(\rs)-A_j(\bp) \le \Mdef{j}
+ [\Act(\bp)-\epsi+\sum_{k\in O\setminus\{j\}}\Mdef{k}] + \tax$, and
\item[(iii)]
$\displaystyle
\Vul_G(\rs) - \Vul_G(\bp) \;\le\; \max_{p} \bigl[A_p(\rs) - A_p(\bp)\bigr]
\;\le\; \max\Bigl( \epsi ,\;\; \Act(\bp) - \epsi +
\textstyle\sum_{q \in O} \Mdef{q} + \tax \Bigr) .$
\end{itemize}
\end{theorem}

By part (i), the resolver's own best-response gain rises by at most its solve error, and its payoff against the blueprint opponents rises by $\Act(\bp)-\epsi$. Any increase of $\Vul_G$ beyond $\epsi$ therefore comes from an opponent becoming more exploitable. Part (ii) identifies the channels as that opponent's own deficit, the value the resolver collected ($\Act-\epsi$) alongside the other opponents' deficits, and the tax $\tax$.

Across \cref{ex:activation,ex:crossopp,ex:leak}, all four terms of
\cref{thm:main}(iii) and both branches of its maximum are attained with
equality. These examples force the per-player granularity of the margin sum
and show that the coefficient $1$ on $\Act$ and on the margin term is
optimal.

\begin{corollary}[Exact post-hoc form]\label{cor:leak}
Let $\Lambda := \sum_p u_p(\bp) - \sum_p u_p(\rs)$ be the realized
total-utility leak of the update. Then \cref{thm:main}(ii) and (iii) hold
with $\tax$ replaced by $\Lambda$, and $\Lambda \le |\Lambda| \le \tax$
(\cref{lem:transfer}). Since $\Lambda$ is signed it stays inside the maximum,
For resolves $t=1,\dots,T$ applied one after another, with every term evaluated at resolve $t$, this gives $\sum_t \max(\epsi[t],\ \Act[t]-\epsi[t]+\sum_q \Mdef{q}[t]+\Lambda[t])$
for sequential resolves.
\end{corollary}

\begin{corollary}[Direct per-player form]\label{cor:direct}
Write $g_p := u_p(\rs)-u_p(\bp)$ for the realized gain of player $p$. Then
\begin{equation}\label{eq:direct}
\Vul_G(\rs) - \Vul_G(\bp) \;\le\; B_{\mathrm{dir}} :=
\max\bigl(-g_i,\ \max_{j\ne i}\,[\Mdef{j}-g_j]\bigr),
\end{equation}
and $B_{\mathrm{dir}}$ is pointwise at most the bound of \cref{cor:leak}
(\cref{app:T1T2}).
\end{corollary}

\begin{corollary}[Baseline-aware form]\label{cor:cache}
With the blueprint best-response gains $A^0_p := A_p(\bp)$,
$\Vul_G(\rs) - \Vul_G(\bp) \le B_{\mathrm{cache}} :=
\max\bigl(A^0_i-g_i,\ \max_{j\ne i}[A^0_j+\Mdef{j}-g_j]\bigr) - \max_p A^0_p
\le B_{\mathrm{dir}}$.
On a constant-sum game $B_{\mathrm{dir}}\le 0$ forces $g_p=0$ and
$\Mdef{j}=0$ for all $p$, $j\ne i$ (\cref{app:T1T2}), whereas $B_{\mathrm{cache}}$ does not.
\end{corollary}

\begin{corollary}[NashConv form]\label{cor:nashconv}
With $\mathrm{NashConv}(\sigma) := \sum_p A_p(\sigma)$ and $\Lambda$ as in
\cref{cor:leak},
$\mathrm{NashConv}(\rs)-\mathrm{NashConv}(\bp) \le \sum_{q\in O}\Mdef{q} +
\Lambda \le \sum_{q\in O}\Mdef{q} + \tax$.
\end{corollary}
The proof (\cref{app:T1T2:nashconv}) sums the per-player identities behind
\cref{thm:main}. The resolver's best-response value is unchanged, each
opponent's rises by at most $\Mdef{j}$, and the realized gains $g_p$ sum to
$-\Lambda$. The activation and solve-error terms cancel in the sum. Thus
the bound on the maximum carries them while the bound on the sum does
not. In \cref{ex:activation} the sum is unchanged while the opponent's gain
doubles. On a constant-sum game a margin-certified update never increases
$\mathrm{NashConv}$.

\begin{corollary}[Constant-sum games pay no structural tax]\label{cor:constant-sum}
If $G$ is pointwise constant-sum ($\sum_p u_p(z) = c$ for every $z \in Z$),
then $\zprof = \zsub = 0$ and hence $\tax = 0$, so \cref{thm:main}(iii)
reads $\Vul_G(\rs)-\Vul_G(\bp) \le \max(\epsi,\ \Act(\bp)-\epsi+
\sum_{q\in O}\Mdef{q})$.
\end{corollary}

\begin{proof}
$U(s) = \mathbb{E}[U(z)] = c$ for every profile and $Z_S \subseteq Z$, so
$\zprof = \zsub = 0$.
\end{proof}

Pointwise constant-sum games include three-player Kuhn, rake-free Hold'em, and
zero-sum-normalized Goofspiel, so none of them pays a
structural tax. Under a rake the tax is positive but small
(\cref{sec:experiments}).

\textbf{Scope.} The theorem covers full-depth public subgames. Margins are computed against exact best responses inside $S$ down to its terminals, and the update changes one resolver at a time. Depth-limited resolving with leaf values \citep{brown2018depth} truncates the subgame and supplies continuation values from a value function. There, an opponent's best response may leave the set of continuation strategies for which the leaf values were computed, so the margin can be underestimated. Extending the certificate to that setting is the main open problem.

\textbf{Cost.} Computing $B_{\mathrm{dir}}$ takes one best-response pass within $S$ per opponent, plus the $n$ expected utilities. The bound $B_{\mathrm{cache}}$ also reads the stored $A^0_p$. Because $\Act(\bp)$ depends only on $(\bp,S)$, it is computed once per subgame. \Cref{app:delta} bounds the tax a priori through the distance from $G$ to constant-sum polymatrix games.

\section{Proof of the Main Theorem and Tightness}\label{sec:proof}

The proof has two steps. When total utility is constant, the players'
realized gains sum to zero. An opponent's payoff drop then equals what the
resolver and other opponents gained, and the certificate bounds each gain
(\cref{lem:delta0}). In general, the gains sum to at most the tax in absolute
value (\cref{lem:transfer}), adding the tax to the bound.

\textbf{The constant-sum case.}
Throughout, $G$ satisfies \cref{asm:F1} and $\rs$ is any update confined
to the resolver's subgame, with $\mhat$, $\Mdef{q}$, $\Act(\bp)$, and $\epsi$
defined for the delivered $\rs$ (\cref{def:margin,def:deficit}).

\begin{lemma}[Safety of gadget resolving under constant total utility]\label{lem:delta0}
Let $G$ satisfy \cref{asm:F1} and have constant total utility across
pure-strategy profiles, $\sum_p u_p(s)=c$ for all $s$ (in particular every
pointwise constant-sum game), and let $\rs$ be any update of the above form. Then:
\begin{enumerate}
\item[(i)] \emph{(resolver)} $\;A_i(\rs) - A_i(\bp) \;=\; \epsi - \Act(\bp)
\;\le\; \epsi$.
\item[(ii)] \emph{(each opponent $j\neq i$)}
$A_j(\rs)-A_j(\bp) \le \Mdef{j} + [\Act(\bp)-\epsi+
\sum_{k\in O\setminus\{j\}}\Mdef{k}] \le \Act(\bp) + \sum_{q\in O}\Mdef{q}$.
\item[(iii)] \emph{(total)}
$\Vul_G(\rs)-\Vul_G(\bp) \le \max\bigl(\epsi,\ \Act(\bp)-\epsi+
\sum_{q\in O}\Mdef{q}\bigr)$.
\end{enumerate}
If $\mhat_q(\iota)\ge 0$ for all $q,\iota$ (a margin-certified update) and
$\epsi\le\varepsilon_{\mathrm{solve}}$, the solve-error budget, then
$\Vul_G(\rs) \le \Vul_G(\bp) + \max\big(\varepsilon_{\mathrm{solve}},\ \Act(\bp)\big)$.
\end{lemma}

The full proof appears in \cref{app:L3}. For every player, $A_p = \mathrm{BRval}_p - u_p$, where $\mathrm{BRval}_p$ is the best-response value against the others. Because the resolver's opponents do not change, (i) follows from the definition of $\epsi$. Perfect recall splits any deviation of opponent $j$ into a trunk part and independent subgame continuations, so $j$'s best-response value rises by at most $\Mdef{j}$. That leaves $j$'s own payoff drop $-g_j$. Constant total utility enters here through $\sum_p g_p=0$, which rewrites $-g_j$ as the resolver's collected value $\Act(\bp)-\epsi$ plus the other opponents' gains, each at most its deficit.

\textbf{The transfer step.}
Constant total utility entered \cref{lem:delta0} at one step, which used
only $\sum_p g_p=0$. When total utility is not constant, this sum is a bounded
drift that the transfer lemma controls.

\begin{lemma}[Transfer]\label{lem:transfer}
For any update $\rs$ as in \cref{lem:delta0}, the realized gains satisfy
$\bigl|\sum_p g_p\bigr| \le \tax = 2\min(\zprof,\rho_S\,\zsub)$.
\end{lemma}

Every profile's total utility lies within $\zprof$ of a common center, so it differs between $\bp$ and $\rs$ by at most $2\zprof$. Because the two profiles differ only inside the subgame, which both reach with probability $\rho_S$, this difference is also at most $2\rho_S\,\zsub$. Replacing the zero sum in \cref{lem:delta0} with $\tax$ yields \cref{thm:main}.

\textbf{Tightness on realized instances.}
The analytic constructions of \cref{ex:activation,ex:crossopp,ex:leak} attain
each term with equality. Three further results exhibit these channels on
resolving instances from the random-game search (\cref{app:constructions}),
where \Cref{prop:delta-nec} shows that margin-only bounds are insufficient.

\begin{proposition}[Negative margins are realized]\label{prop:margin-nec}
There exist three-player games and resolving updates satisfying the
quantifiers of \cref{thm:main} on which strongly negative BR-certified
margins are matched by an increase of $\Vul_G$ of the same order. One
search instance has $\Delta\Vul = 0.2164$ with minimum margin $\mhat = -1.0559$
and $\sum_q \Mdef{q} = 0.252$, while $\Act \approx 0$, $\epsi = 0.002$ and
$\tax = 0.017$. With the margin term removed, the bound is $0.015$
against $\Delta\Vul = 0.216$, so the increase is carried by the
margin channel alone.
\end{proposition}

\begin{proposition}[Margin-only bounds are insufficient]\label{prop:delta-nec}
There is a three-player instance from the search runs in which the resolver's
blueprint subgame strategy is reset to a Dirichlet draw before resolving.
Its resolving update has $\Delta\Vul = 0.0383$ above tolerance $0.02$, while its BR-certified margin deficit aggregates to only
$\sum_q \Mdef{q} = 0.0111$ ($\Act = 0.0382$, $\epsi = 0.0379$). Any margin-only
bound $\Delta\Vul \leq c_2 \sum_q \Mdef{q} + \mathrm{tol}$ needs
$c_2 \approx 1.65$, five times the coefficient fitted over all search
instances (\cref{tab:constants}).
\end{proposition}

\begin{proposition}[Structural margin mismatch near equilibrium for $n > 2$]\label{prop:mismatch}
There exist three-player games and blueprints with full-game gap
$1.4\times10^{-3}$ for which the margins reported by the standard
gadget are strictly negative at zero deviation. In the construction,
the opponents' margins evaluate to $-0.46$ and $-0.74$ on single entry
information sets although the update deviates nowhere. The BR-certified
margins of \cref{def:margin} behave differently. At zero deviation they are
nonpositive at every entry set, and the mass-weighted deficits are
$1.8\times10^{-4}$ and $3.5\times10^{-4}$.
\end{proposition}

The mechanism is a weighting defect. Correct beliefs for opponent $q$ require the reach of all players other than $q$, whereas the gadget's root measure $c_e\,\pi_i^{\bp}(e)$ omits the other opponents' reach. For $n>2$, this omission overweights the alternative even when no player deviates. Only the BR certificate supplies the counterfactual weight $c_e\,\pi_{-q}(e)$ at unreached entry sets. No single root measure serves all opponents at once. On the poker subgames of \cref{sec:experiments}, none of the \eFourCells{} (blueprint, subgame, measure) combinations yields nonnegative gadget margins for every opponent simultaneously. Consequently, the Terminate option, which guarantees nonnegative margins for $n=2$, fails to transfer, and we compute the certificate after the resolve.

\section{Experiments}\label{sec:experiments}

We measure how often resolving is harmful on multi-player subgames, whether the certificate catches that harm, what the certificate costs, and how each scales with player count. Protocol details appear in \cref{app:poker}.

\textbf{Setup.} We solve three-player Kuhn poker \citep{szafron2013parameterized} and three-player Leduc poker \citep{southey2005bayes}
with CFR to obtain blueprints of varying quality. These include runs of $10$
to $1000$ iterations, a blueprint computed under a card abstraction and lifted
to the full game, and blueprints mixed with $5\%$ to $20\%$ uniform noise.
Every public state after the first betting action in Kuhn or at the start of
the second round in Leduc defines a subgame. Each player still acting there
serves as a resolver in turn. This yields $56$ (subgame, resolver, blueprint)
triples per resolver type for Kuhn and $540$ for Leduc.

We compare three resolvers. \emph{Best response} replaces the resolver's
subgame strategy with its exact best response to the blueprint opponents. For
\emph{Standard gadget}, we take the CFR-D gadget \citep{burch2014solving} of
\cref{sec:prelim} with root measure $c_e\pi_i^{\bp}(e)$, solved by CFR.
\emph{Belief-corrected gadget} sets the root measure to $c_e\prod_p\pi_p^{\bp}(e)$,
weighting each entry by the reach of all players as in unsafe subgame solving
\citep{brown2017safe} and Pluribus \citep{brown2019superhuman}. On each
delivered update we compute the exact $\Delta\Vul$ by full-game best responses
and evaluate the baseline-aware certificate $B_{\mathrm{cache}}$ of
\cref{cor:cache} with BR-certified margins. Resolver gain
$g_i = u_i(\rs)-u_i(\bp)$ is averaged over updates. An update is \emph{harmful}
when $\Delta\Vul>0.02$ and \emph{accepted at $t$} when $B_{\mathrm{cache}}\le t$.
The \emph{gated} resolver plays updates accepted at $0.02$ and keeps the
blueprint otherwise, and Max $\Delta\Vul$ after gating records its worst gap
increase across updates. Payoffs are in units of the ante.

\begin{table}[t]
\centering
\caption{Resolving outcomes on three-player Kuhn and Leduc poker, pooled
over blueprints.}
\label{tab:poker}
\footnotesize
\setlength{\tabcolsep}{4.5pt}
\begin{tabular}{llrrrrrr}
\toprule
Game & Resolver & \shortstack{Median\\$\Delta\Vul$} & \shortstack{Harmful\\updates} & \shortstack{Accepted\\at $0$} & \shortstack{Accepted\\at $0.02$} & \shortstack{Resolver\\gain} & \shortstack{Max $\Delta\Vul$\\after gating} \\
\midrule
Kuhn & best response & +0.0620 & 80\% & 0\% & 5\% & +0.0164 & +0.0068 \\
Kuhn & standard gadget & +0.0096 & 29\% & 0\% & 41\% & +0.0082 & +0.0182 \\
Kuhn & belief-corrected gadget & +0.0016 & 20\% & 0\% & 43\% & +0.0087 & +0.0170 \\
Leduc & best response & +0.0000 & 12\% & 53\% & 80\% & +0.0063 & +0.0094 \\
Leduc & standard gadget & +0.0000 & 2\% & 47\% & 86\% & +0.0031 & +0.0135 \\
Leduc & belief-corrected gadget & -0.0000 & 2\% & 52\% & 87\% & +0.0034 & +0.0083 \\
\bottomrule
\end{tabular}

\end{table}

\textbf{Resolving without a certificate is unsafe.}
The best-response resolver raises $\Vul_G$ beyond tolerance in
\harmKuhnUnsafe{} of the Kuhn updates and \harmLeducUnsafe{} of the Leduc
updates (\cref{tab:poker}). Gadgets reduce harm, yet resolving remains unsafe.
The standard gadget is harmful in \harmKuhnStd{} and \harmLeducStd{} of the
updates, and the belief-corrected gadget in \harmKuhnBelief{} and
\harmLeducBelief{}, because neither root measure prices every opponent
correctly (\cref{prop:mismatch}). Across all \nUpdKuhn{} Kuhn and
\nUpdLeduc{} Leduc updates, the bound of \cref{thm:main} and the forms
$B_{\mathrm{dir}}$ and $B_{\mathrm{cache}}$ are violated \violTotal{} times.

\textbf{The certificate flags every harmful update and accepts most
safe ones.} At tolerance $0.02$, the certificate accepts no harmful update
across the two games and three resolvers (\fnTotal{} in total). At the same tolerance, $B_{\mathrm{cache}}$
accepts \accTwoLeducStd{} of the standard-gadget updates on Leduc and
\accTwoKuhnStd{} on Kuhn. The checkable condition for outright safety,
$B_{\mathrm{cache}}\le 0$, holds on \accZeroLeducAll{} of the Leduc updates and
on \accZeroKuhnAll{} of the Kuhn updates. In Kuhn, subgames are so small that
every update moves some opponent's margin. The stricter margin-certified
condition, all $\mhat\ge 0$, is rare on both games because exact best-response
margins against a near-equilibrium blueprint are slightly negative almost
everywhere.

\textbf{Gating removes harm and keeps the gain.} The gated resolver delivers
no harmful update on either game, with a worst update of $\gatedMaxLeducStd$
on Leduc and $\gatedMaxKuhnStd$ on Kuhn, both below tolerance. Across all
resolvers, mean realized gain $g_i$ stays positive (\cref{tab:poker}). The
certificate preserves genuine subgame value while rejecting updates that
collect payoff by making an opponent exploitable. Shrinking the update toward
the blueprint is an alternative to gating. On Kuhn, moving a quarter of the
way reduces the standard gadget's harmful rate to \shrinkHarmKuhnQuarter{}
(\cref{tab:shrink}).

\begin{table}[t]
\centering
\caption{Kuhn poker with $n=3,\dots,6$ players: resolving outcomes and
median wall-clock time per update.}
\label{tab:nkuhn}
\footnotesize
\begin{tabular}{rlrrrrrr}
\toprule
$n$ & Resolver & \shortstack{Median\\$\Delta\Vul$} & \shortstack{Harmful\\updates} & \shortstack{Accepted\\at $0.02$} & \shortstack{Resolve\\(s)} & \shortstack{Certificate\\(s)} & \shortstack{Exact\\evaluation (s)} \\
\midrule
3 & best response & +0.0509 & 80\% & 0\% & 0.010 & 0.004 & 0.003 \\
3 & standard gadget & +0.0097 & 40\% & 40\% & 0.340 & 0.004 & 0.004 \\
3 & belief-corrected gadget & +0.0043 & 20\% & 60\% & 0.340 & 0.004 & 0.004 \\
4 & best response & +0.0489 & 100\% & 0\% & 0.080 & 0.048 & 0.048 \\
4 & standard gadget & +0.0221 & 57\% & 43\% & 9.82 & 0.062 & 0.057 \\
4 & belief-corrected gadget & +0.0028 & 43\% & 57\% & 9.56 & 0.062 & 0.058 \\
5 & best response & +0.0232 & 56\% & 33\% & 1.89 & 1.04 & 1.07 \\
5 & standard gadget & +0.0022 & 0\% & 56\% & 300 & 1.30 & 1.31 \\
5 & belief-corrected gadget & +0.0000 & 11\% & 78\% & 301 & 1.27 & 1.40 \\
6 & best response & +0.0370 & 100\% & 0\% & 47.1 & 30.1 & 24.4 \\
6 & standard gadget & +0.0125 & 36\% & 64\% & 10258 & 26.2 & 32.2 \\
6 & belief-corrected gadget & +0.0112 & 36\% & 64\% & 10181 & 28.5 & 28.2 \\
\bottomrule
\end{tabular}

\end{table}

\Cref{tab:nkuhn} repeats the Kuhn poker comparison for $n=3$ to $6$ players with a deck of $n+1$ cards, running $200$ CFR iterations for the blueprint and $50$ for the gadget. We time the subgame solve, the post-hoc bound evaluation, and the full-game best-response pass for $\Delta\Vul$. At $n=6$, the best-response resolver is harmful in \sixHarmUnsafe{} of the updates and the gadgets in \sixHarmStd{}. The certificate again accepts no harmful update across \nSixRecords{} updates, with \violSix{} violations. For an online system at $n=6$, the median gadget resolve takes \sixResolveMin{} minutes per update, while the certificate takes \sixCertSec{} seconds. Evaluating the bound runs one best-response pass per opponent, whereas the gadget solve iterates CFR. The certificate and the exact full-game evaluation stay within a factor of two of each other on these small games, where the subgame is a large fraction of the tree. As the trunk grows, the certificate's locality pays off.

\begin{table}[t]
\centering
\caption{Cost of the certificate's two parts, with ratios to the
exact full-game evaluation (range over public states).}
\label{tab:cost}
\footnotesize
\begin{tabular}{lrrrr}
\toprule
Game & \shortstack{Activation\\term (s)} & \shortstack{Margin\\pass (s)} & \shortstack{Full certificate /\\exact evaluation} & \shortstack{Margin pass /\\exact evaluation} \\
\midrule
3-player Kuhn & 0.003--0.003 & 0.001--0.002 & 1.34--1.52 & 0.46--0.65 \\
3-player Leduc & 0.450--0.493 & 0.065--0.075 & 1.01--1.11 & 0.13--0.15 \\
4-player Kuhn & 0.030--0.041 & 0.015--0.031 & 1.17--1.26 & 0.32--0.64 \\
5-player Kuhn & 0.745--0.787 & 0.241--0.694 & 0.86--1.12 & 0.20--0.54 \\
6-player Kuhn & 15.1--17.9 & 4.48--16.5 & 0.76--1.24 & 0.17--0.59 \\
\bottomrule
\end{tabular}

\end{table}

\begin{table}[t]
\centering
\caption{Structural tax $\tax$ and realized leak $|\Lambda|$
under a rake on three-player Kuhn and Leduc.}
\label{tab:rake}
\footnotesize
\begin{tabular}{lrrrr}
\toprule
Game & Rake & Max $\tax$ & Max $\tax$ / utility scale & Max $|\Lambda|$ \\
\midrule
Kuhn & 0.0\% & 0.0000 & 0.0\% & 0.0000 \\
Kuhn & 2.5\% & 0.0702 & 1.8\% & 0.0010 \\
Kuhn & 5.0\% & 0.1381 & 3.7\% & 0.0008 \\
Kuhn & 10.0\% & 0.2858 & 8.4\% & 0.0010 \\
Leduc & 0.0\% & 0.0000 & 0.0\% & 0.0000 \\
Leduc & 2.5\% & 0.0730 & 0.5\% & 0.0012 \\
Leduc & 5.0\% & 0.0960 & 0.7\% & 0.0076 \\
Leduc & 10.0\% & 0.0232 & 0.2\% & 0.0029 \\
\bottomrule
\end{tabular}

\end{table}

\begin{table}[ht]
\centering
\caption{Trunk scaling, median
wall-clock time over six seeds with the subgame fixed at $6$ information sets.}
\label{tab:trunk-main}
\footnotesize
\begin{tabular}{rcc}
\toprule
$K$ & \shortstack{Cached\\exact (ms)} & \shortstack{Local /\\cached} \\
\midrule
$1$ & $2.1$ & $0.96\times$ \\
$4$ & $8.1$ & $0.25\times$ \\
$16$ & $32.8$ & $0.061\times$ \\
$64$ & $145$ & $0.014\times$ \\
\bottomrule
\end{tabular}
\end{table}

\Cref{tab:cost} splits the full
certificate into two parts, the activation term $\Act(\bp)$ and the margin
pass. Because $\Act(\bp)$ depends only on the blueprint and the subgame,
precomputing it for reachable subgames pays this cost once. The per-update cost
is then the margin pass, one best response per opponent inside $S$. This pass
takes \costCachedMin{} to \costCachedMax{} of the exact full-game evaluation
across the five games, compared to \costUncachedMin{} to \costUncachedMax{}
when recomputing the full certificate every time. A trunk-scaling experiment shows that the local certificate's cost stays

A rake takes a fraction of the pot, capped at one chip, so poker stops being constant-sum and $\tax>0$. Across rakes of $2.5\%$, $5\%$, and $10\%$ of the pot (\Cref{tab:rake}), the tax reaches at most \rakeThetaKuhnMax{} of the utility scale (the largest absolute terminal payoff) on Kuhn and \rakeThetaLeducMax{} on Leduc. The realized leak $|\Lambda|$ of the exact form is one to two orders of magnitude smaller.

Beyond poker, we searched for counterexamples where
$\Vul_G$ rises while every term of the bound stays silent. Our certified
best-response search covered $162$ random three-player extensive-form games of depth at most
three with random utilities, CFR blueprints, and gadget resolves. Evaluating
every quantity without sampling, we found no case where $\Delta\Vul$ exceeds the
tolerance $0.02$ without a term of the bound accounting for the increase. Numerical checks also corroborate every inequality of \cref{thm:main} on
$97$ and $6{,}157$ finite configurations.

\begin{table}[ht]
\centering
\caption{Acceptance of the $162$ delivered campaign updates. Acc.\ $t$ counts updates with $B\le t$.}
\label{tab:acceptance-main}
\footnotesize
\begin{tabular}{lcc}
\toprule
Certificate & Acc.\ $0$ & Acc.\ $0.02$ \\
\midrule
$B_{\mathrm{dir}}$ (\cref{eq:direct}) & $2$ & $73$ \\
$B_{\mathrm{cache}}$ (\cref{cor:cache}) & $37$ & $104$ \\
$X=\Lambda$ (realized leak) & $0$ & $72$ \\
$X=L_i$ (confined bound) & $0$ & $45$ \\
$X=\tax$ (sharp tax) & $0$ & $19$ \\
$X=2n\dinf$ (\cref{cor:2ndelta}) & $0$ & $2$ \\
\bottomrule
\end{tabular}
\end{table}

The same $162$ delivered updates test the bound's practical selectivity.
Every form of the bound flags all $19$ harmful updates, while the
baseline-aware form $B_{\mathrm{cache}}$ accepts $37$ of the $143$ safe
ones at zero tolerance against $2$ for $B_{\mathrm{dir}}$
(\cref{tab:acceptance-main}).

flat at $2.0$\,ms, whereas exact evaluation grows linearly with the game
(\cref{tab:trunk-main}). The search protocol, a two-term ablation of the
bound, and an $n$-sweep appear in
\cref{app:machine,app:submitted,app:expdetails}.

\section{Conclusion}\label{sec:conclusion}

This paper gives the first safety guarantee for subgame resolving with three
or more independent players, where the two-player identity that one player's
gain is the other's loss fails. A post-hoc certificate takes its place. Any
delivered update raises the profile's best-response gap over the blueprint by
at most four computable terms. These are the blueprint's unkept promise
$\Act$, the structural tax $\tax$, the best-response margin deficits
$\sum_q\Mdef{q}$, and the solve error $\epsi$. Summed over players, the bound
on NashConv keeps only the middle two. Each term is a distinct channel of
harm, and explicit constructions attain all four with equality. The structural
tax vanishes on pointwise constant-sum games, which include every rake-free
poker variant.

On three-player Leduc poker and on Kuhn poker with three to six
players, resolving without the certificate is harmful in \harmLeducUnsafe{} to
\sixHarmUnsafe{} of the updates. The certificate flags every one of them, and
a resolver that keeps the blueprint when the certificate exceeds its tolerance
delivers no harmful update. Because the check needs only best responses inside the subgame, real-time multiplayer
solvers that refine a blueprint can therefore use it as a safety check on each
resolve.

\bibliography{references}
\bibliographystyle{dilab_ref}
\makeappendixtoc

\appendix
\section{Additional Related Work}\label{app:related}

\textbf{Safe exploitation and margin certificates.}
Safe exploitation bounds the value lost when deviating from an equilibrium to
exploit an assumed-weak opponent \citep{ganzfried2015safe} and certifies
the deployed strategy against a reference before play \citep{li2026agents}.
Certificates of solution quality measure how close a computed profile is to
Nash in general-sum games \citep{li2026we}.
In subgame refinement, recent work ensures that the refined strategy never
worsens guaranteed value \citep{liu2022safe} or loses robustness across
opponent classes \citep{ge2024safe}.
Opponent-limited search \citep{liu2023opponent} and adaptation beyond the
depth limit \citep{milec2025adapting} can shrink the subgame itself.
Our margin certificate is a post-hoc verifier of a chosen update.
Built from one exact subgame best response per opponent, it prices every alarm
by counterfactual mass.
Whereas practical exploitability measures for large games give only lower
bounds \citep{lisy2017eqilibrium}, our $\Vul_G$ is exactly computable.

\textbf{Adversarial team games.}
Adversarial team games give multiplayer play solvable structure, as a coordinating team faces a single adversary. Only the team's internal coordination prevents two-sided reasoning. Subgame solving extends to this setting \citep{zhang2022subgame}, and the team belief DAG enables correlated team-maxmin equilibria \citep{carminati2022marriage,zhang2023team}. We consider self-interested players without team structure.

\textbf{Blueprint computation and multiplayer equilibria.} Resolving requires a blueprint produced by a solver. In two-player zero-sum games, CFR converges to an equilibrium \citep{zinkevich2007regret}. Recent analyses explain the linear last-iterate convergence of a scale-invariant regret-matching scheme and why alternating updates accelerate the solver \citep{li2026last,li2026alternation}. In practice, CFR scales via Monte Carlo variants \citep{lanctot2009monte}, discounted regret schedules \citep{brown2019solving}, and neural approximation \citep{brown2019deep}. Search-plus-RL systems preserve this structure \citep{brown2020combining}. Solver engineering continues to scale this pipeline via parallel regret minimization \citep{li2026real} and by compiling the game into a static dataflow that replays on the GPU \citep{li2026gpu}. Correlated chance sampling \citep{li2026correlated} and action abstraction \citep{li2024rl,li2025efficient,li2026effective,li2026abstraction} push this scale further. The belief learning behind depth-limited solving is also being accelerated \citep{liturborebel}. Tournament play solves continuations with tournament equity as the objective \citep{li2026icm}, and language-model agents have been prompted to play without a trained solver \citep{li2026pokerskill}. Our certificate applies to the delivered update regardless of how it was produced. In multiplayer games, the equilibrium target itself deteriorates. Computing an equilibrium is PPAD-complete \citep{daskalakis2009complexity}, and equilibria are not interchangeable. Multiplayer exploitability also lacks a canonical definition, which motivates our profile-level best-response gap. Prior multiplayer work covers regret minimization \citep{abou2010using,gibson2011regret}, with Kuhn poker \citep{kuhn1953extensive} as the standard small testbed, alongside direct Nash algorithms \citep{ganzfried2024fast,ganzfried2026variable,ganzfried2026projected,zhang2023computing}. None of these methods addresses the safety of a refinement update.

\textbf{Polymatrix games.}
Zero-sum polymatrix games admit polynomial-time equilibria
\citep{cai2016zero} and minmax-type theorems beyond two players
\citep{cai2011minmax}. Under a distance-to-CSP measure, self-play guarantees
degrade gracefully \citep{macqueen2023guarantees}. Here that distance serves
as an a priori upper bound on the structural tax in a safety bound for
resolving.

\textbf{Positioning.}
Prior methods each supply one piece of our setting. Two-player safe resolving guarantees safety only in two-player zero-sum games \citep{burch2014solving,brown2017safe}, whereas Pluribus operates in multiplayer games without a safety guarantee \citep{brown2019superhuman}. In adversarial team games, multiplayer safety relies on a team structure \citep{zhang2022subgame}. Correlated resolving certifies a correlation plan \citep{ling2022safe}, and safe exploitation conditions on an assumed opponent model \citep{ganzfried2015safe}. Polymatrix decomposability supplies the distance measure \citep{macqueen2023guarantees}. Our result in \cref{sec:main} combines these elements to guarantee safety in multiplayer play for the delivered profile, tying the distance measure to resolving.

\section{Proof of the Exact Case}\label{app:L3}

This appendix proves \cref{lem:delta0}. It also gives the constructions for \cref{ex:activation,ex:crossopp} and the two in \cref{ex:leak} attaining the structural-tax and solve-error terms of \cref{thm:main}. Notation follows \cref{sec:prelim}. We first restate all objects, give the proof in numbered steps, and close with these four equality constructions. Throughout, $(x)_+ := \max\{x, 0\}$.

\subsection{Setting, Objects, and Assumptions}\label{app:L3:setup}

\paragraph{Game, subgame, entries.} $G$ is an $n$-player finite extensive-form game with perfect recall and chance acting first. Each non-terminal history carries a public label $\mathrm{pub}(h)$, the information visible to all players. Players form $N = \{1, \dots, n\}$, with resolver $i$ and opponents $O = N \setminus \{i\}$. Terminals $Z$ have utilities $u_p(z)$. Fix a public state $\mathrm{pub}^*$. The entry set $E$ consists of the earliest histories $e$ with $\mathrm{pub}(e) = \mathrm{pub}^*$. Each entry carries its chance probability $c_e$ and the per-player blueprint reaches $r_q(e) = \pi_q^{\bp}(e)$. The subgame $S$ is the set of descendants of $E$. Here $Z_S \subset Z$ are the terminals passing through $S$ and $Z_{\mathrm{out}}$ the rest.

\textbf{Entry information sets and counterfactual entry mass.}
For each player $q$, group $E$ by $\iota_q(e) := (\mathrm{priv}_q(e), \mathrm{pub}^*)$ and write $\iota \in I_q$ for the resulting entry information sets. The gadget places $q$'s Terminate or Follow ($T/F$) decision at these sets. The counterfactual entry mass is
\begin{equation}\label{eq:app-L3-wbar}
\bar{W}_q(\iota) := \sum_{e \in \iota} c_e \, \pi_{-q}^{\bp}(e),
\qquad
\pi_{-q}^{\bp}(e) := \prod_{p \neq q} r_p(e) .
\end{equation}

The resolved profile $\rs$ satisfies $\rs_p = \bp_p$ for all $p \neq i$, and $\rs_i$ differs from $\bp_i$ only at information sets of $i$ inside $S$. The lemma holds for \emph{any} such update. Every certificate quantity below is defined post hoc for the delivered $\rs$.

\textbf{Per-player gains.}
For a profile $\sigma$ let
$A_p(\sigma) := \max_{\tau_p} u_p(\tau_p, \sigma_{-p}) - u_p(\sigma)$, so that
$\Vul_G(\sigma) = \max_p A_p(\sigma)$ (\cref{def:vul}). Write
$\mathrm{gain}_p := u_p(\rs) - u_p(\bp)$ and
$\mathrm{drop}_p := -\mathrm{gain}_p$.

We use the CFR-D-style joint gadget \citep{burch2014solving,
moravcik2016refining, brown2017safe}.

\begin{itemize}
\item \textbf{(G1) Joint gadget.} A root chance node enters entry $e$ with
probability proportional to $c_e \cdot r_i(e)$ (opponent reach excluded, as in
robust/maxmargin entry). Below each entry, every opponent $j$ receives, in a
fixed order, a binary $T/F$ decision node with information set
$\iota_j(e)$. An opponent choosing $T$ is frozen to its blueprint behavior inside the subgame copy, and its terminal payoff coordinate is replaced by the constant $\mathrm{CFV}_j(\iota)$. An opponent choosing $F$ acts freely and receives its true payoff, while the resolver always acts freely.
\item \textbf{(G2) Blueprint counterfactual values.}
\begin{equation}\label{eq:app-L3-cfv}
\mathrm{CFV}_j(\iota) :=
\frac{\sum_{e \in \iota} c_e \, \pi_{-j}^{\bp}(e) \cdot
      \mathrm{EV}_j^{\bp}(e)}{\bar{W}_j(\iota)},
\end{equation}
where $\mathrm{EV}_j^{\bp}(e)$ is $j$'s expected value from $e$ under the
all-blueprint profile. Both $\mathrm{CFV}_j$ and $\mhat_j$ are defined only at groups with $\bar{W}_j(\iota) > 0$. A zero-mass group has $c_e\,\pi_{-j}^{\bp}(e) = 0$ at each of its entries, so it contributes $0$ to every entry sum below and is omitted from $\Mdef{j}$ and from the margin-certified condition (\cref{app:calibers}).
\item \textbf{(G4) Subgame best-response values (game-side measure).} For
$j \in O$, $\iota \in I_j$, and a subgame strategy $x_i$ of the resolver,
\begin{equation}\label{eq:app-L3-sbv}
\mathrm{SBV}_j(\iota;\, x_i) :=
\max_{\tau_j^S} \sum_{e \in \iota} c_e \, \pi_{-j}^{\bp}(e) \cdot
\mathbb{E}\!\left[\, u_j \mid e,\ (\tau_j^S,\, x_i,\, \bp_{-ij})
   \text{ inside } S \,\right],
\end{equation}
the maximum ranging over $j$'s behavioral strategies inside $S$. The environment inside $S$ is $x_i$ for the resolver and the blueprint for everyone else, where $\bp_{-ij}$ denotes the blueprint of all players other than $i$ and $j$.
\item \textbf{(G5) BR-certified margin and margin deficit} (\cref{def:margin,def:deficit}).
\begin{equation}\label{eq:app-L3-margin}
\mhat_j(\iota) := \mathrm{CFV}_j(\iota) -
  \frac{\mathrm{SBV}_j(\iota;\, \rs_i)}{\bar{W}_j(\iota)},
\qquad
\Mdef{j} := \sum_{\iota \in I_j} \bar{W}_j(\iota)\,
  \bigl(-\mhat_j(\iota)\bigr)_+ .
\end{equation}
The deficit is weighted by $\bar{W}_j$, never divided by it. Its normalization depends on the entry structure because $\sum_\iota \bar{W}_j(\iota) = \sum_{e \in E} c_e\, \pi_{-j}^{\bp}(e)$ omits $j$'s own reach. If $j$ has a private pre-entry action whose branches all reach $\mathrm{pub}^*$, each branch contributes its full environmental mass and the sum can exceed $1$. When $j$ follows one action sequence into every $e \in E$ (as when all actions before $\mathrm{pub}^*$ are public), let $\tau_j$ play that sequence. Then $\pi_j^{\tau}(e) = 1$ for all $e \in E$ and $\sum_\iota \bar{W}_j(\iota) = \Pr_{(\tau_j, \bp_{-j})}[\text{reach } E] \le 1$, since entries are pairwise non-nested histories. Consequently, $\Mdef{j} \le \bigl(-\min_\iota \mhat_j(\iota)\bigr)_+$ and the weighted-sum convention is directly comparable to a min-margin convention.
\item \textbf{(G7) Activation and solve error.}
\begin{equation}\label{eq:app-L3-act}
\begin{split}
\Act(\bp) &:= \max_{\tau_i \,:\, \tau_i = \bp_i \text{ outside } S}
  u_i(\tau_i, \bp_{-i}) - u_i(\bp)
  \quad\bigl(\le A_i(\bp) \le \Vul_G(\bp)\bigr),\\
\epsi &:= \Act(\bp) - \mathrm{gain}_i .
\end{split}
\end{equation}
Here $\varepsilon_{\mathrm{solve}}$ bounds the solve error the delivered resolver may carry, so $\epsi\le\varepsilon_{\mathrm{solve}}$ by assumption. The corollary below replaces $\epsi$ with this budget when only the solver's accuracy class is known.
Because $\rs_i$ is a feasible point of the maximization defining $\Act$,
\emph{$\epsi \ge 0$ always holds}. The quantity $\epsi$ measures the suboptimality of $\rs_i$ against blueprint opponents inside $S$. It vanishes for an exact gadget solve when all opponents choose $T$ and the root measure induces the resolver's game-side beliefs at each of its entry sets. Both $c_e\,\pi_{-i}(e)$ and the full blueprint reach give the correct beliefs because $i$'s own reach is constant on each of its information sets, whereas the implementation measure $c_e\,r_i(e)$ of \cref{app:calibers} does not in general. For iterative implementations, $\epsi$ comprises solve error plus gadget-environment drift and can be measured exactly post hoc.
\end{itemize}

\begin{remark}[Margin convention and implementation deviation]\label{rem:app-L3-margincaliber}
The margins in \cref{eq:app-L3-margin} are \emph{BR-certified}, meaning the $F$-branch is evaluated by an exact subgame best response of $j$ under the game-side measure $c_e \, \pi_{-j}^{\bp}(e)$ and the environment $(\rs_i, \bp_{-ij})$. Common implementations evaluate the $F$-branch with $j$'s own gadget average strategy under the gadget-internal measure. That average-strategy margin exceeds $\mhat$ on 33 of 40 entry information sets in our numerical regression check, with a worst overstatement of $+1.667$ on an off-path branch. The theorem is stated with $\mhat$ (\cref{app:calibers}).
\end{remark}

\Cref{asm:F1} has two properties. Under \emph{(P1)}, every information set of every player is either entirely inside $S$ or disjoint from it, the standard property of public-state subgames. Under \emph{(P2)}, the gadget's entry grouping $\iota_q = (\mathrm{priv}_q, \mathrm{pub}^*)$ agrees with $q$'s true information at entry, and $q$'s own reach $\pi_q^{\tau}(e)$ is constant across each $\iota$. This is automatic under perfect recall with public actions.

\emph{(F2, exactness)} In the linear programming (LP) profile-table semantics of \cref{def:delta}, $\dinf(G) = 0$ means there exists a table $\{u_{pq}\}$ on reduced pure-strategy pairs with
\begin{equation}\label{eq:app-L3-F2}
\begin{split}
u_p(\rho) &= \sum_{q \neq p} u_{pq}(\rho_p, \rho_q)
\ \ \text{for every pure profile } \rho,\\
u_{pq}(a, b) + u_{qp}(b, a) &= c_{pq}
\ \ \text{for edge constants } c_{pq},
\end{split}
\end{equation}
i.e.\ $G$ is exactly a constant-sum polymatrix (CSP) game
\citep{macqueen2023guarantees}.

\begin{remark}[Two definitional conventions]\label{rem:app-L3-Fdev}
(i) The CSP class is taken with the \emph{per-edge constant-sum} constraint of
\cref{eq:app-L3-F2} (with the $c_{pq}$ free), following
\citet{macqueen2023guarantees} and \citet{cai2016zero}. This class contains
the antisymmetric special case $u_{pq} + u_{qp} = 0$. The only property the
proof uses is that $\sum_p u_p$ is constant on the class, and constants cancel
in the difference taken in Step S7 below.
(ii) The tables $u_{pq}$ are defined on reduced pure-strategy pairs (LP profile-table semantics). A literal terminal-level reading, in which $u_{pq}(z)$ may depend arbitrarily on the full terminal history, would make every pointwise zero-sum game trivially CSP via $u_{pq} := (u_p - u_q)/n$, contradicting the measured lower bound $\dinf \ge 1$ for three-player Kuhn poker. Dependence on third-party randomness is allowed only through the chance expectation, while dependence on third-party strategies is excluded (\cref{app:calibers}).
\end{remark}

\subsection{Proof of \texorpdfstring{\cref{lem:delta0}}{the exact-case lemma}}
\label{app:L3:proof}

We now prove \cref{lem:delta0}. The proof is written under F2 and uses it only in Step S7, through the constant profile sum \cref{eq:app-L3-S7}, so it holds under the lemma's hypothesis of constant total utility. Under \cref{asm:F1} and $\dinf(G) = 0$, for any blueprint $\bp$ and any update $\rs$ that changes only $i$'s strategy inside $S$,
\begin{align}
\text{(i)}\quad & A_i(\rs) - A_i(\bp) = \epsi - \Act(\bp) \le \epsi ;
  \label{eq:app-L3-i}\\
\text{(ii)}\quad & A_j(\rs) - A_j(\bp) \le
  \Mdef{j} + \Bigl[\Act(\bp) - \epsi + \textstyle\sum_{k \in O \setminus \{j\}}
  \Mdef{k}\Bigr]\\
  & \le \Act(\bp) + \textstyle\sum_{q \in O} \Mdef{q}
  \qquad \forall j \in O ; \label{eq:app-L3-ii}\\
\text{(iii)}\quad & \Vul_G(\rs) - \Vul_G(\bp) \le
  \max\Bigl( \epsi,\ \Act(\bp) - \epsi +
  \textstyle\sum_{q \in O} \Mdef{q} \Bigr). \label{eq:app-L3-iii}
\end{align}
In particular, if $\mhat_j(\iota) \ge 0$ for all $j, \iota$
(margin-certified) and $\epsi \le \varepsilon_{\mathrm{solve}}$, then
$\Vul_G(\rs) \le \Vul_G(\bp) +
\max(\varepsilon_{\mathrm{solve}}, \Act(\bp))$. The bound has unit coefficients on all terms. The margin sum $\sum_{q \in O} \Mdef{q}$ carries coefficient $c_2 = 1$ with no inverse reach. The activation term $\Act(\bp)$ carries coefficient $1$, which weakens to $2$ on $\Vul_G(\bp)$ via $\Act \le \Vul_G(\bp)$. The solve-error term carries coefficient $1$ because only the resolver's solve quality enters, while the opponent-side certificates hold exactly for the delivered $\rs$.

\begin{proof}
Fix $\bp$, $\rs$, and $\mathrm{pub}^*$. All maxima below are over finite sets and are attained. Write $\sigma^x$ for $\bp$ if $x = 0$ and $\rs$ if
$x = 1$.

\medskip
\noindent\textbf{Step S0 (terminal decomposition, uses P1).} For any profile $\pi$ the expected utility splits over $Z_{\mathrm{out}}$ and $Z_S$, and a path through $S$ determines its entry $e \in E$ uniquely. For $z \in Z_S$,
$P^{\pi}(z) = c(z) \prod_p \pi_p(z)$, and the trunk segment (before $e$) of
every player's behavior is identical under $\bp$ and $\rs$, because $\rs$
changes only $i$'s information sets inside $S$.

\medskip
\noindent\textbf{Step S1 (trunk-subgame decomposition of opponent strategies, uses P1, P2, perfect recall).} For $j \in O$ and any strategy $\tau_j$, decompose $\tau_j = (\tau_j^{\mathrm{tr}},\, \{\tau_j^S(\iota)\}_{\iota \in I_j})$ into its trunk behavior outside $S$ and its continuations inside $S$. Perfect recall lets $j$ choose its continuation inside $S$ independently at each $\iota$, and P2 makes $j$'s own reach constant on $\iota$, written $\pi_j^{\tau}(\iota) \in [0, 1]$. Then for $x \in \{0, 1\}$,
\begin{equation}\label{eq:app-L3-S1}
u_j(\tau_j, \sigma^x_{-j})
= \mathrm{OUT}_j(\tau_j^{\mathrm{tr}})
+ \sum_{\iota \in I_j} \pi_j^{\tau}(\iota) \cdot
  G_j^x\bigl(\iota;\, \tau_j^S(\iota)\bigr),
\end{equation}
where $\mathrm{OUT}_j$ is the payoff collected on terminals outside $S$ and
\begin{equation}\label{eq:app-L3-Gx}
G_j^x(\iota;\, \tau^S) :=
\sum_{e \in \iota} c_e \, \pi_{-j}^{\bp}(e) \cdot
\mathbb{E}\!\left[\, u_j \mid e,\ (\tau^S,\, \sigma^x_i,\, \bp_{-ij}) \,\right],
\end{equation}
and $\mathrm{OUT}_j$ and the $\pi_j^{\tau}(\iota)$ do not depend on $x$ (the
trunk and $Z_{\mathrm{out}}$ are identical under both profiles).

\medskip
\noindent\textbf{Step S2 (blueprint-value identity, by definition G2).}
$G_j^0(\iota;\, \bp_j|_S) = \bar{W}_j(\iota) \cdot \mathrm{CFV}_j(\iota)$,
by expanding \cref{eq:app-L3-cfv} entry by entry.

\medskip
\noindent\textbf{Step S3 (BR-certificate identity, by definitions G4, G5).}
\begin{equation}\label{eq:app-L3-S3}
\max_{\tau^S} G_j^1(\iota;\, \tau^S)
= \mathrm{SBV}_j(\iota;\, \rs_i)
= \bar{W}_j(\iota) \cdot
  \bigl[\mathrm{CFV}_j(\iota) - \mhat_j(\iota)\bigr].
\end{equation}

\medskip
\noindent\textbf{Step S4 (opponent BR-value change is at most $\Mdef{j}$).}
Let $\mathrm{BRval}_j(x) := \max_{\tau_j} u_j(\tau_j, \sigma^x_{-j})$.
Take $\tau_j$ optimal for $x = 1$. By S1,
\begin{align}
\mathrm{BRval}_j(1)
&= \mathrm{OUT}_j + \sum_\iota \pi_j(\iota)\,
   G_j^1\bigl(\iota;\, \tau^S(\iota)\bigr) \nonumber\\
&\le \Bigl[ \mathrm{OUT}_j + \sum_\iota \pi_j(\iota)\,
   \bar{W}_j(\iota)\, \mathrm{CFV}_j(\iota) \Bigr] \nonumber\\
&\qquad + \sum_\iota \pi_j(\iota)\, \bar{W}_j(\iota)\,
   \bigl(-\mhat_j(\iota)\bigr)_+
   && \text{(S3; } (-\mhat)_+ \ge -\mhat \text{)} \nonumber\\
&\le \mathrm{BRval}_j(0) + \Mdef{j},
   \label{eq:app-L3-S4}
\end{align}
where the last inequality uses $\pi_j(\iota) \le 1$ and the fact that the
bracketed term is, by S2, the value of the feasible strategy
$(\tau_j^{\mathrm{tr}}, \text{blueprint continuations})$ against $\bp_{-j}$,
hence at most $\mathrm{BRval}_j(0)$. Thus $\mathrm{BRval}_j(1) - \mathrm{BRval}_j(0) \le \Mdef{j}$. No CSP structure is used, so this step holds for arbitrary $\dinf$.

\medskip
\noindent\textbf{Step S5 (opponent on-path gain is at most $\Mdef{k}$).}
Here on-path gain means the change in $k$'s expected payoff under the
profile actually played, with $k$ following the blueprint.
For $k \in O$, expand $\mathrm{gain}_k = u_k(\rs) - u_k(\bp)$ via S1 with $x = 1$ versus $x = 0$ and $\tau_k = \bp_k$, where the $\mathrm{OUT}$ and $\pi_k(\iota)$ terms cancel.
\begin{align}
\mathrm{gain}_k
&= \sum_\iota \pi_k^{\bp}(\iota)
   \bigl[ G_k^1(\iota;\, \bp_k|_S) - G_k^0(\iota;\, \bp_k|_S) \bigr]
   \nonumber\\
&\le \sum_\iota \pi_k^{\bp}(\iota)
   \bigl[ \mathrm{SBV}_k(\iota;\, \rs_i)
        - \bar{W}_k(\iota)\, \mathrm{CFV}_k(\iota) \bigr]
   && \text{(S2)} \nonumber\\
&= \sum_\iota \pi_k^{\bp}(\iota)\, \bar{W}_k(\iota)\,
   \bigl(-\mhat_k(\iota)\bigr)
 \;\le\; \Mdef{k}. \label{eq:app-L3-S5}
\end{align}
The blueprint continuation is a feasible point in the maximization defining $\mathrm{SBV}_k(\iota;\rs_i) = \max_{\tau^S} G_k^1(\iota;\tau^S)$, which gives the first inequality. The bracket then equals $-\bar{W}_k(\iota)\,\mhat_k(\iota)$ by S3, and $x \le (x)_+$ with $\pi_k^{\bp}(\iota) \le 1$ gives the bound by $\Mdef{k}$. This step captures value transfers across opponents. If $i$ cedes value to $k$ inside $S$, then $k$'s Follow-and-best-respond value collects at least the ceded amount and $\mhat_k$ turns negative accordingly, ensuring that cross-opponent transfers always leave a margin trace.

\medskip
\noindent\textbf{Step S6 (resolver identity, by definition G7).}
$\mathrm{gain}_i = \Act(\bp) - \epsi$ with $\epsi \ge 0$. Moreover the
maximization term in $A_i$ does not depend on $i$'s own strategy and
$\rs_{-i} = \bp_{-i}$, so
$A_i(\rs) - A_i(\bp) = -\mathrm{gain}_i = \epsi - \Act(\bp)$.
This proves \cref{eq:app-L3-i}.

\medskip
\noindent\textbf{Step S7 (the only use of $\dinf = 0$, constant profile sum).}
By F2, for every pure profile $\rho$,
\begin{equation}\label{eq:app-L3-S7}
\sum_p u_p(\rho)
= \sum_p \sum_{q \neq p} u_{pq}(\rho_p, \rho_q)
= \sum_{p < q} \bigl[ u_{pq}(\rho_p, \rho_q) + u_{qp}(\rho_q, \rho_p) \bigr]
= \sum_{p < q} c_{pq} =: c^*,
\end{equation}
a constant independent of $\rho$. Taking expectations gives $\sum_p u_p(\sigma) = c^*$ for every mixed profile (a product of distributions over pure strategies), and hence for every behavioral profile, since under perfect recall each behavioral strategy is realization-equivalent to a mixed one \citep{kuhn1953extensive}. Hence the constant cancels in the
difference and
\begin{equation}\label{eq:app-L3-S7b}
\sum_p \mathrm{gain}_p = 0 .
\end{equation}

\medskip
\noindent\textbf{Step S8 (pricing the victim's on-path loss).}
For $j \in O$, by \cref{eq:app-L3-S7b},
\begin{equation}\label{eq:app-L3-S8}
\mathrm{drop}_j = -\mathrm{gain}_j
= \sum_{p \neq j} \mathrm{gain}_p
= \mathrm{gain}_i + \sum_{k \in O \setminus \{j\}} \mathrm{gain}_k
\le \bigl(\Act(\bp) - \epsi\bigr)
  + \sum_{k \in O \setminus \{j\}} \Mdef{k},
\end{equation}
using S5 and S6.

\medskip
\noindent\textbf{Step S9 (total change for each opponent).}
Since $A_j(\sigma) = \mathrm{BRval}_j(\sigma) - u_j(\sigma)$,
\begin{equation}\label{eq:app-L3-S9}
A_j(\rs) - A_j(\bp)
= \bigl[\mathrm{BRval}_j(1) - \mathrm{BRval}_j(0)\bigr] + \mathrm{drop}_j
\le \Mdef{j} + \bigl(\Act(\bp) - \epsi\bigr)
  + \sum_{k \neq j} \Mdef{k},
\end{equation}
by S4 and S8. The further weakening to $\Act(\bp) + \sum_{q \in O} \Mdef{q}$ uses $\epsi \ge 0$. This proves
\cref{eq:app-L3-ii}.

\medskip
\noindent\textbf{Step S10 (total bound).}
A difference of maxima is at most the maximum of differences.
\begin{align}
\Vul_G(\rs) - \Vul_G(\bp)
&\le \max_p \bigl[ A_p(\rs) - A_p(\bp) \bigr] \nonumber\\
&\le \max\Bigl( \epsi - \Act(\bp),\
  \max_{j \in O} \bigl[ \Mdef{j} + \Act(\bp) - \epsi +
  \textstyle\sum_{k \neq j} \Mdef{k} \bigr] \Bigr) \nonumber\\
&\le \max\Bigl( \epsi,\ \Act(\bp) - \epsi +
  \textstyle\sum_{q \in O} \Mdef{q} \Bigr).
\label{eq:app-L3-S10}
\end{align}
This proves \cref{eq:app-L3-iii}.
\end{proof}

\textbf{Dependency accounting.}
Steps S1 to S4 use only the entry decomposition, the blueprint counterfactual values, and the game-side best responses defining $\mhat$ (\cref{def:margin}), with no CSP assumption and without executing the gadget. Step S5 uses the same objects plus blueprint feasibility. Step S6 is definitional, Step S7 is the \emph{only} step using $\dinf = 0$, and Steps S8 to S10 are algebra. Consequently, the $\dinf > 0$ extension in \cref{app:T1T2} replaces Step S7 alone.

\subsection{Example 1: the Activation Term Cannot Be Removed}
\label{app:L3:ex1}

We detail the construction of \cref{ex:activation}, a two-player zero-sum game ($\dinf = 0$) with all margins certified strictly positive and $\epsi = 0$. Here $\Delta\Vul = \Act(\bp) > 0$, with equality in \cref{eq:app-L3-iii}.

\textbf{Construction.}
The players are $\{i, j\}$. Player $j$ moves first at the root, where $d$ ends the game with payoffs $(u_i, u_j) = (-1, +1)$ and $\mathit{enter}$ enters the subgame $S$. Inside $S$, only $i$ acts, with $a'$ yielding $(0, 0)$ and $a$ yielding $(+1, -1)$. \Cref{tab:app-L3-ex1} lists all terminals, and the blueprint is $\bp = (a', \mathit{enter})$.

\begin{table}[t]
\centering
\caption{Payoffs of the activation construction (\cref{ex:activation}).}
\label{tab:app-L3-ex1}
\footnotesize
\begin{tabular}{llrr}
\toprule
$j$ (trunk) & $i$ (inside $S$) & $u_i$ & $u_j$ \\
\midrule
$d$              & $\cdot$ & $-1$ & $+1$ \\
$\mathit{enter}$ & $a'$  & $0$  & $0$  \\
$\mathit{enter}$ & $a$   & $+1$ & $-1$ \\
\bottomrule
\end{tabular}
\end{table}

\textbf{Blueprint quantities.}
Under $\bp$, payoffs are $(0, 0)$. Player $i$'s unrealized subgame action $a$ gives $A_i(\bp) = \Act(\bp) = 1$, and player $j$'s unrealized deviation $d$ gives $A_j(\bp) = 1$. Hence $\Vul_G(\bp) = 1$.

\textbf{The update and its certificates.}
Take $\rs_i = a$. Then $\mathrm{CFV}_j = 0$ (blueprint value of the entry for
$j$), and $\mathrm{SBV}_j(\rs_i) = -1$ ($j$ has no action inside $S$), so
\[
\mhat_j = 0 - (-1) = +1 > 0 \qquad \text{(certified)},
\qquad
\epsi = \Act(\bp) - \mathrm{gain}_i = 1 - 1 = 0 .
\]

\paragraph{Outcome and equality.}
Under $\rs$, payoffs are $(+1, -1)$, so
$A_j(\rs) = \max(+1, -1) - (-1) = 2$ and $A_i(\rs) = 0$, giving
$\Vul_G(\rs) = 2 = \Vul_G(\bp) + \Act(\bp)$. The bound in \cref{eq:app-L3-iii} evaluates to $\max(\epsi,\ \Act - \epsi + \sum_q \Mdef{q}) = \max(0, 1 + 0) = 1$. \emph{Equality holds}, so the coefficient $1$ on $\Act(\bp)$ is optimal. All-positive margins cannot prevent this effect. By realizing its own unrealized subgame value, the resolver raises the victim's best-response gain by that exact amount. Two-player CFR-D and maxmargin resolving show the same phenomenon, as their theorems bound the resolver's exploitability while leaving the full profile's best-response gap out of scope. The activation term is therefore intrinsic to the best-response gap of \cref{def:vul} and already appears in two-player zero-sum games.

\subsection{Example 2: Cross-Opponent Transfer and the Necessity of the Sum}
\label{app:L3:ex2}

We detail the construction for \cref{ex:crossopp}, a three-player one-shot game where $\dinf = 0$, $\epsi = 0$, and the deviating opponent's margin is zero. Here the literal per-pair bound fails, while the per-player bound \cref{eq:app-L3-iii} holds with equality.

\textbf{Construction.}
Players $\{i, j, k\}$ act simultaneously once, with $i \in \{a', a\}$, $j \in \{\mathit{stay}, d\}$, and $k$ having no action. We define the game through a pair-local CSP decomposition with antisymmetric edges and edge constants $c_{pq} = 0$.
\begin{equation}\label{eq:app-L3-ex2-witness}
\begin{gathered}
u_{ij}(a, \mathit{stay}) = +1, \quad
u_{ij}(a, d) = 0, \quad
u_{ij}(a', \cdot) = 0, \quad
u_{ik}(a, \cdot) = -1, \\
u_{ik}(a', \cdot) = 0, \quad
u_{jk} \equiv 0,
\end{gathered}
\end{equation}
so $\dinf = 0$ exactly (and the pair-locality assumptions of the per-pair
corollaries in \cref{app:perpair} hold as well). The induced total utilities
are listed in \cref{tab:app-L3-ex2}.

\begin{table}[t]
\centering
\caption{Total payoffs induced by \cref{eq:app-L3-ex2-witness} (\cref{ex:crossopp}), where $k$ has no action.}
\label{tab:app-L3-ex2}
\footnotesize
\begin{tabular}{llrrr}
\toprule
$i$ & $j$ & $u_i$ & $u_j$ & $u_k$ \\
\midrule
$a'$ & $\mathit{stay}$ & $0$  & $0$  & $0$  \\
$a'$ & $d$             & $0$  & $0$  & $0$  \\
$a$  & $\mathit{stay}$ & $0$  & $-1$ & $+1$ \\
$a$  & $d$             & $-1$ & $0$  & $+1$ \\
\bottomrule
\end{tabular}
\end{table}

\textbf{Blueprint quantities.}
The blueprint is $\bp = (a', \mathit{stay})$, with payoffs $(0, 0, 0)$. Every player's best-response gain is $0$, so $\Vul_G(\bp) = 0$. Furthermore, $\Act(\bp) = 0$ because switching to $a$ changes $i$'s own total by $+1 - 1 = 0$.

\textbf{The update and its certificates.}
Take $\rs_i = a$, a legal update (the subgame here is the
whole one-shot interaction, with a single trivial entry of mass
$\bar{W}_q = 1$ for each opponent). Then $\epsi = 0$, and the margins are
\[
\mhat_j = \mathrm{CFV}_j - \mathrm{SBV}_j(\rs_i)
        = 0 - \max(-1, 0) = 0
\qquad (\text{$j$'s subgame BR is } d, \text{ worth } 0 = \mathrm{CFV}_j),
\]
\[
\mhat_k = \mathrm{CFV}_k - \mathrm{SBV}_k(\rs_i) = 0 - (+1) = -1,
\qquad \Mdef{k} = 1, \quad \Mdef{j} = 0 .
\]

\paragraph{Failure of the literal per-pair form.}
Player $j$'s best-response gain on the $(j, i)$ pair table jumps to $\Delta\mathrm{PairVul}_{ji} = \max(0, -1) - (-1) = 1 > 0 = (-\mhat_j)_+$. Here player $i$'s update transfers value across opponents. Taking $1$ from $j$ on the $(i,j)$ interaction and giving $1$ to $k$ on the $(i,k)$ interaction leaves $i$'s total unchanged. Victim $j$'s margin stays silent because $j$'s Follow-and-best-respond value equals $\mathrm{CFV}_j$, whereas the alarm fires at the \emph{beneficiary} $k$ ($\mhat_k = -1$). Under the joint gadget, no per-pair certificate can be keyed to the victim's own margin. The theorem prices the transfer at the per-player level through the beneficiary's margin and the resolver's unrealized subgame value.

\textbf{Equality of the per-player bound.}
With $A_j(\rs) = \max(0, -1) - (-1) = 1$, we have $\Delta\Vul = 1 = \sum_q \Mdef{q}$ because $\Mdef{k} = 1$. The bound in \cref{eq:app-L3-iii} evaluates to $\max(0,\ 0 - 0 + 1) = 1$, achieving \emph{equality again}. This shows that the sum over all opponents in \cref{eq:app-L3-ii,eq:app-L3-iii} cannot be weakened to the victim's own margin and pins the margin coefficient $c_2 = 1$.

\subsection{Example 3: the Tax and Solve-Error Terms Are Attained}
\label{app:L3:ex34}

The two constructions of \cref{ex:leak} complete the tightness results. Together with \cref{ex:activation,ex:crossopp}, they attain every term of \cref{thm:main}(iii) and both branches of its maximum with equality. Both games satisfy \cref{asm:F1}, having one entry and no private information. In both, the certificate quantities are computed for the delivered update.

\paragraph{Leak construction ($\Delta\Vul = \tax = \Lambda$).} Take players $\{i, j\}$, with $j$ moving first. Playing $d$ ends the game at $(0, 0)$, while $\mathit{enter}$ reaches the single entry of $S$. Inside $S$, only $i$ acts, where $a'$ yields $(0, 0)$ and $a$ yields $(0, -1)$. The blueprint is $\bp = (a', \mathit{enter})$ and the update is $\rs_i = a$. \emph{Blueprint quantities.} All payoffs are $0$ under $\bp$, both actions of $i$ are worth $0$ to $i$, and both actions of $j$ are worth $0$ to $j$. This gives $A_i(\bp) = A_j(\bp) = 0$, $\Vul_G(\bp) = 0$, and $\Act(\bp) = 0$. \emph{Certificates.} We have $\mathrm{gain}_i = 0$ and $\epsi = 0$. Also, $\bar{W}_j = c_e \pi_i^{\bp}(e) = 1$, $\mathrm{CFV}_j = 0$, and $\mathrm{SBV}_j(\rs_i) = -1$ since $j$ has no action inside $S$. The resulting $\mhat_j = +1$ and $\Mdef{j} = 0$ make the update margin-certified. \emph{Structure.} The four pure profiles have $U(s) \in \{0, 0, 0, -1\}$, so $\zprof = \tfrac12$, while $Z_S$ has $U(z) \in \{0, -1\}$, giving $\zsub = \tfrac12$. Since $\rho_S = 1$, we obtain $\tax = 2\min(\tfrac12, \tfrac12) = 1$. The realized leak is $\Lambda = \sum_p u_p(\bp) - \sum_p u_p(\rs) = 0 - (-1) = 1$. \emph{Outcome.} Under $\rs$ the payoffs are $(0, -1)$, so $A_i(\rs) = 0$ and $A_j(\rs) = \max(0, -1) - (-1) = 1$, yielding $\Vul_G(\rs) = 1$. The bound evaluates to $\max(\epsi,\ \Act - \epsi + \sum_q \Mdef{q} + \tax) = \max(0,\ 0 + 0 + 1) = 1 = \Delta\Vul$, giving \emph{equality}. The same holds with $\Lambda$ in place of $\tax$ (\cref{cor:leak}). Here the resolver destroys one unit of $j$'s utility at no cost to itself. No margin, activation, or solve-error term detects the change, so only the total-utility oscillation prices it. With two players, the CSP distance is the distance to a constant-sum table, $\dinf = \zprof / 2 = \tfrac14$. Thus $2n\,\dinf = 1 = \tax$, attaining the $2n\,\dinf$ envelope of \cref{eq:app-T1-b} with equality at $n = 2$.

\paragraph{Self-harm construction ($\Delta\Vul = \epsi$).}
Players $\{i, j\}$ play a zero-sum game ($\dinf = 0$ and $\tax = 0$) in which the subgame is the whole game and $i$ alone acts. Action $a$ yields $(+1, -1)$ and $a'$ yields $(0, 0)$. The blueprint is $\bp_i = a$, updated to $\rs_i = a'$.
\emph{Blueprint quantities.} Because $a$ is optimal for $i$, we have $A_i(\bp) = \Act(\bp) = 0$. Player $j$ has no action, giving $A_j \equiv 0$ and $\Vul_G(\bp) = 0$.
\emph{Certificates.} From $\mathrm{gain}_i = -1$, we obtain $\epsi = \Act(\bp) - \mathrm{gain}_i = 1$. Player $j$ has $\bar{W}_j = 1$, $\mathrm{CFV}_j = -1$, and $\mathrm{SBV}_j(\rs_i) = 0$, so $\mhat_j = -1$ and $\Mdef{j} = 1$.
\emph{Outcome.} Since $A_i(\rs) = \max(1, 0) - 0 = 1$, we get $\Vul_G(\rs) = 1$ and $\Delta\Vul = 1 = \epsi$. Claims (i) and (ii) hold with equality, as $A_i(\rs) - A_i(\bp) = 1 = \epsi - \Act$ and $0 \le \Mdef{j} + \Act - \epsi = 0$. Bound (iii) evaluates to $\max(\epsi,\ \Act - \epsi + \Mdef{j} + \tax) = \max(1,\ 0) = 1$, attaining the $\epsi$ branch. The resolver's own loss raises the best-response gap, while the opponent's margin deficit records the handed-over value and cancels against $-\epsi$ in the second branch.

\section{Proof of the Transfer Lemma and Main Theorem}\label{app:T1T2}

This appendix proves \cref{lem:transfer}. With the constant-sum proof in
\cref{app:L3}, this gives \cref{thm:main} and \cref{cor:2ndelta}. It closes by
explaining why $\delta$ enters only through the total utility, without passing
through a proxy constant-sum polymatrix (CSP) game. Entries $E$, subgame $S$,
terminals $Z_S$ and $Z_{\mathrm{out}}$, and entry information sets $\iota$ keep
their earlier definitions. So do masses $\bar{W}$, values $\mathrm{CFV}$,
margins $\mhat$, deficits $\Mdef{q}$, activation $\Act$, solve error $\epsi$,
gains $\mathrm{gain}_p$ and $\mathrm{drop}_p$, best-response gains $A_p$, and
$\Vul_G$.

\subsection{Additional Objects}\label{app:T1T2:setup}

\paragraph{Profile tables and the CSP subspace.}
Let $s = (s_1, \dots, s_n)$ range over reduced pure-strategy
profiles and let $u_p(s)$ denote $p$'s utility after taking the
expectation over chance (the entries of the profile utility tensor). Write
\begin{equation}\label{eq:app-T1-S}
U(s) := \sum_p u_p(s), \qquad
U(z) := \sum_p u_p(z) \ \ \text{for terminals } z .
\end{equation}
Let $V$ be the set of profile utility tables admitting a decomposition
$u_p(s) = \sum_{q \neq p} u_{pq}(s_p, s_q)$ with per-edge constant-sum
constraints $u_{pq}(a, b) + u_{qp}(b, a) = c_{pq}$ for edge constants
$c_{pq}$ \citep{cai2016zero,macqueen2023guarantees}. The set $V$ is a
\emph{linear subspace}. The decomposition is componentwise linear and the
constraints are homogeneous linear, so a sum or scalar multiple of two
decompositions is again a decomposition. By \cref{def:delta},
$\dinf(u) = \min_{v \in V} \|u - v\|_\infty$ is the sup-distance to $V$ in
the LP profile-table semantics.

\textbf{Distances to constant-sum.}
Define
\begin{equation}\label{eq:app-T1-zeta}
\begin{split}
\zprof &:= \min_{c \in \mathbb{R}} \max_s |U(s) - c|
        = \tfrac{1}{2}\bigl(\max_s U(s) - \min_s U(s)\bigr),\\
\zeta_{\mathrm{term}} &:= \min_c \max_{z \in Z} |U(z) - c|,
\qquad
\zsub := \min_c \max_{z \in Z_S} |U(z) - c| .
\end{split}
\end{equation}
Clearly $\zprof \le \zeta_{\mathrm{term}}$ (each $U(s)$ is a convex
combination of the $U(z)$) and $\zsub \le \zeta_{\mathrm{term}}$.

Let $\rho_S := \sum_{e \in E} c_e \prod_p \pi_p^{\bp}(e) \in [0, 1]$. Since
$\rs$ changes strategy only inside $S$, the reach probability of $Z_S$ under
$\rs$ remains $\rho_S$.

As in \cref{app:L3:setup}, a legal update satisfies $\rs_p = \bp_p$ for
all $p \neq i$ and changes $\rs_i$ only at $i$'s information sets inside
$S$. All certificate quantities are defined post hoc for the delivered $\rs$.
We assume \cref{asm:F1} (P1, P2) and let $\dinf$ be arbitrary.

\subsection{Proof of the Transfer Lemma
(\texorpdfstring{\cref{lem:transfer}}{T1})}\label{app:T1T2:T1}

We prove \cref{lem:transfer} together with its a priori envelope. With $\delta := \dinf(G)$ the distance of \cref{def:delta},
\begin{itemize}
\item[(a)] \emph{(profile level)} $\zprof \le n \cdot \delta$,
\item[(b)] \emph{(update level)} for any legal update $\rs$,
\begin{equation}\label{eq:app-T1-b}
\begin{split}
\Bigl| \sum_p \mathrm{gain}_p \Bigr|
&= \Bigl| \sum_p \bigl[u_p(\rs) - u_p(\bp)\bigr] \Bigr|
 \;\le\; \tax := 2 \cdot \min\bigl( \zprof,\ \rho_S \cdot \zsub \bigr)\\
&\le 2\,\zprof \;\le\; 2n\,\delta .
\end{split}
\end{equation}
\end{itemize}

\begin{proof}[Proof of (a)]
Take an optimal decomposition $\{u_{pq}\}, \{c_{pq}\}$ attaining $\delta$.
For each pure $s$ set
$e_p(s) := u_p(s) - \sum_{q \neq p} u_{pq}(s_p, s_q)$, so
$|e_p(s)| \le \delta$ for every $p$. Summing over players,
\begin{equation}\label{eq:app-T1-a}
U(s)
= \sum_p \sum_{q \neq p} u_{pq}(s_p, s_q) + \sum_p e_p(s)
= \sum_{p < q} c_{pq} + \sum_p e_p(s),
\end{equation}
where the pairwise terms are eliminated by pairing each edge and applying
the constant-sum constraint. Hence
$|U(s) - c^*| \le n\delta$ with $c^* := \sum_{p < q} c_{pq}$, and taking
the minimum over centers, $\zprof \le n\delta$.
\end{proof}

\begin{proof}[Proof of (b), first leg ($\zprof$)]
Under perfect recall, behavioral and mixed strategies are
realization-equivalent \citep{kuhn1953extensive}. Against every opposing
profile they reach each terminal history with the same probability. The
expected utility of any behavioral profile $\sigma$ is therefore a mixture
of pure profiles under a product distribution,
$\mathbb{E}_\sigma[U] = \sum_s P_\sigma(s)\, U(s)
\in [c - \zprof,\ c + \zprof]$, where $c$ is the optimal center attaining
$\zprof$. Both $\rs$ and $\bp$ lie in this interval, so their difference is
at most $2\,\zprof$.
\end{proof}

\begin{proof}[Proof of (b), second leg ($\rho_S \cdot \zsub$)]
Let $P^x(z)$ be the terminal distribution under $\sigma^x$
($x \in \{0,1\}$, $\sigma^0 = \bp$, $\sigma^1 = \rs$). Since $\rs$ changes
strategy only inside $S$, we have $P^1 = P^0$ pointwise on
$Z_{\mathrm{out}}$, and $P^1(Z_S) = P^0(Z_S) = \rho_S$ (trunk and chance
are unchanged). Therefore
\begin{equation}\label{eq:app-T1-signed}
\sum_p \mathrm{gain}_p
= \sum_{z \in Z_S} \bigl[P^1(z) - P^0(z)\bigr]\, U(z)
= \sum_{z \in Z_S} \bigl[P^1 - P^0\bigr](z)\, \bigl[U(z) - c\bigr],
\end{equation}
where the second equality holds for \emph{any} constant $c$ because the
signed measure $P^1 - P^0$ has total mass $0$ on $Z_S$. Choosing $c$ as the
optimal center attaining $\zsub$, and writing $\|P^1-P^0\|_1(Z_S) :=
\sum_{z\in Z_S}|P^1(z)-P^0(z)|$,
\[
\Bigl| \sum_p \mathrm{gain}_p \Bigr|
\le \bigl\| P^1 - P^0 \bigr\|_1(Z_S) \cdot \zsub
\le 2\,\rho_S \cdot \zsub . \qedhere
\]
\end{proof}

\paragraph{Explicit accounting for $c_1 = 2n$.}
Two factors link $\delta$ to the structural tax. Part (a) contributes a
factor $n$ because the $n$ players' decomposition errors add while the
constant-sum constraint eliminates the pairwise terms. Part (b) contributes
a factor $2$ from evaluating both $\rs$ and $\bp$. Hence $c_1 = 2n$.

\begin{remark}[The tax can be far below $n\delta$]\label{rem:app-T1-min}
The pairpot family is pointwise zero-sum on terminals, so
$\zeta_{\mathrm{term}} = 0$ and $\tax = 0$ for the whole family, even though
$\dinf = \lambda/16 > 0$ because the coupling term is a pure transfer
payment that preserves the total. Games with $\dinf > 0$ can thus carry zero
structural tax. The converse is impossible because $\zprof \le n\delta$ by
part (a). On the adversarial random family of \cref{sec:experiments},
$2\,\zprof$ is often smaller than $2n\delta$ (for instance $0.058$ versus
$0.064$ on one instance with $n = 3$) and $\rho_S \zsub$ and $\zprof$
alternate as the smaller leg. The numerical regression check
(\cref{app:machine}) also found random instances on which part (a) is
attained with equality ($\zprof / (n\delta) = 1.000000$ at $n = 3$).
\end{remark}

\subsection{Proof of the Main Theorem
(\texorpdfstring{\cref{thm:main} and \cref{cor:2ndelta}}{T2})}
\label{app:T1T2:T2}

We now prove \cref{thm:main}. Under \cref{asm:F1}, for any blueprint
$\bp$ and any legal update $\rs$, with $\mhat$, $\Mdef{q}$, $\Act$, and
$\epsi$ defined for $\rs$ and $\tax$ as in \cref{eq:app-T1-b},
\begin{align}
\text{(i)}\quad & A_i(\rs) - A_i(\bp) = \epsi - \Act(\bp) \le \epsi
  \qquad (\epsi \ge 0 \text{ always}), \label{eq:app-T2-i}\\
\text{(ii)}\quad & A_j(\rs) - A_j(\bp) \le
  \Mdef{j} + \Bigl[ \Act(\bp) - \epsi +
  \textstyle\sum_{k \in O \setminus \{j\}} \Mdef{k} \Bigr] + \tax
  \qquad \forall j \neq i, and \label{eq:app-T2-ii}\\
\text{(iii)}\quad & \Vul_G(\rs) - \Vul_G(\bp)
  \le \max_p \bigl[ A_p(\rs) - A_p(\bp) \bigr]\\
  & \le \max\Bigl( \epsi,\ \Act(\bp) - \epsi +
  \textstyle\sum_{q \in O} \Mdef{q} + \tax \Bigr). \label{eq:app-T2-iii}
\end{align}

\begin{proof}
Steps S0 through S6 of the proof in \cref{app:L3:proof} carry over
unchanged. Steps S0 and S1 use only \cref{asm:F1} and perfect recall, while
S2 and S3 are definitional identities. Step S4 bounds the opponent BR-value
change by $\Mdef{j}$, and S5 bounds the opponent on-path gain by $\Mdef{k}$.
Both use only the entry decomposition, the blueprint counterfactual values,
blueprint feasibility, and the game-side best responses defining $\mhat$
(\cref{def:margin}). The gadget is never executed in the proof, which
therefore applies to any legal update $\rs$. Step S6 is the definition of
$\Act$ and $\epsi$. None of these steps involves $\delta$. The remaining
steps are replaced as follows.

\medskip
\noindent\textbf{Step S7$'$ (structural tax, replaces S7).}
By \cref{lem:transfer}, $\bigl| \sum_p \mathrm{gain}_p \bigr| \le \tax$.
(When $\delta = 0$, part (a) gives $\zprof = 0$, so $\tax = 0$ and
$\sum_p \mathrm{gain}_p = 0$, recovering the original S7.)

\medskip
\noindent\textbf{Step S8$'$ (pricing the victim's on-path loss).}
For $j \in O$,
\begin{equation}\label{eq:app-T2-S8}
\mathrm{drop}_j = -\mathrm{gain}_j
= \sum_{p \neq j} \mathrm{gain}_p - \sum_p \mathrm{gain}_p
\le \bigl(\Act(\bp) - \epsi\bigr)
  + \sum_{k \in O \setminus \{j\}} \Mdef{k} + \tax,
\end{equation}
by S5, S6, and S7$'$.

\medskip
\noindent\textbf{Step S8$''$ (exact form, proves \cref{cor:leak}).}
The first line of \cref{eq:app-T2-S8} is an identity, and
$-\sum_p \mathrm{gain}_p = \sum_p u_p(\bp) - \sum_p u_p(\rs) = \Lambda$.
Bounding only the first two groups by S5 and S6 gives
$\mathrm{drop}_j \le (\Act(\bp) - \epsi) + \sum_{k \in O \setminus \{j\}}
\Mdef{k} + \Lambda$, and S9$'$ and S10$'$ below go through verbatim with
$\Lambda$ in place of $\tax$. Step S7$'$ states that $|\Lambda| \le \tax$,
so the $\Lambda$-form implies the $\tax$-form pointwise. Because $\Lambda$
may be negative, it enters S10$'$ inside the maximum and stays there
(compare \cref{cor:telescoping} for sequential resolves).

\medskip
\noindent\textbf{Step S9$'$.}
$A_j(\rs) - A_j(\bp) = \bigl[\mathrm{BRval}_j(1) - \mathrm{BRval}_j(0)\bigr]
+ \mathrm{drop}_j \le \Mdef{j} + \bigl(\Act(\bp) - \epsi\bigr)
+ \sum_{k \neq j} \Mdef{k} + \tax$, by S4 and S8$'$. This is
\cref{eq:app-T2-ii}.

\medskip
\noindent\textbf{Step S10$'$.}
Claim (i) is S6 ($\rs_{-i} = \bp_{-i}$ leaves the maximization term
of $A_i$ unchanged). A difference of maxima is at most the maximum of
differences, so combining (i) and (ii),
\[
\begin{split}
\Vul_G(\rs) - \Vul_G(\bp)
&\le \max\Bigl( \epsi - \Act(\bp),\
  \max_{j} \bigl[ \Mdef{j} + \Act(\bp) - \epsi +
  \textstyle\sum_{k \neq j} \Mdef{k} + \tax \bigr] \Bigr)\\
&\le \max\Bigl( \epsi,\ \Act(\bp) - \epsi +
  \textstyle\sum_{q} \Mdef{q} + \tax \Bigr). \qedhere
\end{split}
\]
\end{proof}

\paragraph{The corollary form.}
Since $\tax \le 2n \cdot \dinf$ by \cref{eq:app-T1-b}, the bound of
\cref{eq:app-T2-iii} weakens to
\begin{equation}\label{eq:app-T2-cor}
\Vul_G(\rs) \le \Vul_G(\bp) + \Act(\bp) + 2n \cdot \dinf
  + \sum_{q \in O} \Mdef{q} + \epsi ,
\end{equation}
with constants $c_1 = 2n$ ($6$ at $n = 3$), $c_2 = 1$, and $D = 1$.
Only the resolver's solve quality enters. Under an exact solve in a
Terminate-compatible environment whose root measure induces the resolver's
game-side beliefs at each of its entry sets, $\epsi \le
\varepsilon_{\mathrm{solve}}$ (\cref{app:L3:setup}). Using
$\Act(\bp) \le \Vul_G(\bp)$ gives the weaker bound
$\Vul_G(\rs) \le 2\,\Vul_G(\bp) + c_1 \dinf + c_2 \sum_q \Mdef{q} + \epsi$.
If in addition all margins are nonnegative and the solve is exact, then
$\Vul_G(\rs) \le \Vul_G(\bp) + \Act(\bp) + 2n \cdot \dinf$.

\paragraph{A blueprint-specific a priori bound on the leak.}
Let $U := \sum_p u_p$ and
$L_i := U(\bp) - \min_{\tau_i \,:\, \tau_i = \bp_i \text{ outside } S}
U(\tau_i, \bp_{-i})$. The blueprint is feasible, so $L_i \ge 0$. Every
update $\rs = (\rs_i, \bp_{-i})$ with $\rs_i = \bp_i$ outside $S$ is
feasible, giving $\Lambda = U(\bp) - U(\rs) \le L_i$. Since $L_i$ is the
leak of one such update, $L_i \le \tax$ by \cref{lem:transfer}, and
$\Lambda \le L_i \le \tax$. Computing $L_i$ takes one best response of
player $i$ confined to $S$ against $\bp_{-i}$ with payoff $-U$, without
using $\rs$. The tax $\tax$ is a structural upper envelope for $L_i$ that
depends on the blueprint only through $\rho_S$. Its coefficient is attained
on \cref{ex:leak}, yet it can stay loose even after maximizing $L_i$ over
blueprints and resolvers. With $u_p = a_p$ for independent binary actions
and $S$ the whole game, $\tax = n$ while every unilateral update moves the
total by at most one.

\paragraph{The direct per-player form.}
With $g_p := u_p(\rs) - u_p(\bp)$, claim (i) reads
$A_i(\rs) - A_i(\bp) = -g_i$, and S4 gives
$A_j(\rs) - A_j(\bp) \le \Mdef{j} - g_j$ for $j \ne i$. Hence
$\Vul_G(\rs) - \Vul_G(\bp) \le \max_p[A_p(\rs)-A_p(\bp)] \le
B_{\mathrm{dir}}$, which gives \cref{eq:direct}. Every summand in
$\Act - \epsi + \sum_q \Mdef{q} + \Lambda = \sum_{j \ne i} (\Mdef{j} - g_j)$
is nonnegative by S5, and $-g_i = \epsi - \Act \le \epsi$. Consequently,
$B_{\mathrm{dir}}$ is pointwise at most the $\Lambda$ form of
\cref{cor:leak}. It needs only the BR-certified margins and the $n$ expected
utilities of $\bp$ and $\rs$, with no activation term or structural
quantity. The instantiations of \cref{thm:main} do not exhaust the useful
post-hoc certificates (\cref{app:expdetails:usefulness}).

The a priori instantiation of \cref{thm:main}(iii) is stated here for
reference.

\begin{corollary}[$2n \cdot \dinf$ form]\label{cor:2ndelta}
Since $\tax \le 2\zprof \le 2n \cdot \dinf$ (\cref{eq:app-T1-b}),
\cref{thm:main}(iii) implies
$\Vul_G(\rs) \le \Vul_G(\bp) + \Act(\bp) + 2n\,\dinf +
\sum_{q\in O}\Mdef{q} + \epsi$,
with constants $(c_1, c_2, D) = (2n, 1, 1)$ ($c_1 = 6$ for $n = 3$).
\end{corollary}

\paragraph{The baseline-aware form.}
Write $A^0_p := A_p(\bp)$. Claim (i) gives $A_i(\rs) = A^0_i - g_i$
exactly, and S4 gives $A_j(\rs) \le A^0_j + \Mdef{j} - g_j$ for $j \ne i$.
Taking the maximum over $p$ and subtracting $\Vul_G(\bp) = \max_p A^0_p$
yields \cref{cor:cache}. Since $\max_p (A^0_p + x_p) - \max_p A^0_p \le
\max_p x_p$ for any reals $x_p$, setting $x_i = -g_i$ and
$x_j = \Mdef{j} - g_j$ gives $B_{\mathrm{cache}} \le B_{\mathrm{dir}}$. The
quantity $B_{\mathrm{cache}}$ bounds $\Vul_G(\rs) - \Vul_G(\bp)$ directly,
whereas every other form of this appendix bounds the middle term
$\max_p [A_p(\rs) - A_p(\bp)]$ of claim (iii). The two differ whenever the
player whose advantage rises most is distinct from the most exploitable
player after the update.

\paragraph{Zero tolerance on constant-sum games.}
When $\sum_p u_p$ is constant on the game, $\sum_p g_p = 0$. Suppose
$B_{\mathrm{dir}} \le 0$. Then $-g_i \le 0$ and $\Mdef{j} - g_j \le 0$ for
every $j \ne i$, so $g_i \ge 0$ and $g_j \ge \Mdef{j} \ge 0$. Since
$\sum_p g_p = 0$, this forces $g_p = 0$ for every $p$ and $\Mdef{j} = 0$ for
every $j \ne i$. Conversely, these equalities give $B_{\mathrm{dir}} = 0$.
On a constant-sum game, $B_{\mathrm{dir}} \le 0$ therefore holds if and only
if the update changes no expected utility and leaves every opponent's
certified deficit at zero. Every form of \cref{thm:main} is pointwise at least
$B_{\mathrm{dir}}$ and inherits this restriction, whereas the baseline-aware
form $B_{\mathrm{cache}}$ escapes it. Consider an update with $g_i > 0$,
$g_j = -g_i$ for one opponent $j$, $g_k = 0$ for the others, and every
$\Mdef{q} = 0$. If $A^0_j + g_i \le \max_p A^0_p$, then
$B_{\mathrm{cache}} \le 0$ while $B_{\mathrm{dir}} = g_i > 0$.

\textbf{Degeneration at $\delta = 0$.}
If $\dinf = 0$, then $\zprof \le n \cdot 0 = 0$ and $\tax = 0$. Claims (i),
(ii), and (iii) thus reduce term by term to \cref{lem:delta0}. The numerical
check (\cref{app:machine}) confirms that the bound of \cref{thm:main} equals
the constant-sum bound pointwise on all constant-sum configurations.

\paragraph{Tightness status.}
The coefficient $1$ on $\Act$ is optimal (\cref{ex:activation}), and the
sum over opponents cannot be weakened to a single-margin per-pair
certificate (\cref{ex:crossopp}). These two equality constructions pin
$c_2 = 1$. In \cref{ex:leak}, the leak instance attains the coefficient $1$
on $\tax$ with $\Delta\Vul = \tax = \Lambda = 1$ and every other term zero.
The self-harm instance attains the $\epsi$ branch of the maximum
(\cref{app:L3:ex34}). With $n = 2$ and $\dinf = \tfrac14$ in the leak
construction, $2n\,\dinf = \tax$, so \cref{cor:2ndelta} is tight at $n = 2$.
For $n \ge 3$, adversarial random instances attain the $n$ factor in part (a)
of \cref{app:T1T2:T1} (\cref{rem:app-T1-min}). Tightness in $n$ of the
end-to-end response of $\Delta\Vul$ to $\delta$ remains open. Because the
single-instance tax can vanish when $\delta > 0$ (as on the pairpot family,
where $\tax \equiv 0$), a worst-case lower bound in $\delta$ requires a
construction different from the leak.

\subsection{Why \texorpdfstring{$\delta$}{delta} Enters Only Through the Total Utility}
\label{app:T1T2:routeB}

An alternative route would project $G$ onto $V$ to obtain the nearest
CSP game $\tilde{G}$, bound
$|\Vul_G(\sigma) - \Vul_{\tilde{G}}(\sigma)| \le 2\delta$, run the
constant-sum proof inside $\tilde{G}$, and transfer the bound back. Two
obstacles block this approach.

First, $\tilde{G}$ is a profile utility table with no tree structure.
Steps S1 through S6 of \cref{app:L3:proof} rely on the trunk-subgame
decomposition and the gadget certificates, which are undefined on
$\tilde{G}$.

Second, suppose $\tilde{G}$ is realized as an EFG on the same tree with
modified terminal utilities. The entry-level certificate quantities $\mhat$
and $\mathrm{CFV}$ are defined on $G$. Game-side activation $\Act$ moves by at
most $2\delta$, yet the profile-table distance $\delta$ does not control the
transport of $\mhat$ and $\mathrm{CFV}$ to $\tilde{G}$. Because these
quantities are conditioned on reaching $S$, their discrepancy between $G$ and
$\tilde{G}$ can scale as $\delta/\rho_S$. Suppose chance enters $S$ with
probability $r$ and the resolver alone acts inside ($a \mapsto (0,0)$,
$b \mapsto (0,-1)$). Then $\delta = r/4$, whereas any realization of the
nearest constant-sum table on the same tree shifts $\mathrm{CFV}_j$ by $1/4$
and $\mhat_j$ by $1/2$ for every $r$. Without restrictions on the terminal
realization, the discrepancy is unbounded even at fixed $r$. Shifting $j$'s
payoff by $H$ on $Z_S$ and by $-rH/(1-r)$ outside leaves every profile
utility, $\delta$, and $\rho_S$ unchanged while moving $\mathrm{CFV}_j$ by
$H$. The profile-table distance therefore fails to transport the entry-level
certificate quantities uniformly in the blueprint.

In our proof, $\delta$ enters only at step S7$'$ through the
constant-sum defect. This explains the gap between $\tax$ and $2n\dinf$
noted in \cref{sec:proof}, because the distance to constant-sum can be far
smaller than $n \cdot \dinf$.

\subsection{Proof of the NashConv Form (\texorpdfstring{\cref{cor:nashconv}}{NashConv corollary})}\label{app:T1T2:nashconv}

For every player $p$ write $\mathrm{BRval}_p(\sigma) := \max_{\tau_p}
u_p(\tau_p, \sigma_{-p})$, so that $A_p(\sigma) = \mathrm{BRval}_p(\sigma) -
u_p(\sigma)$ and, with $g_p := u_p(\rs) - u_p(\bp)$,
\[
A_p(\rs) - A_p(\bp) \;=\; \bigl[\mathrm{BRval}_p(\rs) - \mathrm{BRval}_p(\bp)\bigr] - g_p .
\]
For the resolver, $\rs_{-i} = \bp_{-i}$, so $\mathrm{BRval}_i(\rs) =
\mathrm{BRval}_i(\bp)$. For an opponent $j \neq i$, step S4 of
\cref{app:L3} (\cref{eq:app-L3-S4}) gives $\mathrm{BRval}_j(\rs) -
\mathrm{BRval}_j(\bp) \le \Mdef{j}$, using only perfect recall and
\cref{asm:F1}. Summing over all players,
\[
\mathrm{NashConv}(\rs) - \mathrm{NashConv}(\bp)
\;\le\; \sum_{j \in O} \Mdef{j} - \sum_p g_p
\;=\; \sum_{j \in O} \Mdef{j} + \Lambda ,
\]
by the definition $\Lambda = \sum_p u_p(\bp) - \sum_p u_p(\rs)$, and
$\Lambda \le |\Lambda| \le \tax$ by \cref{lem:transfer}. The activation
and solve-error terms do not appear because they enter the per-player
bounds of \cref{thm:main} only through the re-pricing of $-g_j$ by
$\sum_p g_p = -\Lambda$, which the sum performs directly.

\section{Linearity and the Explicit Constant}\label{app:linearity}

This appendix proves \cref{lem:linear} on the exact linearity of $\dinf$ along
affine perturbation families anchored in the constant-sum polymatrix (CSP)
class. We also prove \cref{prop:K}, obtaining the explicit constant $K = 1/4$
for the pairpot family by hand case analysis with an equality construction.
Notation for the CSP subspace $V$ follows \cref{app:T1T2:setup}. Here $V$ is
the set of profile utility tables admitting a decomposition
$u_p(s) = \sum_{q \neq p} u_{pq}(s_p, s_q)$ with per-edge
constant-sum constraints $u_{pq}(a,b) + u_{qp}(b,a) = c_{pq}$. Finally,
$\dinf(u) = \mathrm{dist}_\infty(u, V) = \min_{v \in V} \|u - v\|_\infty$
in the linear program (LP) profile-table semantics of \cref{def:delta}.

\subsection{Proof of the Linearity Lemma
(\texorpdfstring{\cref{lem:linear}}{L1b})}\label{app:lin:L1b}

The pairpot game has three players. Chance deals each player $p$ a private card $t_p$, i.i.d.\ uniform on $\{0,1\}$, and a public signal $X$ uniform on $\{A,B\}$. The effective card is $h_p := t_p \oplus [X=B]$ (exclusive or with the indicator of $X=B$), so $X=B$ reverses the card order. Each player sees only $(t_p,X)$ and either checks or bets. Because the three actions are mutually unobserved, a pure strategy is a map $\{0,1\}\times\{A,B\}\to\{\text{check},\text{bet}\}$, yielding $16^3=4096$ pure profiles. The base payoff $u^0$ is one independent side pot per pair $\{p,q\}$. If both bet, the higher $h$ receives $1$ from the other, and a tie pays nothing. When one bets, the bettor receives $1/2$ from the checker. Mutual checks pay nothing. The coupling $C$ is a shared pot among the set $\mathcal{B}$ of players who bet. If $|\mathcal{B}|\ge 2$ and one member holds the unique highest $h$ within $\mathcal{B}$, that member collects $1/2$ from every other member of $\mathcal{B}$. A tie for the highest card pays nothing.

The perturbation family is $u_\lambda = u^0 + \lambda \cdot C$, where the base table $u^0$ of the pairpot family consists of independent per-pair side pots. This table is CSP, with $\dinf(u^0) = 0$ verified by exact LP. The shared-pot coupling tensor $C$ equals the difference between the $\lambda = 1$ and $\lambda = 0$ profile utility tables. By construction, the family is affine in $\lambda$, verified numerically to residual $10^{-12}$ in \cref{app:machine}.

\begin{proof}[Proof of \cref{lem:linear}]
The set $V$ is a linear subspace because the decomposition map is componentwise linear and
the per-edge constant-sum constraints are homogeneous linear. A sum or a
scalar multiple of two decomposable tables is again decomposable, with the
edge constants adding or scaling accordingly. Since $u^0 \in V$, for any
$\lambda \ge 0$,
\begin{equation}\label{eq:app-lin-chain}
\dinf(u^0 + \lambda C)
= \mathrm{dist}_\infty(u^0 + \lambda C,\, V)
= \mathrm{dist}_\infty(\lambda C,\, V)
= \lambda \cdot \mathrm{dist}_\infty(C,\, V)
= \lambda \cdot \dinf(C).
\end{equation}
The second equality is translation invariance of the distance along the
subspace ($u^0 \in V$), and the third is positive homogeneity of the sup
distance to a linear subspace.
\end{proof}

Consequently, the measured pairpot behavior has slope $1/16$ and zero
intercept, matching pointwise across all recorded $\lambda$ values
(\cref{app:delta}). This follows from
$\dinf(u_\lambda) = \lambda \cdot \dinf(C)$, with $\dinf(C) = 1/16$ computed
by exact LP. Linearity holds for any affine family with base point in $V$.
Only the numerical constants below are specific to the pairpot family.

\subsection{Proof of the Explicit Constant
(\texorpdfstring{\cref{prop:K}}{Proposition K})}\label{app:lin:K}

\begin{proof}[Proof of \cref{prop:K}]
Define $K := \max_{p, s} |C_p(s)|$. Since $0 \in V$, the
zero-decomposition bound gives
$\dinf(C) = \mathrm{dist}_\infty(C, V) \le \|C - 0\|_\infty = K$, hence by
\cref{lem:linear},
\begin{equation}\label{eq:app-lin-K}
\dinf(u_\lambda) = \lambda \cdot \dinf(C) \le \lambda \cdot K .
\end{equation}
It remains to show $K = 1/4$.

The quantity $C_p(s)$ is player $p$'s expected coupling payoff at $\lambda = 1$ under
the pure profile $s$. Among the set of bettors $\mathcal{B}$ (with coupling active
only when $|\mathcal{B}| \ge 2$), the holder of the unique highest
$X$-adjusted card collects $1/2$ from each other member of $\mathcal{B}$. Fix the public signal $X$ and
take conditional expectations. A pure strategy is a function of
$(t_p, X)$, so conditionally on $X$ the betting action of each player is a
function of $h_p := t_p \oplus [X = B]$, and $(h_p, h_q, h_r)$ are i.i.d.\
uniform on $\{0, 1\}$.

\emph{Receipt side.} Player $p$ collects if and only if $h_p = 1$,
$p \in \mathcal{B}$, and every other bettor has $h = 0$. Since $h$ takes only the
values $0$ and $1$, a player with $h_p = 0$ can never be the unique highest
within any bettor set of size at least $2$. The amount collected is
$(|\mathcal{B}| - 1)/2 \le 1$. We bound this case by case over $(h_q, h_r)$. The pair
$(0,0)$ gives at most $1$ because both opponents may bet. The pairs $(0,1)$ and $(1,0)$
give at most $1/2$ each, since an opponent holding $1$ destroys $p$'s uniqueness
when it bets, so $p$ collects from at most the opponent holding
$0$. Finally, $(1,1)$ gives $0$. Hence
\begin{equation}\label{eq:app-lin-receipt}
\mathbb{E}[\text{receipt} \mid X]
\le \Pr(h_p = 1) \cdot \tfrac{1}{4}\,
    \bigl( 1 + \tfrac{1}{2} + \tfrac{1}{2} + 0 \bigr)
= \tfrac{1}{2} \cdot \tfrac{1}{2}
= \tfrac{1}{4} .
\end{equation}

\emph{Payment side.} Player $p$ pays if and only if $h_p = 0$, $p \in \mathcal{B}$,
and some opponent with $h = 1$ is the unique highest within $\mathcal{B}$, in which case the payment
is $1/2$. Across $(h_q, h_r)$, the pair $(0,0)$ gives $0$, while
$(0,1)$, $(1,0)$, and $(1,1)$ give at most $1/2$ each. Hence
\begin{equation}\label{eq:app-lin-payment}
\mathbb{E}[\text{payment} \mid X]
\le \tfrac{1}{2} \cdot \tfrac{3}{4} \cdot \tfrac{1}{2}
= \tfrac{3}{16} .
\end{equation}

Averaging over $X$ leaves both bounds unchanged, so
$|C_p(s)| \le \max(1/4,\ 3/16) = 1/4$, i.e.\ $K \le 1/4$.

\emph{Equality construction.} Let $p$ bet if and only if $h_p = 1$, and let
$q$ and $r$ bet if and only if their $h = 0$. Then $p$'s payment is $0$, and $p$'s receipt is
\[
\tfrac{1}{8} \cdot 1
+ \tfrac{1}{8} \cdot \tfrac{1}{2}
+ \tfrac{1}{8} \cdot \tfrac{1}{2}
= \tfrac{1}{4}.
\]
These terms correspond to $h = (1,0,0), (1,0,1), (1,1,0)$, with bettor sets of sizes
$3, 2, 2$ respectively. For $h = (1,1,1)$ only $p$ bets and the
coupling is inactive. Hence $K = 1/4$.
\end{proof}

Exhaustive enumeration over all $4096$ pure profiles and all $3$ players
reproduces $\max_{p,s} |C_p(s)| = 0.25$ exactly
(\cref{app:machine}).

\begin{remark}[The factor-$4$ gap]\label{rem:app-lin-gap}
The exact LP value is $\dinf(C) = 1/16 = 0.0625$, matching the
measured slope of the pairpot family (\cref{app:delta}), so the analytic bound
$K = 1/4$ has a factor-$4$ gap. Because $K = 1/4$ is a zero-decomposition bound,
it does not use nonzero elements of $V$. The LP uses the optimal
decomposition to reduce the constant from $1/4$ to $1/16$, whose analytic
characterization remains open.
\end{remark}

\section{Distance to Constant-Sum Polymatrix Games}\label{app:delta}

This appendix studies the distance $\dinf$ from a game to the constant-sum polymatrix (CSP) class. That distance bounds the structural tax a priori ($\tax\le 2\zprof\le 2n\dinf$) and links the bound to the polymatrix literature, where self-play guarantees degrade with distance to the CSP class \citep{macqueen2023guarantees}. No result in the main text depends on this distance. \Cref{thm:main} uses the sharp tax $\tax$, and \cref{lem:transfer} proves $|\sum_p g_p|\le\tax$ directly.

\textbf{The distance.} Let $s = (s_1,\dots,s_n)$ range over profiles of
reduced pure strategies (one action per reachable information set).
Here $u_p(s)$ is $p$'s expected payoff over chance, and
$U(s) := \sum_p u_p(s)$ is the player sum.

\begin{definition}[$\dinf$-CSP distance]\label{def:delta}
Let $V$ denote the set of profile-utility tables $v$ admitting a
\emph{constant-sum polymatrix decomposition}: pairwise tables
$\{v_{pq}\}_{p \neq q}$ and edge constants $\{c_{pq}\}_{p < q}$ such that
$v_p(s) = \sum_{q \neq p} v_{pq}(s_p, s_q)$ for every profile $s$, and
$v_{pq}(a, b) + v_{qp}(b, a) = c_{pq}$ for every edge $\{p, q\}$ and all
pure strategies $a$ of $p$ and $b$ of $q$. The \emph{$\dinf$-CSP distance}
of $G$ is $\dinf(G) := \min_{v \in V} \|u - v\|_\infty =
\min_{v \in V} \max_{p,s} |u_p(s) - v_p(s)|$.
\end{definition}

The class is the \emph{per-edge constant-sum} class of
\citet{macqueen2023guarantees} and \citet{cai2016zero} (see also
\citet{cai2011minmax}), which contains the
antisymmetric subclass. Our proofs use only that $\sum_p v_p$ is constant on
$V$. Defining $\dinf$ on profile tables lets a
pointwise zero-sum game have $\dinf>0$, as in three-player Kuhn below. $V$ is a
linear subspace, so $\dinf$ is an exact LP. Since $\sum_p u_p$ is
constant on $V$, a game with $\dinf=0$ has constant total utility across
pure profiles, which is the hypothesis of \cref{lem:delta0}.

\textbf{The $2n$ envelope.} For an optimal LP witness, summing the residuals
$|e_p(s)|\le\dinf$ cancels the pairwise terms via the per-edge
constant-sum constraint. This leaves $|U(s)-c^\ast|\le n\dinf$, which gives
$\zprof\le n\dinf$ and $\tax\le 2n\dinf$ (\cref{cor:2ndelta} in
\cref{app:T1T2}). The factor $n$ in this bound is tight on individual
instances, with $\zprof/(3\dinf)=1.000000$ on two screening games
(\cref{app:expdetails}). Whether the end-to-end response of $\Delta\Vul$
to $\dinf$ is tight in $n$ remains open, while the sharp form of
\cref{thm:main} is independent of $n$.

The $\min$ inside $\tax$ is essential. On the pairpot family below, $\tax=0$ while $\dinf=\lambda/16>0$. Since $\zprof\le n\dinf$ rules out the reverse, the $\tax$-form is strictly stronger than the $2n\dinf$ corollary. On the screening set, $\tax<2n\dinf$ on $266$ of $296$ points, with $\zprof$ the smaller leg on $204$ and $\rho_S\zsub$ on $92$ (\cref{app:machine}).

\begin{lemma}[Linear perturbation law]\label{lem:linear}
The set $V$ of CSP-decomposable profile utility tables is a linear subspace
of the space of utility tables. Consequently, for every $u^0 \in V$, every
utility table $C$, and every $\lambda \ge 0$,
$\dinf(u^0+\lambda C) = \lambda\,\dinf(C)$.
\end{lemma}

Since $V$ is a subspace, the sup-distance is
translation-invariant and positively homogeneous (\cref{app:linearity}). This confirms the measured
pairpot slope of $1/16$ analytically.

The \emph{pairpot family} is $u_\lambda := u^0 + \lambda C$. Here $u^0$ is a
three-player table of independent per-pair side pots, which is CSP. The
coupling tensor $C$ adds a shared pot that is a pure transfer between the
players, so every $u_\lambda$ is pointwise zero-sum on terminals (rules in
\cref{app:linearity}).

\begin{proposition}[Explicit slope constant for the pairpot family]\label{prop:K}
For the coupling tensor $C$, $K := \max_{p,s} |C_p(s)| = 1/4$, and
hence $\dinf(u^0 + \lambda C) \leq \lambda/4$ for all $\lambda \geq 0$.
An explicit pure profile attains the bound $|C_p(s)| = 1/4$.
\end{proposition}

The proof is a direct case analysis (\cref{app:linearity}), verified by
exhaustive enumeration over all $4096$ pure profiles. Since $\dinf(C)=1/16$
by exact LP, \cref{prop:K} bounds the true slope within a factor of four.

\textbf{A positive class with natural rules.}
Pairwise-showdown games satisfy $\dinf = 0$. Across all $4096$ pure profiles, exact LP gives $\dinf = 0$. Correlated dealing preserves this result because each edge utility enters expectations only through that pair's marginal of the deal. Payoff coupling through opponents' \emph{actions} breaks CSP structure (\cref{app:constructions:psd}).

\begin{proposition}[Poker-style games are far from CSP]\label{prop:pokerlb}
The following lower bounds hold, each certified by an LP dual certificate.
Three-player Kuhn poker with a four-card deck has $\dinf \ge 1.000$
($\max|u| = 4.0$, relative distance $\ge 0.250$). Blind Goofspiel with
$K{=}3$ prizes has $\dinf = 2.500$ exactly ($216$ pure profiles, relative
$0.500$), and with $K{=}4$ it has $\dinf = 4.625$ exactly
($13824$ pure profiles, relative $0.514$). Observed Goofspiel with $K{=}3$
has $\dinf \ge 3.000$ (relative $\ge 0.500$).
\end{proposition}

Across the two Goofspiel sizes, the relative distance rises from $0.500$ at $K{=}3$ to $0.514$ at $K{=}4$. Each of the four lower bounds comes with an LP dual certificate (\cref{app:constructions}).

For poker-style games, the $2n\,\dinf$ bound is valid but vacuous. Since
$\dinf \ge 1$, the structural term $2n\,\dinf$ exceeds the utility scale of
three-player Kuhn. The sharp $\tax$-form charges rake-free poker no tax
(\cref{cor:constant-sum}). Under a rake of up to $10\%$ of the pot, it charges
at most \rakeThetaKuhnMax{} of the utility scale (\cref{tab:rake}). Pointwise
constant-sum payoffs ensure $\tax=0$, while the pairpot family shows
$\dinf=0$ is not required.

\section{Construction Details for the Tightness and Necessity Results}\label{app:constructions}

This appendix details the constructions behind the negative results (\cref{prop:margin-nec,prop:delta-nec,prop:mismatch}) and the poker-class $\dinf$ lower bounds (\cref{prop:pokerlb} in \cref{app:delta}). Two single-core checks (runtimes 1\,min\,57\,s and under 1\,s) re-derive every number quoted here with tolerance $10^{-9}$, keeping preliminary figures at their recorded two-decimal precision. Throughout, we use constant-sum polymatrix (CSP) in the sense of \cref{def:delta}.

\subsection{Margin-term necessity: the initial and mutated search instances (\cref{prop:margin-nec})}\label{app:constructions:margin}

Both instances come from the adversarial screen in \cref{sec:experiments}. On random small extensive-form games (depth $\le 3$, 2 to 3 actions per node), the screen runs the pipeline of blueprint, corruption, resolve, and exact best-response evaluation. The screening objective is $\Delta\Vul_{\max} - 0.03\,\dinf - 0.03\,(-\mathrm{margin})_+$, with the $0.03$ constants from the preliminary run's benign-family fit.

\textbf{Scan original.} This instance has a sampled $\dinf$ lower bound of $0.2188$, a realized $\Delta\Vul = 0.2164$, and a minimum margin of $-1.0559$. It attains $+0.178$, the scan-maximal objective over all 433 points. Under a deep re-solve with $7.5\times$ blueprint and $5\times$ resolve iterations, the screening objective persists and grows slightly. Specifically, $\Delta\Vul$ increases from $0.216$ to $0.224$, while the top objectives shift from $0.216$ to $0.235$, $0.208$ to $0.222$, and $0.225$ to $0.255$. Three companion originals behave similarly, with scores moving from $0.148$ to $0.193$, $0.147$ to $0.151$, and $0.194$ to $0.152$. Their margins remain stably and strongly negative between $-0.27$ and $-1.07$. Because the scan original's utility tensor exceeds the exact-enumeration cap, its $\dinf$ certificate is a restricted-support \emph{lower bound} sharing the screening figure's support and value. A lower bound on $\dinf$ shrinks the right-hand side of any bound with a positive coefficient on $\dinf$, so the objective serves as a screening score.

This instance certifies that deleting the margin term breaks the structural theorem, because the sharp tax is exact. Computing that tax requires only the range of total utility over the $1\times64\times256$ profile table. Here $\zprof = 0.817$, $\rho_S = 0.0057$, and $\zsub = 1.458$, giving $\tax = 2\rho_S\zsub = 0.0166$. The realized leak is $\Lambda = -0.0025$, as the update slightly raises total utility. With $\Act = 0.00005$ and $\epsi = 0.0019$, deleting $\sum_q\Mdef{q}$ gives $\max(\epsi, \Act-\epsi+\tax) = 0.015$ in \cref{thm:main} and $0.002$ in the $\Lambda$ form of \cref{cor:leak}. Both fall far below $\Delta\Vul = 0.216$, whereas the full bound $\Act-\epsi+\sum_q\Mdef{q}+\Lambda = 0.248$ covers it.

\textbf{Hill-climbed variant.} Starting from the scan point, 25 Gaussian leaf-utility mutation steps (9 accepted) produce a hill-climbed variant with $\Delta\Vul = 0.249829$ and minimum margin $-0.578777$. This variant has objective $0.224797$ at the climb-time restricted-support lower bound $\dinf \ge 0.255645$. Exact full profile enumeration yields $\dinf = 0.303275$, matching the deep-dive value within $10^{-7}$. At this exact value, the objective remains positive at $0.223367$.

In both instances, the margin certificate fires with strongly negative $\mhat$, matched by a real gap increase of the same order. Setting $c_2 = 0$ in the bound of \cref{thm:main} fails on the scan original, where we measure the sharp tax. On the full 162-instance set, $c_2 = 0$ also forces $c_1 \ge 2.64$.

\subsection{Margin-only bounds are insufficient: a Dirichlet-reset instance (\cref{prop:delta-nec})}\label{app:constructions:delta}

The complementary construction refutes a margin-only bound. When we reset the resolver's blueprint subgame strategy to a Dirichlet draw before resolving (\cref{app:expdetails}), exact full enumeration on one instance gives $\dinf = 0.4292$. That resolve realizes $\Delta\Vul = 0.0383$, above the $0.02$ noise floor, while the BR-certified margin deficit sums to only $\sum_q \Mdef{q} = 0.0111$. Here $\Act = 0.0382$, $\epsi = 0.0379$, and the minimum margin is $-0.0446$. Any margin-only bound $\Delta\Vul \le c_2 \sum_q \Mdef{q} + \mathrm{tol}$ needs $c_2 \approx 1.65$, five times the fitted $c_2 = 0.31$ (\cref{tab:constants}). Because the solve-error term carries most of the increase, this instance isolates the non-margin channels jointly. The bound of \cref{thm:main} covers the point through $\epsi$ and the $\dinf$ term.

\subsection{Structural margin mismatch near equilibrium (\cref{prop:mismatch})}\label{app:constructions:mismatch}

The preliminary mismatch experiment resolves from a CFR blueprint with \emph{zero} deviation in a three-player game (3p Kuhn, deck 4, $6000$ iterations) and still observes gadget margins of $-0.4606$ and $-0.7448$ (reported as $-0.46$/$-0.74$ in the main text). The run records the blueprint strategy, its per-player full-game best-response gains $0.0012/0.0014/0.0004$ (giving a near-equilibrium $\Vul_G(\bp) = 1.4\times10^{-3}$), and the complete margins table. The BR-certified margins of \cref{def:margin} at the same zero deviation are nonpositive at all $16$ entry information sets of the two preliminary public states. At the two entries where the gadget reports $-0.46$ and $-0.74$, the certified margins are $-0.093$ and $-0.0026$, with mass-weighted deficits of $1.8\times10^{-4}$ and $3.5\times10^{-4}$. This happens because the gadget's root measure $c_e\,r_i(e)$ omits the other opponents' reach and can misweight opponents' beliefs when $n > 2$. The two measures coincide only in the special entry structures of \cref{app:calibers}. Elsewhere, the gadget margin can fire as a false positive even when nothing has changed. A negative gadget margin in multiplayer resolving is an alarm that does not by itself establish damage. The previous subsection shows the converse regime, where the alarm is real.

\subsection{Certificate robustness under utility perturbation}\label{app:constructions:gid183}

The \emph{fixed-certificate} convention freezes certificates before a utility perturbation. Because $\rs$ is then not the resolve of the perturbed game, it falls outside the quantifier range of \cref{thm:main}. Across the 162-instance attack campaign, this convention produces exactly one strict-filter candidate, with $\dinf = 0.063$, $\mhat \approx 0$, and $\Delta\Vul = 0.077$. Its $2n\dinf = 0.38 > 0.077$ dominates the observed increase, so re-measuring $\Act \ge 0$ under the pipeline convention places the candidate under the theorem. Certificates must therefore be issued under the same convention as the deployed pipeline.

\subsection{Pairwise showdown: a positive class with $\dinf = 0$}\label{app:constructions:psd}

This appendix presents the analysis behind the positive class of \cref{app:delta}. In the pairwise-showdown family, three players are dealt distinct private cards from a shared deck and act once. At showdown, players compare cards and each \emph{pair} settles an independent side pot ($\pm 2$ when both bet, $\pm 1$ when one bets, and $0$ when both check). Solving the exact LP over all $4096$ pure profiles gives $\dinf = 0$, even under correlated dealing without replacement. Because each edge utility $u_{ij}$ enters the profile table only through the $(i,j)$ marginal of the deal, chance expectation integrates out any third-party card. The pairwise witness $\{u_{pq}\}$ thus absorbs that card without depending on any third-party \emph{strategy}. By \cref{def:delta}, third-party randomness is in scope while third-party strategies are excluded.

The same computation isolates what breaks CSP structure. Settling a shared pot by the joint ranking \emph{and} the actions couples players through $(\text{private card}, \text{action})$ pairs, so a player's utility depends on an opponent's strategy-dependent behavior. In an early variant whose shared pot depended only on the joint ranking, independent of any action, $\dinf$ is identically zero (exact LP measures $0$ even at $\lambda = 0.5$). Action-passing coupling is the sole obstruction, yielding a checkable design criterion. If every settlement is pairwise and third parties enter a pair's payoff only through chance, then $\dinf = 0$.

\subsection{Poker-class $\dinf$ lower bounds: LP dual certificates (\cref{prop:pokerlb})}\label{app:constructions:duals}

The four lower bounds quoted in \cref{app:delta} are established by explicit dual solutions of the $\dinf$ LP. Each certificate records the utility tensor, the LP primal-dual pair, the claimed value, and the verification conditions (\cref{tab:duals}).

\begin{table}[ht]
\centering
\caption{Lower bounds on $\dinf$ certified by LP dual solutions.}
\label{tab:duals}
\footnotesize
\begin{tabular}{@{}llll@{}}
\toprule
Game & Claimed bound & $\dinf$ convention & Relative scale \\
\midrule
3-player Kuhn (deck 4) & $\dinf \ge 1.0$ & restricted-support lower bound & $\ge 0.250$ \\
blind Goofspiel $K{=}3$ & $\dinf = 2.5$ & exact (216 profiles) & $0.500$ \\
blind Goofspiel $K{=}4$ & $\dinf = 4.625$ & exact (13{,}824 profiles) & $0.514$ \\
observed Goofspiel $K{=}3$ & $\dinf \ge 3.0$ & restricted-support lower bound & $\ge 0.500$ \\
\bottomrule
\end{tabular}
\end{table}

The Kuhn certificate is small enough to state in full. Its dual puts weight $\tfrac14$ on four pure profiles fixing player $3$'s strategy and varying players $1$ and $2$ over two strategies each, with the signs of a mixed second difference. Player $3$'s payoffs at the four profiles are $2$, $-1.0833$, $-0.6667$, and $0.25$ in the order $(s_1,s_2)=(a,a),(a,b),(b,a),(b,b)$. Their mixed second difference $u_3(a,a)-u_3(a,b)-u_3(b,a)+u_3(b,b)$ equals $4.0$. For every $v\in V$, the pairwise tables give $v_3(s)=v_{31}(s_3,s_1)+v_{32}(s_3,s_2)$, whose mixed second difference in $(s_1,s_2)$ is zero. This yields $4.0\le 4\,\|u-v\|_\infty$ and hence $\dinf\ge 1.0$ for the full game, beyond the restricted support where the LP was solved. Here the certificate's verification conditions (nonnegative dual weights summing to one, symmetric aggregated pair tables summing to zero, dual objective equal to the claimed value) reduce to this one identity.

In Goofspiel, the unique highest bid wins the prize, ties void it, and utilities are zero-sum normalized as $u_i = \mathrm{score}_i - \tfrac{1}{2}\sum_{j \ne i}\mathrm{score}_j$. Opponents' bids are unobserved in the blind variant and revealed after each round in the observed variant. In both, the prize order is fixed and known. Verifying each certificate requires only arithmetic. One checks nonnegativity of the dual variables, sum-to-one normalization, aggregate symmetry, aggregate zero-sum, and the dual-objective match against the claimed value. The Kuhn certificate has dual support of size $4$. Before the check, each utility tensor is rebuilt from the game definition. For restricted-support entries, the certified quantity is a lower bound by LP duality.

\section{Numerical Regression and Property Tests}\label{app:machine}

Numerical suites verify the stated inequalities on finite configurations for the proofs in \cref{app:L3,app:T1T2}. Across \textbf{97 checks} for the exact-case lemma and \textbf{6{,}157 checks} for the transfer lemma and main theorem, both suites record \textbf{0 violations}. All runs use seed 42 on a single core. Best responses, $\dinf$ linear programs (LPs), and utility tensors are computed \emph{exactly}. CFR generates only the input profiles $\bp,\rs$, so the checked inequalities are exact up to a $10^{-9}$ floating-point tolerance.

\textbf{Hard-gate semantics.} A \emph{hard gate} is an inequality or
identity that must hold on every checked configuration. Any excess beyond the
$10^{-9}$ tolerance counts as a violation and fails the suite.

\subsection{The 97-check suite for the exact case (\cref{lem:delta0})}\label{app:machine:L3}

We test the two $\dinf = 0$ families of \cref{app:delta}. The first is
pairwise-showdown with a four-card deck, public state ``o'', and resolver in
$\{0,1,2\}$. The second is pairpot at $\lambda = 0$, taking public signal A
with resolver 0 and public signal B with resolver in $\{0,1,2\}$. This yields
7 CFR configurations (500 blueprint iterations, 1600 resolve iterations),
which we combine with the two equality constructions. Eight gates mirror the
proof steps.
\begin{itemize}
  \item \textbf{G1}. Exact-LP recomputation confirms $\dinf = 0$ for both families (4{,}096 pure profiles each) and pointwise terminal zero-sum $\sum_p u_p(z) = 0$, the premise of the single step where $\dinf = 0$ is used.
  \item \textbf{G2} (each opponent $j$). The best-response value change is at most $\Mdef{j}$.
  \item \textbf{G3} (each opponent $k$). The on-path gain (the change in $k$'s expected payoff under the profile actually played) is at most $\Mdef{k}$.
  \item \textbf{G4}. $\epsi \ge 0$.
  \item \textbf{G5} (each $j$). $\Delta A_j \le \Mdef{j} + (\Act - \epsi) + \sum_{k \ne j} \Mdef{k}$, matching \cref{lem:delta0}(ii).
  \item \textbf{G6}. $\Delta\Vul \le \max(\epsi,\ \Act - \epsi + \sum_q \Mdef{q})$, matching the total bound.
  \item \textbf{G7} (each $j$). The accounting identity $\mathrm{drop}_j = \sum_{p \ne j} \mathrm{gain}_p$ holds.
  \item \textbf{G8}. The two constructions reproduce their claimed values with equality. In \cref{ex:activation}, $\mhat_j = +1$ and $\Delta\Vul = 1 = \Act$. In \cref{ex:crossopp}, $\mhat_j = 0$, $\mhat_k = -1$, the pairwise best-response gain rises by $1 > (-\mhat_j)_+$, and $\Delta\Vul = 1 = \sum_q \Mdef{q}$.
\end{itemize}
The 97 checks finish in 170\,s with 0 violations. On the 7 CFR
configurations, the realized $\Delta\Vul$ ranges from $-6.0 \times 10^{-5}$ to
$0$, while the bound ranges from $3.4 \times 10^{-4}$ to $6.1 \times 10^{-2}$.
The suite also reports, as an ungated diagnostic, the implementation-margin
versus BR-margin comparison quantified in \cref{app:calibers}.

\subsection{The 6{,}157-check suite for the general case (\cref{lem:transfer,thm:main})}\label{app:machine:T1T2}

The suite is layered into five parts, progressing from theoretical
constants to a full experimental replay.

\begin{itemize}
  \item \textbf{Part A, perturbation law and explicit constant} (\cref{lem:linear,prop:K}, 2.3\,s). This part checks six gates. The pairpot family is affine in $\lambda$ (residual $\le 10^{-12}$), and $\dinf(u^0) = 0$. Enumerated $\max_{p,\rho}|C_p(\rho)| = 0.25$ holds, and $\dinf(C) = 0.0625$ by exact LP. Exact linearity $|\dinf(\lambda) - \lambda\,\dinf(C)| \le 10^{-9}$ holds on all 5 recorded $\lambda$ points, with coverage $K_{\mathrm{theory}} \ge K_{\mathrm{measured}}$ and a gap of $4.0\times \le 10\times$.
  \item \textbf{Part B, $\dinf = 0$ degeneration hard gate} (99\,s). All 7 exact-case CFR configurations and both examples are re-run under the \cref{thm:main} machinery. Here $\tax = 0$ on every configuration. The general bound and the best-response gap differences match those of \cref{lem:delta0} \emph{pointwise} within $10^{-9}$. The general proof degenerates to the exact one term by term.
  \item \textbf{Part C, fresh $\dinf > 0$ resolves} (141\,s). We evaluate 16 new configurations (pairpot $\lambda \in \{0.1, 0.25, 0.5, 1\}$ crossed with 4 resolve configurations at 500 blueprint and 1600 resolve CFR iterations). All theorem gates pass alongside the transfer gate $\zprof \le n\dinf$ and the composite $2n\dinf$ gate, while recomputed $\dinf$ matches the recorded measurement 4/4. Because the coupling is a transfer payment that preserves the utility sum, $\zsub$ and $\tax$ are numerically zero throughout, giving an entire family with $\dinf > 0$ and zero structural tax.
  \item \textbf{Part D, adversarial search points}. All 312 theorem-convention points from the certified adversarial search (\cref{sec:experiments}) are checked against the composite bound $\Delta\Vul \le \max(\epsi,\ \Act - \epsi + \sum_q \Mdef{q} + 2n\dinf)$. All pass, with a minimum slack of $7.4 \times 10^{-5}$ on game 12 in the unperturbed condition. The 155 fixed-certificate-convention points are skipped because their $\Act$ and $\epsi$ were not recorded and the convention lies outside the theorem's quantifier range (\cref{app:constructions:gid183}).
  \item \textbf{Part E, full screening replay} (1{,}665\,s). All \textbf{420} stage-1 screening configurations (210 games across 2 perturbation conditions) are re-run from the original random-number stream under a \emph{fidelity gate}. Replayed $\Delta\Vul$, margins, $\dinf$, and the exact or lower-bound convention flags must match the recorded screen values pointwise within $10^{-9}$. All 420 configurations match with zero replay deviation, and all theorem gates pass with a core-bound minimum slack of $3.4 \times 10^{-5}$ on game 7 under the uniform perturbation.

  \item \textbf{Part E, exact-$\dinf$ subset}. For the 148 games (296 points) whose tensors admit exact $\dinf$ enumeration, the transfer gate passes everywhere and is \emph{attained}. Specifically, $\max \zprof/(n\dinf) = 1.000000$ on games 49 and 41, showing that the $n$ factor in part (a) of \cref{app:T1T2:T1} is instance-level tight, while the composite gate passes with a minimum slack of $7.5 \times 10^{-3}$ on game 188. Across the 296 points, $\zprof$ is the smaller leg on 204 points and $\rho_S \zsub$ on 92, with $\tax < 2n\dinf$ on 266 points (relative slack beyond $10^{-9}$). The composite inequality $\tax \le 2n\dinf$ holds on all 296 points with 0 violations and is \emph{attained} ($\tax = 2n\dinf$) on the remaining 30 points (15 instances across 2 perturbation conditions). Each of those 30 points has $\zprof = n\dinf$ and $\rho_S\,\zsub \ge n\dinf$, so the $\min$ binds on the $\zprof$ leg.
\end{itemize}

In total, 6{,}157 checks pass with 0 violations in 1{,}908\,s. The replay covers the 420 stage-1 base points and omits the 13 mutated hill-climb points. For the 62 games (124 points) whose tensors exceed the enumeration cap, only the $\tax$-form core gate is checked, leaving the composite and transfer gates unverified there.

\textbf{Sample-free measurement.}
We measure every theorem-level quantity in the main text without sampling.
All negative results rest on eight exact finite constructions (\cref{app:constructions}), computed by full-table LP or enumeration.
Estimators for sampled evaluation, variance reduction \citep{burch2018aivat}, and anytime-valid stopping \citep{li2026av} are therefore unnecessary.

\section{Nested Resolving: a Telescoping Corollary and Supporting Data}\label{app:nested}

\textbf{Scope of \cref{asm:F1}.}
Condition (P1) requires that no information set straddle the boundary of $S$. Cut $S$ at a public state and (P1) holds automatically, since every player knows whether play has entered $S$. A cut below a private signal can break (P1) when a player who did not observe that signal owns information sets with nodes on both sides of the boundary. Condition (P2) requires that the gadget's entry grouping $\iota_q=(\mathrm{priv}_q,\mathrm{pub}^*)$ match $q$'s actual information at entry, with $q$'s own reach constant on each group. Under perfect recall, this holds whenever the entries follow a public signal and $\mathrm{priv}_q$ records $q$'s full private history. Suppose two entries share $(\mathrm{priv}_q,\mathrm{pub}^*)$ while differing in an earlier action of $q$. Player $q$ can then reach them with different probabilities, causing the margin of \cref{def:margin} to average over strategies $q$ can tell apart, so (P2) fails. Both conditions depend only on the cut.

The main theorem bounds a \emph{single} resolve. Practical systems resolve repeatedly as play descends the tree of public states. Under \cref{asm:F1}, these repeated resolves telescope. Indeed, \cref{thm:main} holds for \emph{any} update modifying only the resolver's strategy inside the current subgame, with all certificate quantities defined a posteriori for the delivered profile.

\begin{corollary}[Telescoping bound for sequential resolves]\label{cor:telescoping}
Let $\sigma^{(0)} = \bp$ and let $\sigma^{(t)}$, $t = 1, \dots, T$, be a sequence of profiles where each $\sigma^{(t)}$ is obtained from $\sigma^{(t-1)}$ by an update that changes only resolver $i$'s strategy inside a public subgame $S_t$, with \cref{asm:F1} holding for each $(\sigma^{(t-1)}, S_t)$. Let $\Act[t]$, $\Mdef{q}[t]$, $\epsi[t]$, and $\tax[t]$ denote the certificate quantities of \cref{thm:main} evaluated at step $t$ (blueprint role played by $\sigma^{(t-1)}$). Then
\[
\begin{gathered}
\Vul_G(\sigma^{(T)}) \;\le\; \Vul_G(\bp) \;+\; \sum_{t=1}^{T} B_t \;+\; \sum_{t=1}^{T} \tax[t],\\
B_t := \max\Bigl(\epsi[t],\; \Act[t] - \epsi[t] + \textstyle\sum_{q} \Mdef{q}[t]\Bigr),
\end{gathered}
\]
and in the $2n\dinf$ form of \cref{cor:2ndelta} the tax sum is at most $T \cdot 2n\,\dinf$.
\end{corollary}

\begin{proof}
Apply \cref{thm:main}(iii) at each step with $\sigma^{(t-1)}$ as blueprint and $\sigma^{(t)}$ as resolved profile, and sum the resulting one-step inequalities. The intermediate $\Vul_G(\sigma^{(t)})$ terms cancel. The tax bound $\tax[t] \le 2n\dinf$ is \cref{eq:app-T1-b} applied at each step (the game, and hence $\dinf$, is fixed across steps).
\end{proof}

In the exact form of \cref{cor:leak}, each one-step bound is $\max(\epsi[t],\ \Act[t]-\epsi[t]+\sum_q \Mdef{q}[t]+\Lambda[t])$, and their sum bounds $\Vul_G(\sigma^{(T)})-\Vul_G(\bp)$. The signed leak cannot be pulled outside the maximum. For instance, with one acting player, blueprint $a\mapsto(1,-1)$ and update $a'\mapsto(0,\tfrac12)$ give $B_1=1$ and $\Lambda[1]=-\tfrac12$, yet $\Delta\Vul=1>B_1+\Lambda[1]$. Although the leaks $\Lambda[t]$ telescope to $\sum_p u_p(\bp)-\sum_p u_p(\sigma^{(T)})$, only their positive parts $\Lambda[t]_+$ can be pulled out, and those do not telescope.

The corollary is linear in $T$. Empirically, the depth sweep below measures whether the \emph{certificate quantities} degrade as resolving proceeds, for instance through margins drifting negative after repeated local updates. The corollary covers CFR-D-style nested resolving, where each step edits only the resolver's strategy inside the current subgame.

\textbf{Supporting data.}
We ran a depth sweep on 14 instances (7 random, 7 near constant-sum polymatrix (CSP), from the same generator family and seed as the adversarial search), with the resolver fixed and $T = 3$ sequential resolves. The sequence resolves the depth-1 public subgame, then a depth-2 subgame below it, and then a sibling depth-2 subgame. At each step we recorded $B_t$ from \cref{lem:delta0} and the cumulative change $\mathrm{cum}_T := \Vul_G(\sigma^{(T)}) - \Vul_G(\bp)$. The check
\[
\mathrm{cum}_T \;\le\; \sum_{t \le T} B_t + T \cdot c_1 \dinf + \mathrm{tol}, \qquad \mathrm{tol} = 0.02,
\]
was evaluated for all 42 $(\text{instance}, T)$ pairs and for $c_1 \in \{0, 0.34, 1.0\}$, yielding \textbf{0 failures}. The pure-margin form $c_1 = 0$ passes everywhere, so the $\dinf$ term was unneeded on this family. Single-step violations ($\Delta\Vul_t > B_t + c_1\dinf + \mathrm{tol}$) were also zero. Per-step $\Delta\Vul$ has a median around $10^{-5}$ with no systematic growth in $t$. The one large single step ($t = 3$, $\Delta\Vul = 0.230$) is covered by its own concurrent certificate ($B_3 = 0.303$). The certificate quantities ($\Mdef{q}$, $\Act$, $\epsi$) show no drift amplification when later resolves start from the updated profile. In one boundary instance, $\mathrm{cum}_3 = 0.0163$ against $\sum_t B_t = 0.0050$ is absorbed by the tolerance at $c_1 = 0$ (any $c_1 \ge 0.04$ covers it without tolerance). The other 13 instances are covered by the margin term alone. The $\dinf$ values used are sampled \emph{lower} bounds.

\textbf{Tree depth.}
The instance family has public-tree depth $\le 3$ (decision depths 0 to 2). No depth-3 subgames exist, so the sweep combines subgame depths 1 and 2 with sequential step counts $T \in \{1, 2, 3\}$. The data support \cref{cor:telescoping} on trees of this depth.

\section{Per-Pair Statements: What Survives and What Fails}\label{app:perpair}

The main theorem operates at per-player granularity, summing the margin deficits of \emph{all} opponents. This appendix asks whether stronger structural conditions yield a \emph{per-pair} safety statement that bounds each pairwise best-response gain by that pair's own certificate. Under two additional assumptions, half of the literal per-pair form is a theorem. The other half is false, and \cref{ex:crossopp} is the minimal counterexample.

\textbf{Additional assumptions.}
Fix a constant-sum polymatrix (CSP) witness $\{u_{pq}\}$ attaining $\dinf = 0$.
\begin{itemize}
  \item \textbf{(F3) Pair-local witness.} The witness can be chosen so that $u_{pq}$ is a function $f_{pq}(\omega_c, s_p, s_q)$ of the chance outcome and the private type and action sequence of players $p$ and $q$ only.
  \item \textbf{(F4) Pair blindness.} For every pair $(p,q)$ not containing the resolver $i$, the joint distribution of $(\omega_c, s_p, s_q)$ does not depend on $i$'s strategy inside $S$. This holds when $p$ and $q$ cannot observe $i$'s in-subgame actions before completing their own, and no chance move in $\omega_c$ follows an in-subgame action of $i$.
\end{itemize}
Both $\dinf = 0$ families used in this paper (pairpot at $\lambda = 0$ and pairwise-showdown) satisfy F3 and F4 because pairwise settlement is per-edge independent and actions are mutually unobserved.

\textbf{Pairwise best-response gain.}
For a witness $\{u_{pq}\}$, define
\[
\mathrm{PairVul}_{pq}^{\sigma_{-(p,q)}}(\sigma_p, \sigma_q) \;:=\; \max_{\tau_p}\, \mathbb{E}_{(\tau_p, \sigma_q, \sigma_{-(p,q)})}[u_{pq}] \;-\; \mathbb{E}_{(\sigma_p, \sigma_q, \sigma_{-(p,q)})}[u_{pq}],
\]
and write $X_{ij} := \mathbb{E}_{\rs}[u_{ij}] - \mathbb{E}_{\bp}[u_{ij}]$ for the resolver's on-path value change on board $(i,j)$ (the \emph{board} $(p,q)$ is the pair table $u_{pq}$). Throughout, $\rs$ is an update confined to resolver $i$'s subgame strategy, $\dinf = 0$, and the certificate quantities $\mhat$, $\Mdef{q}$, $\Act$, and $\epsi$ are as in \cref{sec:prelim,sec:main}.

\begin{proposition}[Per-pair statements under F3 and F4]\label{prop:perpair}
Under \cref{asm:F1}, $\dinf = 0$, F3, and F4:
\begin{itemize}
  \item[\textup{(C1)}] \textbf{Identity.} For every opponent $j$, $\mathrm{gain}_j = -X_{ij}$ (boards not containing $i$ are frozen by F3/F4, and the $(i,j)$ board is per-edge constant-sum under the $\dinf=0$ witness), and $\sum_j X_{ij} = \mathrm{gain}_i = \Act - \epsi$. The edge constants $c_{pq}$ may be nonzero. They cancel in every difference used below, so the identities are unaffected.
  \item[\textup{(C2)}] \textbf{Deviator-$i$ pairs: the literal form holds with constant $1$.} We have $\Delta\mathrm{PairVul}_{ij} = -X_{ij} \le (X_{ij})_- \le \Mdef{j}$. The max term of $\mathrm{PairVul}_{ij}$ does not depend on $\sigma_i$, and $j$'s pairwise on-path gain equals its full-utility on-path gain under F3/F4, which is priced by $j$'s own margin deficit.
  \item[\textup{(C3)}] \textbf{Telescoping identity.} $\Delta\mathrm{PairVul}_{ij} + \Delta\mathrm{PairVul}_{ji}$ equals the change in $j$'s pair best-response value, $\max_{\tau_j} U_{ji}(\tau_j, \rs_i) - \max_{\tau_j} U_{ji}(\tau_j, \bp_i)$.
  \item[\textup{(C4)}] \textbf{Deviator-$j$ pairs: the literal form is false.} There is a game satisfying all the hypotheses above (\cref{ex:crossopp}) with $\epsi = 0$ and $\mhat_j(\iota) = 0$ for all $\iota$, yet $\Delta\mathrm{PairVul}_{ji} = 1 > 0 = (-\mhat_j)_+$. No bound of the form $\Delta\mathrm{PairVul}_{ji} \le C_m(-\mhat_j)_+ + C_\varepsilon\,\epsi$ can hold with any finite constants.
\end{itemize}
\end{proposition}

Cross-opponent value transfer drives (C4). In \cref{ex:crossopp}, the resolver's update moves value from the $(i,k)$ board to the $(i,j)$ board while leaving its own total unchanged. The \emph{victim} $j$'s margin stays silent because $j$'s follow-best-response is held at its blueprint counterfactual value. Meanwhile, the alarm fires at the \emph{beneficiary} $k$ ($\mhat_k = -1$). A per-pair bound keyed to the victim's own certificate is therefore unobtainable. \Cref{thm:main}(ii) prices every cross-opponent trade at the per-player level through the cross terms $\sum_{k \ne j} \Mdef{k} + \Act$, charging it in full to the beneficiary's margin and the resolver's activation term.

The 97-check suite of \cref{app:machine} verifies (C2) and (C4) numerically. On \cref{ex:crossopp}, it confirms $\mhat_j = 0$, $\mhat_k = -1.0$, and $\Delta\mathrm{PairVul}_{ji} = 1.0 > (-\mhat_j)_+ = 0$, while the aggregate bound of \cref{lem:delta0}(iii) holds with equality ($\Delta\Vul = 1.0 = \sum_q \Mdef{q}$).

\section{Poker Experiments: Protocol and Additional Tables}\label{app:poker}

\textbf{Games.} In $n$-player Kuhn poker, a deck of $n+1$ cards gives one private card to each player. Each player pays an ante of $1$, and one betting round allows at most one bet of $1$. At $n=3$, this matches the three-player Kuhn tree of \cref{prop:pokerlb}. We checked this tree against the OpenSpiel implementation \citep{lanctot2019openspiel}, confirming identical utilities and best-response gains under the uniform profile within $10^{-15}$. Three-player Leduc uses a six-card deck with two copies of three ranks. Players pay an ante of $1$ and receive one private card each. A public card is dealt from the three remaining cards. The two betting rounds use bet sizes of $2$ and $4$ with at most one raise per round, producing \leducDecisionNodes{} decision nodes and \leducInfosets{} information sets. The rake experiment charges the winner $\min(r\cdot\mathrm{pot}, 1)$ at every terminal.

\textbf{Blueprints.} We take CFR average strategies after $10$, $50$, $200$, and
$1000$ iterations on Kuhn and $10$, $100$, and $1000$ on Leduc. An abstracted
blueprint merges the two lowest cards on Kuhn or the two highest ranks on Leduc
during $1000$ CFR iterations and lifts the result to the full game. At every
information set, noisy blueprints mix the $1000$-iteration blueprint with $5\%$
and $20\%$ of the uniform strategy on Kuhn, or $10\%$ on Leduc. The $n$-player
experiment uses the $200$-iteration CFR blueprint for every $n$.

\textbf{Subgames and resolvers.} On Kuhn, subgames start at the public states after the
first bet and after the first check, and also after two checks at $n=3$.
On Leduc, every second-round public state (defined by the first-round betting sequence and
board card) forms a subgame. Every player with an information set inside the subgame serves as the
resolver in turn. This yields $8$ (subgame, resolver) pairs per blueprint
on three-player Kuhn, $108$ on Leduc, and $5$, $7$, $9$, and $11$ on Kuhn
with $n=3,4,5,6$.

\textbf{Arms and certificates.} For the three resolvers of \cref{sec:experiments}, CFR solves the gadget in $800$ (Kuhn), $200$ (Leduc), or $50$ ($n$-player Kuhn) iterations. The shrink conditions deliver $(1-\lambda)\bp_i+\lambda\rs_i$ on the resolver's subgame information sets, where $\rs$ is the standard-gadget update. For each delivered update, exact best response gives $A_p(\bp)$ and $A_p(\rs)$ for all $p$. We also compute $\Delta\Vul=\max_pA_p(\rs)-\max_pA_p(\bp)$, $\Delta\mathrm{NashConv}$, $\Act(\bp)$, $\epsi$, the BR-certified deficits $\Mdef{q}$ for every opponent, $\Lambda$, $\tax$, $B_{\mathrm{dir}}$, and $B_{\mathrm{cache}}$. Finally, we record single-core CPU wall-clock times for the resolve, the local certificate (best-response passes inside $S$, the activation term, and expected utilities), and the exact full-game evaluation.

\textbf{Per-blueprint results.} The pooled rows of \cref{tab:poker} aggregate across blueprint qualities. Across individual blueprints, the standard gadget is harmful in $12\%$ to $50\%$ of updates on Kuhn and $1\%$ to $5\%$ on Leduc. The certificate accepts no harmful update for any blueprint-condition pair. Complete per-blueprint tables appear in the supplementary material.

\textbf{Why no root measure can serve all opponents.} \Cref{tab:measure} reports each opponent's minimum gadget margin on three-player Kuhn across ten combinations of blueprint, subgame, and resolver under four root measures. These are the standard $c_e\pi_i(e)$, the belief-corrected $c_e\prod_p\pi_p(e)$, and the per-opponent counterfactual measures $c_e\pi_{-q}(e)$ for opponent $q$. Counterfactual measures price each opponent's Terminate option correctly, but they differ across opponents. One gadget cannot use them all. None of the \eFourCells{} cells makes all opponents' margins nonnegative, blocking a Terminate-style guarantee for $n>2$ (\cref{prop:mismatch}). The blueprints use $200$ and $1000$ CFR iterations, the public states follow a bet (b) or a check (k), and players are numbered from $1$. Each cell lists the two opponents in increasing player order.

\begin{table}[ht]
\centering
\caption{Minimum gadget margin $v(\mathrm{T})-v(\mathrm{F})$ of the two
opponents (separated by a slash) on three-player Kuhn under four root
measures.}
\label{tab:measure}
\footnotesize
\begin{tabular}{llrllll}
\toprule
blueprint & pub & $i$ & $c_e\pi_i$ & $c_e\prod_p\pi_p$ & $c_e\pi_{-q}$, opp.\ 1 & $c_e\pi_{-q}$, opp.\ 2 \\
\midrule
cfr200 & b & 2 & -0.432 / -1.097 & -0.386 / -0.102 & -0.432 / -1.097 & -0.386 / -0.102 \\
cfr200 & b & 3 & -0.433 / -0.688 & -0.404 / -0.032 & -0.433 / -0.688 & -0.404 / -0.032 \\
cfr200 & k & 1 & -0.025 / -0.016 & -0.025 / -0.016 & -0.025 / -0.016 & -0.025 / -0.016 \\
cfr200 & k & 2 & -0.023 / -0.027 & -0.037 / -0.006 & -0.023 / -0.027 & -0.037 / -0.006 \\
cfr200 & k & 3 & +0.004 / -0.014 & +0.004 / -0.016 & +0.004 / -0.014 & +0.004 / -0.016 \\
cfr1000 & b & 2 & -0.422 / -0.991 & -0.015 / -0.029 & -0.422 / -0.991 & -0.015 / -0.029 \\
cfr1000 & b & 3 & -0.423 / -0.669 & -0.040 / -0.016 & -0.423 / -0.669 & -0.040 / -0.016 \\
cfr1000 & k & 1 & -0.026 / -0.011 & -0.026 / -0.011 & -0.026 / -0.011 & -0.026 / -0.011 \\
cfr1000 & k & 2 & -0.034 / -0.072 & -0.011 / -0.076 & -0.034 / -0.072 & -0.011 / -0.076 \\
cfr1000 & k & 3 & -0.021 / -0.007 & -0.002 / -0.004 & -0.021 / -0.007 & -0.002 / -0.004 \\
\bottomrule
\end{tabular}

\end{table}

\begin{table}[ht]
\centering
\caption{Shrinking the standard-gadget update toward the blueprint,
$(1-\lambda)\bp_i+\lambda\rs_i$.}
\label{tab:shrink}
\footnotesize
\begin{tabular}{lrrrrr}
\toprule
Game & $\lambda$ & \#matched & mean $\Delta\Vul$ & harmful & mean $g_i$ \\
\midrule
Kuhn & 0.25 & 56 & +0.0001 & 2\% & +0.0020 \\
Kuhn & 0.50 & 56 & +0.0038 & 9\% & +0.0040 \\
Kuhn & 0.75 & 56 & +0.0092 & 18\% & +0.0061 \\
Kuhn & 1.00 & 56 & +0.0159 & 29\% & +0.0082 \\
Leduc & 0.25 & 540 & -0.0007 & 0\% & +0.0007 \\
Leduc & 0.50 & 540 & -0.0010 & 1\% & +0.0015 \\
Leduc & 0.75 & 540 & -0.0012 & 1\% & +0.0023 \\
Leduc & 1.00 & 540 & -0.0009 & 2\% & +0.0031 \\
\bottomrule
\end{tabular}

\end{table}

\begin{table}[ht]
\centering
\caption{NashConv replay of the $162$ random-game updates.}
\label{tab:nashconv}
\footnotesize
\begin{tabular}{lr}
\toprule
Quantity & count \\
\midrule
replayed updates & 162 \\
violations of $\Delta\mathrm{NashConv}\le\sum_q\Mdef{q}+\Lambda$ & 0 \\
violations of $\Delta\mathrm{NashConv}\le\sum_q\Mdef{q}+\tax$ & 0 \\
updates with $\Delta\Vul\le0$ & 53 \\
updates with $\Delta\mathrm{NashConv}\le0$ & 119 \\
$\Delta\Vul\le0$ but $\Delta\mathrm{NashConv}>0$ & 0 \\
$\Delta\mathrm{NashConv}\le0$ but $\Delta\Vul>0$ & 66 \\
harmful, $\Delta\Vul>0.02$ & 19 \\
harmful, $\Delta\mathrm{NashConv}>0.02$ & 15 \\
\bottomrule
\end{tabular}

\end{table}

\textbf{NashConv versus the maximum gap.} The NashConv bound holds on all $162$
random-game updates of \cref{sec:experiments} (\Cref{tab:nashconv}). On
\eOneNashSafeVulHarm{} updates, NashConv does not increase while the maximum
gap $\Vul_G$ does. The reverse happens on \eOneVulSafeNashHarm{} updates.
Whenever the maximum gap is safe, the sum NashConv is safe too, whereas the sum
hides increases in a single player's gain. In the poker games, the gadget
resolvers keep the mean $\Delta\mathrm{NashConv}$ within $\pm0.04$ of zero.

\textbf{Timing.} The $n$-player wall-clock time in \cref{tab:nkuhn} is the
median over updates. For the gadget resolvers, gadget CFR dominates resolve
time because its cost grows with the gadget tree ($n-1$ Terminate/Follow layers
above a six-player subgame at $n=6$). On three-player Leduc, the median local
certificate takes $0.6$\,s versus $0.7$\,s for the exact full-game evaluation
and $4.4$\,s for the resolve. The two evaluations take similar time because the
second-round subgame is a large fraction of the tree. \Cref{tab:cost}
decomposes this cost into the activation and margin passes.

\section{Adversarial Evaluation on Random Games (original protocol)}\label{app:submitted}

This appendix reproduces the random-game evaluation of the original experimental protocol, with the symbol $\Vul_G$ renamed. The poker experiments appear in \cref{sec:experiments}.

Numerical regression checks corroborate every inequality in \cref{thm:main}. This appendix searches for theorem-level counterexamples, meaning instances where the gap rises while every term of the bound remains silent, and measures how small the coefficients of a two-term $(\dinf,\sum_q\Mdef{q})$ law can be made. Three instruments drive this analysis (\cref{app:expdetails}). First, a certified cutting-plane best-response search explores small $n$-player EFGs, evaluating $162$ instances at $n=3$ without sampling. A heuristic sweep then covers $433$ screened points. Finally, we apply LP fits alongside a theorem-level counterexample filter at $\mathrm{tol}=0.02$.

\textbf{Zero theorem-level counterexamples.}
On the full $162$-instance campaign set, the strict filter fires \emph{zero} times
under both margin conventions. Every $\Delta\Vul$ above tolerance is explained by a
large $\dinf$, a margin deficit, $\Act$, or $\epsi$. Relaxing the filter to
($\dinf\le 0.15$, $\sum_q\Mdef{q}\le 0.05$) and then ($\dinf\le 0.30$,
$\sum_q\Mdef{q}\le 0.10$) yields $3$ and $6$ hits, none theorem-level. The
screening sweep agrees, reaching $\max\Delta\Vul=0.039$ at $\dinf\le 0.1$ across
$156$ configurations.

\textbf{A two-term ablation.}
An LP over all campaign measurements finds the minimum $(c_1,c_2)$ satisfying
$\Delta\Vul \le c_1\dinf + c_2\sum_q\Mdef{q} + \mathrm{tol}$ at every point
(\cref{tab:constants}). This gives $(1.76, 0.31)$ at $\mathrm{tol}=0.02$ and $(2.35, 0.17)$
at $\mathrm{tol}=0$. Both terms are needed. Setting $c_2=0$ forces
$c_1\ge 2.64$, and setting $c_1=0$ is infeasible. The resulting pair differs from
the $(2n,1)$ coefficients of the $2n\dinf$ bound, because that bound also
carries $\Act$ and $\epsi$.

\textbf{Scaling in the number of players.}
Varying $n$ fits the $c_1$ coordinate of the two-term law at $0.162$, $0.563$, and $0.916$ for $n=3,4,5$. The slope is $0.377$ per player (\cref{app:expdetails:nsweep}).

\textbf{Which terms carry the bound.}
We split the bound into $B=\max(\epsi,\ \Act-\epsi+\sum_q \Mdef{q})$
and the $\dinf$ term (\cref{tab:constants}). Under the nominal convention, $B$ alone covers all $162$ points.
Under the pipeline convention, $B$ covers $89.3\%$ of $150$ points, and $\dinf$ covers the rest.

\textbf{What the certificate does for delivered updates.} On the $162$ delivered campaign updates, every certificate form covers $\Delta\Vul$. At tolerance $0.02$, every form flags all $19$ harmful updates (\cref{tab:acceptance-main}). At tolerance $0$, the structural forms accept none of the $143$ safe updates. A form is useful when its bound is small enough to \emph{accept} a safe update. The baseline-aware form $B_{\mathrm{cache}}$ accepts $37$ safe updates at zero tolerance. It costs $1.5\times$ a cached exact evaluation, which already accepts every safe update on these small games. The local certificate's cost remains constant as the trunk grows (\cref{tab:trunk-main}).

\section{Experimental Details}\label{app:expdetails}

This appendix details the protocol behind \cref{sec:experiments}, covering the search procedure, the affine encoding of the attack, the filter tiers, the $n$-sweep, and the fitting LP for the minimal-constant analysis. All runs use a single-core CPU with seed 42. Payoffs and best responses are evaluated without sampling. On the full table, LPs are solved to the stated tolerances, and restricted-support distances are reported as lower bounds.

\subsection{Heuristic screen and certified attack campaign}\label{app:expdetails:harness}

\textbf{Screen.} We sample 210 random small extensive-form games across three families of 70 each (random, near-CSP with coupling $\lambda \sim U[0, 0.3]$, and synthetic with $\lambda \sim U[0, 0.3]$). The trees have depth $\le 3$, 2 or 3 actions per node, and random utilities. Crossing these with 2 corruption conditions (none, or a uniform-or-Dirichlet reset of the resolver's subgame strategy) yields 420 configurations. Each run computes a 400-iteration CFR blueprint, applies corruption, resolves the gadget for 800 iterations, and measures the exact best-response $\Delta\Vul$. The screening objective is $\Delta\Vul_{\max} - 0.03\,\dinf - 0.03\,(-\mathrm{margin}_{\min})_+$ with tolerance $0.02$ (the preliminary run's CFR noise floor) and benign-family fit constants of $0.03$. The top-5 points undergo 25 steps of Gaussian leaf-utility hill climbing. The top-3 receive a deep re-solve ($7.5\times$ blueprint and $5\times$ resolve iterations, $\times 3$), giving 433 data points in total. We compute $\dinf$ as an exact LP for small games and a restricted-support lower bound for large ones, which overestimates the objective. Deterministic replay completed all 420 stage-1 records with zero deviation (\cref{app:machine}).

\textbf{Affine encoding and certified search.} The exact campaign replaces heuristic best responses with exact evaluations. Margins enter the model through an affine encoding. Against brute-force enumeration, its values show a maximum deviation of $1.1 \times 10^{-16}$ on the smoke-test instance. The mixed-integer linear program (MILP) optimum also matches exact brute force on a perturbation grid. Across the 162 instances of the full campaign, the encoding check reports 0 errors and a maximum absolute deviation of $9.1 \times 10^{-10}$.

With $(c_1, c_2) = (1, 0.5)$ in the search objective, the $486$ (instance, player) searches finish with 380 certified, 78 exceeding the trial bound, and 28 reaching the call budget. Exceeding the trial bound means a point exceeds the searched bound for those trial constants, distinct from a theorem violation. Every such point passes through the filters below. When the budget is reached, the cutting-plane attack has exhausted its $\dinf$-oracle allowance of $60$ exact LP calls. Its incumbent (the best value reached) remains below its upper bound. For those $28$ searches, only the exactly evaluated incumbent enters any table, leaving the attack interval above it open. The campaign is therefore complete as a certificate of every reported point and incomplete as a search on those $28$.

We evaluate payoffs, reach probabilities, and best responses in floating point without sampling, using backward induction over the enumerated pure strategies ($\le 128$ per player). Where the profile table fits the enumeration cap, $\dinf$ is the optimum of the full-table LP. Otherwise, we report a restricted-support lower bound. Each attack search bounds individual pure strategies via an LP over globally valid $\dinf$ cuts, closing at an absolute gap of $10^{-7}$ or an upper bound below tolerance. HiGHS \citep{huangfu2018parallelizing} solves all LPs at its default primal and dual feasibility tolerances ($10^{-7}$). Here, ``exact'' means full-table and sample-free within those tolerances, distinct from exact arithmetic. The reported numbers come from the certified cutting-plane search. For the usefulness study, a replay run with Python 3.12.3, NumPy 1.26.4, and SciPy 1.11.4 reproduced all $162$ $\Delta\Vul$ values and the minimum zero-tax slack of $4.49\times10^{-6}$ exactly. Library versions from the original cluster run were not logged.

\subsection{Counterexample filter tiers}\label{app:expdetails:kill}

A theorem-level counterexample must combine small $\dinf$, a silent margin certificate, and a real best-response gap increase unexplained by $\Act$ or $\epsi$. The screen applied four nested threshold tiers across its 433 points (\cref{tab:tiers}).

\begin{table}[ht]
\centering
\caption{Counterexample filter tiers on the $433$ screened points.}
\label{tab:tiers}
\footnotesize
\begin{tabular}{@{}llll@{}}
\toprule
Tier & Margin condition & $\dinf$ condition & Hits ($\Delta\Vul > 0.02$) \\
\midrule
1 (strict) & $\mathrm{margin} \ge 0$ & $\dinf \le 0.10$ & \textbf{0} \\
2 & $\mathrm{margin} \ge -0.05$ & $\dinf \le 0.10$ & \textbf{0} \\
3 & $\mathrm{margin} \ge -0.05$ & $\dinf \le 0.15$ & \textbf{0} \\
4 (very loose) & $\mathrm{margin} \ge -0.10$ & $\dinf \le 0.30$ & 3 (each with $\dinf \ge 0.05$ or margin $\le -0.09$) \\
\bottomrule
\end{tabular}
\end{table}

In the small-$\dinf$ region ($\dinf \le 0.1$, 156 configurations), the maximum $\Delta\Vul$ is $0.039$. Only 2 points exceed tolerance, both with margin $\le -0.09$. Across all 56 above-tolerance points, the smallest $\dinf$ is $0.051$ and the largest implementation-convention margin is $+0.031$, attained at $\dinf = 0.43$ by the construction of \cref{prop:delta-nec} in \cref{app:constructions:delta}.

We run the certified attack campaign across three filter tiers in two margin conventions, BR margin $\mhat$ and implementation margin. The \emph{strict} tier requires small $\dinf$, quiet $\mhat$, and $\Delta\Vul > \mathrm{tol}$, excluding points explained by $\Act$. In the \emph{relaxed} tier we allow $\dinf \le 0.15$ and $\sum_q\Mdef{q} \le 0.05$, and in the \emph{loose} tier $\dinf \le 0.30$ and $\sum_q\Mdef{q} \le 0.10$. Across the 162 instances, the strict filter yields \textbf{0 hits} in both conventions. The relaxed and loose filters yield 3 and 6 hits, respectively, each explained by a positive $\sum_q\Mdef{q}$ or by $\Act/\epsi$, leaving no theorem-level candidates. Under the non-theorem fixed-certificate convention, a single strict candidate appears, analyzed in \cref{app:constructions:gid183}.

\subsection{$n$-sweep protocol and restricted convention}\label{app:expdetails:nsweep}

The sweep runs the certified attack pipeline at $n \in \{3, 4, 5\}$, targeting 32 instances per $n$ and fitting the $c_1$ coordinate of the minimum-sum LP (\cref{app:expdetails:fitlp}). Runs completed 32/32 instances (98 points) at $n = 3$ and 31/32 (91 points, after one worker terminated early) at $n = 4$. At $n = 5$, \textbf{15/32} instances (36 points, 7 binding) finished before a 24-hour compute cutoff. The $n = 5$ results reflect the 15 fastest instances. Because slower unfinished instances could be more adversarial, the $n = 5$ figures carry a fast-instance selection bias.

\textbf{$\dinf$ convention under the enumeration cap.} Exact $\dinf$ computation enumerates up to $60{,}000$ pure-strategy profiles. We skip instances above this cap to avoid restricted-support lower bounds. All 15 completed $n = 5$ instances have $1{,}024$ profiles, so every $n = 5$ $\dinf$ value in the sweep is an exact LP optimum.

\textbf{Results.} Under the main joint LP fit, $c_1$ grows across $n = 3, 4, 5$ as $0.162$, $0.563$, and $0.916$, which gives a slope of $0.377$ per player and a second difference of $-0.049$. Pinning $c_2 = 0$ yields $c_1 = (0.190,\ 0.563,\ 0.916)$ with second difference $-0.020$, whereas pinning $c_2 = 0.5$ yields $c_1 = (0.162,\ 0.260,\ 0.606)$ with second difference $+0.247$. These exploratory ablation fits omit $\Act$ and $\epsi$ in the two-term law, estimate $c_1$ necessity only on the discovered points, and inherit the fast-instance bias on the $n = 5$ subset. Consequently, the sequences do not measure the scaling of the theorem's $2n$ coefficient. On this family at $n = 3$, a margin-only fit ($c_1 = 0$) is infeasible within the two-term law because the uncovered points require $\dinf$, $\Act$, or $\epsi$. Pinning $c_2 = 0$ there inflates the required $c_1$ by a factor of $1.17$, from $0.162$ to $0.190$. We report these numbers separately from the more adversarial MILP attack family.

\subsection{Fitting LP for minimal constants}\label{app:expdetails:fitlp}

Given a point set with per-point data $(\Delta\Vul_r, \delta_r, M_r)$, where $\delta_r$ is the instance's $\dinf$ and $M_r$ its margin-deficit aggregate, the minimal-constant analysis solves the LP
\[
\min\; c_1 + c_2 \quad \text{s.t.} \quad \Delta\Vul_r \;\le\; c_1\,\delta_r + c_2\,M_r + \mathrm{tol} \;\; \forall r, \qquad c_1, c_2 \ge 0,
\]
with $\mathrm{tol} = 0.02$ (and a $\mathrm{tol} = 0$ variant), alongside solves pinned at $c_2 = 0$, $c_2 = 0.5$, and $c_1 = 0$. The output estimates a necessity lower bound on the constants required for the hardest discovered points, as opposed to a sufficiency certificate. \Cref{tab:constants} summarizes the main attack campaign fits across all 162 instances, where 17 points are binding at $\mathrm{tol}=0.02$ and the trial convention $(1, 0.5)$ leaves 2 points uncovered. An earlier 118-instance snapshot gave $(0.34, 0.81)$. On the 433-point heuristic screen, the fit gives $(0.39, 0.30)$ with tolerance and $(0.11, 0.69)$ without. Pinning $c_2 = 0$ on the screen forces $c_1 \ge 1.33$ with tolerance and $1.99$ without, while $(1, 0.5)$ covers all 433 points. Because the two-term law omits the $\Act$ and $\epsi$ terms of \cref{cor:2ndelta}, the fitted pair does not measure the tightness of the proven coefficients $(2n, 1) = (6, 1)$.

\begin{table}[ht]
\centering
\caption{Minimum feasible constants of the two-term ablation on the $162$-instance attack set ($n=3$).}
\label{tab:constants}
\footnotesize
\begin{tabular}{llcc}
\toprule
Source & Condition & $c_1$ & $c_2$ \\
\midrule
\cref{cor:2ndelta} & coefficients of $\dinf$, $\sum_q\Mdef{q}$ ($n=3$) & $6$ & $1$ \\
Fitted (LP) & $\mathrm{tol}=0.02$ & $1.76$ & $0.31$ \\
Fitted (LP) & $\mathrm{tol}=0$ & $2.35$ & $0.17$ \\
Fitted (LP) & $c_2=0$ pinned & $\ge 2.64$ & $0$ \\
Fitted (LP) & $c_1=0$ pinned (margin-only) & \multicolumn{2}{c}{infeasible} \\
\bottomrule
\end{tabular}
\end{table}

The ablation in \cref{tab:constants} fits $\Delta\Vul \le c_1\dinf + c_2\sum_q \Mdef{q} + \mathrm{tol}$. Its first row lists the theoretical coefficients from \cref{cor:2ndelta}, whose bound also includes $\Act$ and $\epsi$.

\subsection{Certificate usefulness on the gadget's own outputs}\label{app:expdetails:usefulness}

The campaigns above test the inequality. This experiment evaluates the certificate on updates delivered by the gadget. For each instance, the attack campaign records one \emph{nominal} update $\bp\to\rs$ on the unperturbed utility table. Each uses a 400-iteration CFR blueprint and an 800-iteration gadget resolve across the \emph{none} and \emph{corrupt} conditions. Where the attack finds an argmax, it also records one \emph{pipeline} update from a fresh blueprint and resolve on the perturbed table. Together, the $162$ nominal and $150$ pipeline points form the $312$ theorem-convention points of \cref{app:machine}.

We re-run the $162$ nominal updates with identical seeds, games, conditions, tables, and blueprints, recomputing $\Delta\Vul$ by exact best response. All $162$ values and the minimum slack of the zero-tax bound ($4.49\times10^{-6}$) match the recorded campaign values exactly. Coverage is tested at an absolute tolerance of $10^{-9}$. We exclude the $30$ skipped pool records.

For each update we evaluate the four instantiations of \cref{thm:main},
\[
B_X \;=\; \max\bigl(\epsi,\ \Act-\epsi+\textstyle\sum_q\Mdef{q}+X\bigr),\qquad X\in\{\Lambda,\ L_i,\ \tax,\ 2n\dinf\},
\]
the zero-tax bound $X=0$ (which drops the structural term while keeping activation and solve error), and the direct per-player form $B_{\mathrm{dir}}=\max(-g_i,\max_{j\ne i}[\Mdef{j}-g_j])$ of \cref{eq:direct}, all with BR-certified margins.

Computing $\Lambda$ takes two expected-utility evaluations. For $L_i$, we compute one subgame-confined best response of the resolver on the total utility (\cref{app:T1T2}). Evaluating $\tax=2\min(\zprof,\rho_S\zsub)$ enumerates the full pure-profile table (at most $6{,}561$ profiles here), while $2n\dinf$ solves the exact $\dinf$ LP. The direct bound $B_{\mathrm{dir}}$ needs only the margins and the $n$ expected utilities.

\Cref{tab:usefulness} summarizes coverage, slack, acceptance rates, and runtime across the $162$ instances. Every instance satisfies $\Lambda\le L_i\le\tax\le 2n\dinf$ and $L_i\ge 0$. On $91$ instances, $\Lambda<0$ because the update raises total utility. Moreover, $B_{\mathrm{dir}}\le B_\Lambda$ holds on all $162$ instances, and strictly so on $147$. Because every form is a valid upper bound, each variant flags all $19$ harmful updates ($\Delta\Vul>0.02$). The bounds differ on the $143$ safe updates, accepting $104$ under $B_{\mathrm{cache}}$ and $73$ under $B_{\mathrm{dir}}$, down to $2$ under $2n\dinf$. Although uncertified for $\dinf>0$, the zero-tax bound covers all $162$ nominal points, yet fails on the $150$ pipeline points.

On the same single-core replay ($n=3$, small trees), we measure each certificate's total wall-clock time against exact recomputation of $\Vul_G$ for both profiles. Recomputation requires six full-game best responses (median $2.2$\,ms). Because we report the median of paired per-instance ratios, ratios and median times need not multiply out. The direct forms $B_{\mathrm{dir}}$, $B_{\mathrm{cache}}$, and the realized-leak form $B_\Lambda$ run in $1.2$ to $2.3$\,ms, comparable to full recomputation. By contrast, $\tax$ and $2n\dinf$ are design-time quantities because enumerating the pure-profile table or solving the exact LP takes orders of magnitude longer. Worst-case ratios reach $9.3\times$ for $B_{\mathrm{dir}}$, $11.9\times$ for $\Lambda$, and $68\times$ for $\tax$.

A cached exact evaluation is the natural runtime baseline for $B_{\mathrm{dir}}$. Because the update is unilateral, $\mathrm{BRval}_i(\rs_{-i}) = \mathrm{BRval}_i(\bp_{-i})$, so the exact $\Delta\Vul$ needs only the blueprint's stored best-response values and $n-1=2$ full-game best responses on $\rs$. Timing the methods in a single process over the $162$ updates (minimum of five repeats) puts the cached evaluation at $0.37\times$ the full recomputation. The bound $B_{\mathrm{dir}}$ costs $1.49\times$ the cached evaluation at the paired median ($2.16\times$ worst, $0.55\times$ the full recomputation) and is slower on all $162$ updates. On trees this small, subgame-confined best responses save little time. The benefit of $B_{\mathrm{dir}}$ is locality, since it never leaves $S$ and its entry sets while the cached baseline traverses the full game for the $n-1$ opponents.

The cached baseline also highlights the selectivity gap at zero tolerance. Although $73$ of the $162$ updates satisfy $\max_p[A_p(\rs)-A_p(\bp)]\le 0$, $B_{\mathrm{dir}}$ accepts only $2$ at tolerance $0$ and the forms of \cref{thm:main} accept none, matching the constant-sum behavior derived in \cref{app:T1T2} (here the median $|\Lambda|$ is $8\times10^{-4}$). The baseline-aware form $B_{\mathrm{cache}}$ of \cref{cor:cache} closes much of this gap while running in the same time as $B_{\mathrm{dir}}$ ($1.47\times$ the cached evaluation at the paired median, $2.06\times$ worst). It satisfies $B_{\mathrm{cache}} \le B_{\mathrm{dir}}$ on all $162$ updates, strictly on $112$, and accepts $37$ updates at tolerance $0$. No accepted update has $\Vul_G(\rs)>\Vul_G(\bp)+\mathrm{tol}$. At tolerance $0$, $B_{\mathrm{cache}}$ accepts $8$ updates where the per-player maximum increase is positive while $\Vul_G$ does not rise ($105$ of the $162$ updates have $\Vul_G(\rs)\le\Vul_G(\bp)$). Finally, the gadget's proxy margin is negative on $148$ updates, whereas the BR-certified deficit $\sum_q\Mdef{q}$ is positive on $145$. They disagree on $5$ instances (four false proxy alarms and one missed deficit), consistent with \cref{prop:mismatch}.

\begin{table}[ht]
\centering
\caption{Certificate usefulness on the $162$ nominal updates.}
\label{tab:usefulness}
\footnotesize
\begin{tabular}{lcccc}
\toprule
Form & Cover & \shortstack{Slack\\med.\,/\,q90} & \shortstack{Acc.\\$0$\,/\,$0.02$\,/\,$0.05$\,/\,$0.10$} & Time (ratio) \\
\midrule
$B_{\mathrm{dir}}$ (\cref{eq:direct}) & $162/162$ & $0.018$ / $0.22$ & $2$ / $73$ / $85$ / $113$ & $1.3$\,ms ($0.57\times$) \\
$B_{\mathrm{cache}}$ (\cref{cor:cache}) & $162/162$ & $0.020$ / $0.22$ & $37$ / $104$ / $122$ / $143$ & $1.2$\,ms ($1.0\times B_{\mathrm{dir}}$) \\
$X=\Lambda$ (realized leak) & $162/162$ & $0.031$ / $0.31$ & $0$ / $72$ / $82$ / $105$ & $2.3$\,ms ($1.05\times$) \\
$X=L_i$ (confined bound) & $162/162$ & $0.094$ / $0.55$ & $0$ / $45$ / $57$ / $78$ & $3.7$\,ms ($1.6\times$) \\
$X=\tax$ (sharp tax) & $162/162$ & $0.273$ / $1.22$ & $0$ / $19$ / $29$ / $42$ & $123$\,ms ($28\times$) \\
$X=2n\dinf$ (\cref{cor:2ndelta}) & $162/162$ & $1.007$ / $2.34$ & $0$ / $2$ / $3$ / $6$ & $2.16$\,s ($381\times$) \\
$X=0$ \shortstack[l]{(zero-tax $B$\\of Result~4)} & $162/162$ & $0.037$ / $0.32$ & $0$ / $70$ / $80$ / $99$ & $1.8$\,ms ($0.82\times$) \\
\midrule
$\Delta\Vul$ (cached exact) & $162/162$ & $0$ / $0$ & $73$ / $143$ / $150$ / $157$ & $0.83$\,ms ($0.37\times$) \\
\bottomrule
\end{tabular}
\end{table}

In \cref{tab:usefulness}, Cover counts the instances with $B\ge\Delta\Vul$, Slack is $B-\Delta\Vul$, and Acc.\ $t$ counts the instances with $B\le t$. Time is the median wall-clock time with its paired ratio to full recomputation ($2.2$\,ms). Here $\Delta\Vul$ is $\max_p[A_p(\rs)-A_p(\bp)]$, except for $B_{\mathrm{cache}}$, which bounds $\Vul_G(\rs)-\Vul_G(\bp)$ and reports its time ratio against $B_{\mathrm{dir}}$.

\subsection{Trunk scaling: certificate cost with a fixed subgame in a growing game}\label{app:expdetails:trunk}

To measure when locality pays off, we hold the subgame fixed and grow the surrounding trunk. The base game comes from the synthetic family of \cref{sec:experiments} ($\lambda=0.1$, six seeds, $12$ information sets, resolver $i=\text{seed}\bmod 3$). At the root, a public chance move with $K$ equiprobable outcomes selects one of $K$ distinct copies of this base game. In the first copy, the subgame $S$ is a fixed public subtree with $6$ information sets and $8$ entry histories. The full game scales to $12K$ information sets. We replicate the base game's 400-iteration CFR blueprint across all copies. Each run applies a fixed Dirichlet corruption to the resolver's blueprint inside $S$, followed by an 800-iteration gadget resolve.

We time four evaluation methods in a single process across $K\in\{1,2,4,8,16,32,64\}$, taking the median of seven repeats. The exact evaluations are full recomputation of $\Vul_G$ ($2n$ full-game best responses) and cached exact evaluation ($n-1$ full-game best responses and one expected-utility pass). Next, the whole-tree $B_{\mathrm{dir}}$ implementation scans the full tree to locate entry histories and compute the $n$ expected utilities. Reading cached entry reaches, a local $B_{\mathrm{dir}}$ evaluates the realized gains $g_p=u_p(\rs)-u_p(\bp)$ strictly inside $S$. The two $B_{\mathrm{dir}}$ variants agree to within $10^{-9}$ on all $42$ (seed, $K$) pairs, as do the two exact evaluations, and $B_{\mathrm{dir}}\ge\Delta\Vul$ holds throughout.

In \Cref{tab:trunk}, the local certificate's runtime stays flat at $2.0$\,ms across all $K$. Cached exact evaluation grows linearly with the full game, running slower than the local certificate on all $42$ instances. The whole-tree $B_{\mathrm{dir}}$ grows linearly too. Its entry search and utility passes traverse the full tree, though its slope is smaller than the cached evaluation's because an expected-utility pass is cheaper than a best response. Since the subgame carries $1/(2K)$ of the blueprint's reach, $\Delta\Vul$ and the bounds scale down as $1/K$, so we report runtime only.

\begin{table}[ht]
\centering
\caption{Trunk scaling, median wall-clock time over six seeds for four evaluation methods.}
\label{tab:trunk}
\footnotesize
\begin{tabular}{rrrrrrrr}
\toprule
$K$ & Infosets & Full & \shortstack{Cached\\exact} & $B_{\mathrm{dir}}$, whole tree & $B_{\mathrm{dir}}$, local & \shortstack{Whole tree\\/ cached} & \shortstack{Local\\/ cached} \\
\midrule
$1$ & $12$ & $5.5$\,ms & $2.1$\,ms & $2.8$\,ms & $2.0$\,ms & $1.35\times$ & $0.96\times$ \\
$2$ & $24$ & $11.1$\,ms & $4.1$\,ms & $4.3$\,ms & $2.0$\,ms & $1.06\times$ & $0.49\times$ \\
$4$ & $48$ & $22.4$\,ms & $8.1$\,ms & $7.6$\,ms & $2.0$\,ms & $0.93\times$ & $0.25\times$ \\
$8$ & $96$ & $46.4$\,ms & $17.1$\,ms & $13.8$\,ms & $2.0$\,ms & $0.83\times$ & $0.12\times$ \\
$16$ & $192$ & $90.9$\,ms & $32.8$\,ms & $27.6$\,ms & $2.0$\,ms & $0.82\times$ & $0.061\times$ \\
$32$ & $384$ & $189$\,ms & $70.6$\,ms & $54.4$\,ms & $2.0$\,ms & $0.77\times$ & $0.029\times$ \\
$64$ & $768$ & $401$\,ms & $145$\,ms & $111$\,ms & $2.0$\,ms & $0.76\times$ & $0.014\times$ \\
\bottomrule
\end{tabular}
\end{table}

In \cref{tab:trunk} the subgame is fixed at $6$ information sets and the full game grows to $12K$ information sets. Ratios are medians of paired per-seed ratios against the cached exact evaluation.

\section{Definitional Conventions: Implementation versus Theorem}\label{app:calibers}

\Cref{tab:calibers} lists six conventions where the experimental implementation and the theorem definitions of \cref{sec:prelim} differ. Re-implementations of the certificates and of $\dinf$ follow the theorem column, while comparisons against the implementation follow the implementation column.

\begin{table}[ht]
\centering
\caption{Definitional conventions of the implementation and of the theorem.}
\label{tab:calibers}
\footnotesize
\begin{tabular}{@{}lp{5.0cm}p{5.0cm}p{4.3cm}@{}}
\toprule
ID & Implementation & Theorem convention & Consequence \\
\midrule
D-a & Counterfactual action values \emph{normalized} by the weight sum & $\Mdef{q}$ aggregates with unnormalized weights $\bar W_q(\iota)$ & Per-infoset margin signs agree, while cross-infoset aggregation differs \\
D-b & Margin's Follow-branch valued under the opponent's own \emph{average strategy} & Margin $\mhat$ valued against the exact subgame \emph{best response} & Implementation margin optimistic on 33/40 infosets, worst overestimate $+1.667$ \\
D-c & Follow-branch evaluated under the gadget-internal measure, which depends on the opponent \emph{ordering} in the gadget chain (opponents' gadgets nested one inside the next) & Both branches evaluated under the game-side measure $c_e\,\pi^{\bp}_{-j}(e)$ and environment $(\rs_i, \bp_{-ij})$ & Measures coincide only when entries follow a public chance node with no trunk actions and the other opponents play Terminate purely (the benchmark families) \\
D-d & Zero-denominator entry infosets fall back to a raw mean (flagged) & Assumes $\bar W_q(\iota) > 0$ & Such $\iota$ carry weight $0$ in $\Mdef{q}$ \\
D-e & $\dinf$ LP defined on the reduced pure-strategy \emph{profile table} & Same & A literal terminal-history-dependent witness would force $\dinf \equiv 0$ on all zero-sum games, contradicting the Kuhn bound \\
D-f & LP constraint is per-edge \emph{constant-sum}: $u_{pq}(a,b) + u_{qp}(b,a) = c_{pq}$ & Same (constant-sum) & Proofs use only that $\sum_p u_p$ is constant on the class \\
\bottomrule
\end{tabular}
\end{table}

\textbf{D-a (normalization).}
The implementation divides counterfactual action values by the weight sum. Within each information set, this leaves the sign of the margin unchanged because the Terminate and Follow branches are divided by the same positive number. Across information sets, the aggregate differs. The deficit $\Mdef{q}$ is the unnormalized $\bar W_q(\iota)$-weighted sum. When all pre-entry actions are public, each opponent's own action sequence is identical across entries, so $\sum_\iota \bar W_q(\iota) \le 1$ and $\Mdef{q} \le (-\min_\iota \mhat_q(\iota))_+$ connects the deficit to the min-margin convention (\cref{app:L3}). Private pre-entry branching can push the sum above $1$.

\textbf{D-b (average-strategy margins are optimistic).} The implementation values an opponent's Follow branch under that opponent's CFR \emph{average strategy}, while the certificate requires the Follow-branch \emph{best-response} value. With environment and measure fixed, the average-strategy value can only understate the best response. Implementation margins are thus optimistic. The effect is worst when the opponent plays Terminate almost purely in the gadget, leaving its Follow-branch behavior off-path and untrained. Across the benchmark families, the implementation margin meets or exceeds the BR margin on 33 of 40 entry information sets. The largest overestimate is $+1.667$, occurring at the information set holding the highest card in pairwise-showdown. The 7 reversals come from the D-c measure gap and have magnitude $\le 1.2 \times 10^{-3}$, so the net direction is not pointwise constant. We recompute $\mhat$ a posteriori via one exact subgame best response per opponent.

\textbf{D-c (gadget-internal measure depends on opponent order).}
The implementation evaluates Follow-branch values under the gadget's internal measure. Root chance is proportional to $c_e \cdot \pi_i^{\bp}(e)$ (excluding other opponents' trunk reach) times the Terminate and Follow mixing probabilities of earlier opponents in the nesting chain. This measure therefore depends on the opponent chaining order, and the other opponents play their gadget strategies in the subgame. The theorem uses the game-side measure $c_e\,\pi^{\bp}_{-j}(e)$ with environment $(\rs_i, \bp_{-ij})$. The two coincide when entries immediately follow a public chance node, trunk actions are absent, and other opponents play Terminate purely, as in the benchmark families. They differ in general. For the Terminate-branch value (the blueprint counterfactual value), the implementation uses the game-side measure, and $\mhat$ puts both branches on that measure.

\textbf{D-d (zero-denominator entries).}
Entry information sets with zero counterfactual weight fall back to a raw unweighted mean and are flagged. They contribute weight $0$ to $\Mdef{q}$.

\textbf{D-e ($\dinf$ is defined by LP profile-table semantics).}
If the pairwise witness $u_{ij}$ depended arbitrarily on the terminal history $z$, every pointwise zero-sum game would admit $u_{ij} := (u_i - u_j)/n$ and $\dinf \equiv 0$. This contradicts the 3-player Kuhn bound $\dinf \ge 1$. \Cref{def:delta} places the witness table on reduced pure-strategy \emph{pairs} $(\rho_p, \rho_q)$ after taking expectation over chance and third parties. The witness may still depend on third-party randomness through the chance expectation, while remaining independent of third-party \emph{strategies}.

\textbf{D-f (per-edge constant-sum).}
The LP implements the constant-sum polymatrix (CSP) class of the literature, $u_{pq}(a, b) + u_{qp}(b, a) = c_{pq}$, with edge constants $c_{pq}$ as free LP variables. This class contains the antisymmetric class $u_{pq} + u_{qp} = 0$. The proofs use only that $\sum_p u_p$ is constant on the class. At $\dinf = 0$, the conclusion $\sum_p g_p = 0$ follows because the constant cancels in differences. The benchmark $\dinf = 0$ families are also pointwise zero-sum on terminals.

\end{document}